\documentclass[fleqn,usenatbib]{mnras}

\usepackage{newtxtext,newtxmath}
\usepackage[T1]{fontenc}
\usepackage{ae,aecompl}

\usepackage{graphicx}	% Including figure files
\usepackage{amsmath}	% Advanced maths commands
\usepackage{amssymb}	% Extra maths symbols
\usepackage{multirow}
\usepackage{array,booktabs}
\usepackage{nccmath}
\usepackage{gensymb}
\usepackage[normalem]{ulem}
\usepackage[dvipsnames]{xcolor}
\newcommand{\Msun}{{M}_{\odot}}
\title[ASL for multi-D simulations]{A multi-dimensional
implementation of the Advanced Spectral neutrino Leakage scheme}
\author[D.Gizzi]{
D. Gizzi,$^{1}$\thanks{E-mail: davide.gizzi@astro.su.se}
E. O'Connor,$^{1}$
S. Rosswog,$^{1}$
A. Perego,$^{2,3}$
R. M. Cabez\'on,$^{4}$
L. Nativi$^{1}$
\\
$^{1}$Department of Astronomy \& The Oskar Klein Centre, Stockholm University, AlbaNova, 106 91, Stockholm, Sweden\\
$^{2}$Dipartimento di Fisica, Universit\`a degli Studi di Trento, via Sommarive 14, 38123 Trento, Italy\\
$^{3}$INFN, Sezione di Milano-Bicocca, Piazza della Scienza 3, 20126 Milano, Italy\\
$^{4}$Center for Scientific Computing - sciCORE, University of Basel, Klingelbergstrasse 61, CH-4056 Basel, Switzerland
}

\date{Accepted XXX. Received YYY; in original form ZZZ}

\pubyear{2019}

\begin{document}
\label{firstpage}
\pagerange{\pageref{firstpage}--\pageref{lastpage}}
\maketitle

\begin{abstract}
We present a new, multi-dimensional 
implementation of
the Advanced Spectral Leakage (ASL) scheme with
the purpose of modelling neutrino-matter interactions in
neutron star mergers. A major challenge is the neutrino
absorption in the semi-transparent regime, which 
is responsible for driving winds from the merger remnant.
The composition of such winds is crucial in the understanding of the electromagnetic emission in the recently observed macronova following GW170817. 
Compared to the original version, 
we introduce an optical-depth-dependent flux factor to
model the average angle of neutrino propagation, and
a modulation that accounts for flux anisotropies in
non-spherical geometries. We scrutinise our approach by
first comparing the new scheme against the original one
for a spherically symmetric core-collapse supernova
snapshot, both in 1D and in 3D, and additionally
against a two-moment (M1) scheme 
as implemented in 1D into the code GR1D. 
The luminosities and
mean energies agree to a few percents in most tests.
Finally, for the case of a binary merger remnant snapshot we compare the new ASL 
scheme with the M1 scheme that is implemented in the 
Eulerian adaptive mesh refinement
code FLASH.
We find that the neutrino absorption distribution
in the semi-transparent regime is overall well reproduced.
Both approaches agree to within $\lesssim 15\%$ for the
average energies and to better than $\sim 35 \%$ in the total luminosities.
\end{abstract}


%%%%%%%%%%%%%%%%%%%%%%%%%%%%%%%%%%%%%%%%%%%%%%%%%%

%%%%%%%%%%%%%%%%% BODY OF PAPER %%%%%%%%%%%%%%%%%%

\section{Introduction}

The first multi-messenger detection of a neutron star merger
\citep{Abbott2017c} has brought major leaps forwards for many areas
of (astro)physics. For example, the 1.7s delay between the 
gravitational wave (GW) peak and the gamma-rays from an event
detected by the Fermi satellite \citep{Goldstein2017b} allowed to
constrain the deviations of the GW propagation speed from the 
speed of light to 1 part in $10^{15}$ \citep{Abbott2017c}. 
The detection further allowed for an independent measure 
of the Hubble parameter \citep{Abbott2017b} as suggested by
\cite{Schutz86}. The GW signal was followed by emission all 
across the electromagnetic (EM) spectrum
\citep[e.g.][]{Arcavi2017,Chornock17,Kasliwal17,Kilpatrick17,Kasen2017,Pian17,Smartt17,Abbott2017c,Abbott2017a,Goldstein2017b,Savchenko2017,Coulter2017,Troja2017,Margutti2017,Haggard2017,Alexander2017,Hallinan2017,Tanvir17}.
The intensity of the EM emission 
detected in the aftermath of the event
decayed with a power-law exponent close to
$-1.3$  \citep{Kasliwal17,Rosswog18a} as expected for
a distribution of freshly synthesised r-process elements
\citep{Metzger10b,Korobkin12a}.  Estimates of the involved ejecta
masses point to $\sim 0.02\:\Msun$ for the early blue emission 
component and $\sim 0.04\:\Msun$ for the later emerging red component \citep{Villar2017,Kasen2017,Perego2017,Rosswog18a}.
The early blue component requires lanthanide-free ejecta which, in turn,
are the r-process nucleosynthesis result of matter with $Y_e \gtrsim 0.25$ \citep{Korobkin12a,Kasliwal19} (ejected
at velocities of $\sim 0.3$c).  The later emerging, red component 
stems from matter with electron fractions below this threshold value.
Since the original neutron stars are in $\beta$-equilibrium
they contain only about $10^{-4}\:\Msun$ of matter with 
$Y_e > 0.25$. Therefore, the
observed $\sim$ 2\% of a solar mass in the
blue component point to a major re-processing of a large
fraction of the ejecta by weak interactions, raising 
$Y_e$ via $e^+ + n \rightarrow p + \bar{\nu}_e$
and $\nu_e + n \rightarrow p + e^-$. With GW170817 and its EM emission we have thus
witnessed weak interaction "in flagranti".
This underlines the paramount importance of
carefully modelling weak interactions and neutrino 
physics in a neutron star merger
for reliable predictions of their EM
signature.\\
Addressing the neutrino transport problem by solving the full 
multi-dimensional Boltzmann equation \citep{Lindquist1966} is 
computationally very demanding and for most astrophysical problems
it is prohibitively expensive. Therefore, most multi-dimensional 
hydrodynamic studies, both for supernovae and compact binary mergers
resort to transport approximations
\citep[e.g.][]{Thorne1981,Bruenn1978,Mezzacappa1999,Bruenn1985,Rosswog2003,Buras2006,Oconnor2018,Dessart2009,Foucart2016a,Perego2017b,Ardevol19,Cabezon2018}.
Our particular focus here is on 
the Advanced Spectral Leakage (hereafter ASL)
\citep{Perego2016}, that has 
recently been scrutinised against more 
expensive neutrino treatments \citep{pan19} in
a core-collapse supernova context. 
Since supernovae are roughly spherically
symmetric, they allow for approximations that
are not admissible in a neutron star merger
context. In this paper we extend the original 
ASL scheme to multi-dimensional applications,
while keeping the general structure 
of the equations as presented in
the original papers
\citep{Perego2014,Perego2016}. We 
examine the modified scheme in typical core-collapse supernova and
neutron star merger remnant
snapshots. We focus on the modelling of the
absorption in the semi-transparent regime, 
which is responsible for the neutrino-driven winds 
\citep{Perego2014,Radice2018b}, 
one of the possible
ejection channels related to
the observed blue EM component.
We rely on a spectral treatment of 
the neutrino-matter interactions, a 
key ingredient for capturing the
composition of the polar ejecta 
\citep{Foucart2016a}. The ASL 
presented here allows for a 
computationally-inexpensive, 
spectral treatment of the neutrino
absorption in the semi-transparent
regime, and it is therefore suitable
for long-term binary merger simulations,
where more detailed neutrino treatments
require larger computational resources.\\
The paper is structured as follows: in Sec.~\ref{sec:meth} we 
describe the ASL methodology
both in its original 1D version 
(Sec.~\ref{sec:1Dmeth})
and in our new 
multi-D implementation (Sec.~\ref{sec:3Dmeth}). 
Simulation results are presented 
in Sec.~\ref{sec:results}. In
Sec.~\ref{sec:ASLvsGR1D} we start with
a one-dimensional core-collapse 
profile.
We then move to three-dimensional configurations
in Sec.~\ref{sec:ASL3D} where we first use the same
core-collapse supernova profile to inspect our 
multi-D implementation of the 
ASL scheme. 
Finally, we apply the ASL to 
a neutron star merger remnant. 
In all cases, we scrutinise the ASL scheme by
comparison with a two-moment (M1) scheme, 
and we neglect relativistic effects
everywhere.
In Sec.~\ref{sec:conclusion} we summarise our results.

\section{The Advanced Spectral Leakage}
\label{sec:meth}
\subsection{1D implementation}
\label{sec:1Dmeth}
We first summarise the most relevant 
features of the ASL scheme for spherically 
symmetric systems, as they are described in \cite{Perego2016}. In their work,
the ASL scheme is explored both in 1D 
and multi-D spherically symmetric
core-collapse setups,
showing flexibility and overall agreement
with other neutrino transport models.\\
At the heart of the ASL approach is 
a spectral (i.e. energy-dependent) 
description of 
neutrino transport in 
which a neutrino energy spectrum
is initially set up to 
account for the energy-squared  
dependence of neutrino-matter
interactions.
As in most approximate treatments, 
we model neutrinos as three independent species:
electron neutrinos $\nu_{\rm{e}}$, electron
anti-neutrinos $\Bar{\nu}_{\rm{e}}$, and
a collective species for 
heavy-lepton neutrinos and 
anti-neutrinos $\nu_{\rm{x}}$.
For the interactions between neutrinos and 
matter we consider the production and absorption
of electron neutrinos and anti-neutrinos via charged current processes involving 
nucleons and nuclei, neutrino emission
by bremsstrahlung and pair processes,
and finally the scattering off nucleons and
nuclei. These reactions enter the 
computation of
the local optical depth $\tau_{\nu}(E,\textbf{x})$
for an energy $E$ at position \textbf{x}, 
which is a measure of the 
average number of interactions
a neutrino experiences before 
escaping to infinity
and defined as integral of the inverse
local mean free path $\lambda_{\nu}(E,\textbf{x'})$
over a path $\gamma$
\begin{ceqn}
\begin{equation}
    \label{eq:opdep}
    \tau_{\nu}(E,\textbf{x})= \int_{\gamma:\textbf{x}\rightarrow+\infty}
    \frac{1}{\lambda_{\nu}(E,\textbf{x'}(s))}{\rm d}s.
\end{equation}
\end{ceqn}
Two different optical depths are defined: 
the first is the \textit{total optical depth} 
$\tau_{\rm{\nu,tot}}$ 
where both absorption and elastic scattering 
interactions are equally considered in the inverse mean free 
path calculation.
The second is the \textit{energy optical depth}
$\tau_{\rm{\nu,en}}$, which is related to
the mean free path over which neutrinos
can exchange energy with the fluid. 
An analytical estimate of the latter is given 
by computing the geometric mean between the total
and the absorption inverse mean free paths:
\begin{ceqn}
\begin{equation}
\lambda_{\nu,\mathrm{en}}(E,\textbf{x})= \sqrt{\frac{c\:\lambda_{\nu,\mathrm{tot}}(E,\textbf{x})}{\sum_{\mathrm{s}} \chi_{\nu,\mathrm{ab,s}}(E,\textbf{x})}},
\end{equation}
\end{ceqn}
where $\chi_{\nu,\rm{ab,s}}(E,\textbf{x})$ is the absorpitivity 
of the absorption process 's' and $c$ is the speed of light.
Each optical depth defines 
a \textit{neutrino surface} at 
$\tau_{\rm{\nu}}= 2/3$, where neutrinos begin to
decouple from matter.

The net specific\footnote{To be explicit:
we always use "specific" for quantities 
on a per mass basis.},
spectral\footnote{We always use
"spectral" for quantities in units of $E^2dE$.}
neutrino emission rate
(units of $\rm{s^{-1}g^{-1}erg^{-3}}$)
is initially calculated as a smooth 
interpolation between the 
production $r_{\rm{\nu,prod}}(E,\textbf{x})$ 
and diffusion
$r_{\rm{\nu,diff}}(E,\textbf{x})$ rate
\begin{ceqn}
\begin{equation}
\label{eq:emrate}
    \Tilde{r}_{\nu}(E,\textbf{x})= \frac{r_{\mathrm{\nu,prod}}(E,\textbf{x})\: r_{\mathrm{\nu,diff}}(E,\textbf{x})}{r_{\mathrm{\nu,prod}}(E,\textbf{x})+r_{\mathrm{\nu,diff}}(E,\textbf{x})}.
\end{equation}
\end{ceqn}
where $r_{\rm{\nu,prod}}(E,\textbf{x})$ 
depends on the production timescale $t_{\nu,\rm{prod}}$,
which in turn is set by the local 
emissivity, while 
$r_{\rm{\nu,diff}}(E,\textbf{x})$
depends on the timescale over
which neutrinos diffuse out
of the system, $t_{\nu,\rm{diff}}$. 
This timescale
is set by the local opacity via 
$\sim \tau_{\rm{\nu,tot}}^2(E,\textbf{x})$.
Eq.~(\ref{eq:emrate}) 
favours $r_{\rm{\nu,diff}}(E,\textbf{x})$ 
in optically thick conditions
($\tau_{\nu,\rm{tot}}(E,\textbf{x}) \gg 1$) and 
$r_{\rm{\nu,prod}}(E,\textbf{x})$ in
optically thin conditions 
($\tau_{\nu,\rm{tot}}(E,\textbf{x}) \lesssim 1$).
We add two further corrections. 
First, when a large amount of neutrinos
is emitted at the neutrino surface or 
is locally produced, Pauli blocking
occurs as a consequence
of the fermionic nature of neutrinos.
Second, emission in optically
thin regimes provided by $r_{\rm{\nu,prod}}$
is assumed isotropic, and a fraction 
of neutrinos are emitted toward the optically
thick regime. 
To account for these effects,
neutrino emission is
reduced by introducing a Pauli 
blocking parameter $\alpha_{\nu,\rm{blk}}$:
$\Tilde{r}_{\nu} \rightarrow (1-\alpha_{\nu,\rm{blk}})\Tilde{r}_{\nu}$.
Second, during the 
diffusion process in the optically thick 
regime, neutrinos thermalize to lower energies 
and therefore the spectrum at the neutrino
surface is softened. The softening of the spectrum 
is included via the term $\frac{1}{\Psi_{\nu}(\textbf{x})}
\rm{exp}(-\tau_{\nu,\rm{en}}(E,\textbf{x})/\tau_{\rm{cut}})$,
with $\Psi_{\nu}(\textbf{x})$ defined as
\begin{ceqn}
\begin{equation}
    \Psi_{\nu}(\textbf{x})= \frac{\int_0^{+\infty}\Tilde{r}_{\nu}(E,\textbf{x})e^{-\tau_{\nu,\mathrm{en}}(E,\textbf{x})/\tau_{\mathrm{cut}}}E^2{\rm d}E}{\int_0^{+\infty}\Tilde{r}_{\nu}(E,\textbf{x})E^2{\rm d}E},
\end{equation}
\end{ceqn}
where $\tau_{\rm{cut}}$ parametrizes the typical number of interactions required to thermalize neutrinos.
The equation for the neutrino emission rate finally becomes
(we will occasionally refer to this
emission as \textit{cooling})
\begin{equation}
\label{eq:ser1D}
    r_{\nu}(E,\textbf{x})= (1 - \alpha_{\nu,\mathrm{blk}})\Tilde{r}_{\nu}(E,\textbf{x})\frac{1}{\Psi_{\nu}(\textbf{x})}\mathrm{exp}(-\tau_{\nu,\mathrm{en}}(E,\textbf{x})/\tau_{\mathrm{cut}}).
\end{equation}
The values of 
$\alpha_{\nu,\rm{blk}}$ and $\tau_{\rm{cut}}$
are geometry-dependent, and must ideally be calibrated
every time the system geometry changes significantly over
the time of the simulation. A good trade-off is to provide
fixed values for these parameters that are able to approximately
reproduce the neutrino properties dynamically in comparison to 
other transport approaches. So far, 
$\alpha_{\nu,\rm{blk}}$ and 
$\tau_{\rm{cut}}$ have been calibrated in 
the context of spherically symmetric 
core-collapse supernovae simulations 
against full Boltzmann neutrino transport
\citep{Perego2016}.
Electron neutrinos and anti-neutrinos 
have $\alpha_{\nu,\rm{blk}} \sim 0.55$.
Heavy-lepton neutrinos are generally
subdominant and their emission in optically
thin regime is negligible, therefore 
$\alpha_{\nu,\mathrm{blk}} \sim 0$.
For the thermalization coefficient we adopt $\tau_{\rm{cut}}=20$
for all neutrino species.
Although a new calibration in the context of binary merger simulations would be preferred, for the time being we
assume the same values adopted for core-collapse
simulations, leaving the detailed 
binary merger calibration task to a future work.\\
The absorption of neutrinos in the optically thin
regime is hereafter referred to as \textit{heating}, 
and the specific, spectral absorption rate
(units of $\rm{s^{-1}g^{-1}erg^{-3}}$)
is defined as
\begin{ceqn}
\begin{equation}
    h_{\mathrm{\nu}}(E,\textbf{x})= \frac{1}{\rho(\textbf{x})}\:n_{\mathrm{\nu,\tau \lesssim 1}}\:\chi_{\mathrm{\nu,ab}}\:F_{\mathrm{e^{\mp}}}\:H,
    \label{eq:heating}
\end{equation}
\end{ceqn}
where $\rho(\textbf{x})$ is the mass density of the fluid 
at position \textbf{x}, $\chi_{\rm{\nu,ab}}$ the 
absorpitivity, $n_{\rm{\nu,\tau \lesssim 1}}$ the 
neutrino number density in optically thin regime, $H=
\rm{ exp}(-\tau_{\rm{\nu,tot}})$ 
ensures the heating to be applied only in 
the optically thin regime.
All quantities on the RHS of Eq.~(\ref{eq:heating}) are functions
of energy \textit{E} and position
\textbf{x}.
The Pauli blocking factor
for final state electrons and positrons 
is given by
\begin{ceqn}
\begin{equation}
    F_{\mathrm{e^{\mp}}}= 1 - \frac{1}{\mathrm{exp}((E\pm Q \mp \mu_e)/k_BT)+1},
\end{equation}
\end{ceqn}
where $k_B$ is the 
Boltzmann constant, $Q \approx 1.293$ MeV
is the difference between 
neutron and proton rest mass energy, 
$\mu_e$ is the
electron chemical potential and $T$ is the
fluid temperature.
The form of $n_{\rm{\nu,
\tau \lesssim 1}}$ for a spherically symmetric 
heating is
\begin{ceqn}
\begin{equation}
    n_{\mathrm{\nu,\tau \lesssim 1}}(E,R)= \frac{l_{\mathrm{\nu}}(E,R)}{4\pi R^2\:c\: \mu_{\mathrm{\nu}}(E,R)},
    \label{eq:nudensity}
\end{equation}
\end{ceqn}
where $l_{\rm{\nu}}(E,R)$ is the total, 
spectral number 
rate (in $\rm{s^{-1}erg^{-3}}$)
at radius $R$ obtained as solution of a differential 
equation that accounts for both 
emission and absorption of neutrinos while they 
propagate from the centre of the system to a distance $R$
\begin{equation}
    \frac{dl_{\nu}(E,R)}{dR}= 4\pi R^2\rho(R)r_{\nu}(E,R) - \frac{\chi_{\rm{ab}}(E,R)}{c}H(E,R)l_{\nu}(E,R).
    \label{eq:spartlm}
\end{equation}
In Eq.~(\ref{eq:nudensity}) $\mu_{\rm{\nu}}(E,R)$ 
is called flux factor.
It corresponds to the average 
of the cosine of the propagation angle for the
free streaming neutrinos. 
An analytic approximation is given by
\cite{Liebendorfer2009}
\begin{ceqn}
\begin{equation}
    \mu_{\mathrm{\nu}}(E,R)= \frac{1}{2}\Bigg(1+\sqrt{1-\Big(\frac{R_{\mathrm{\nu}}(E)}{\mathrm{max}(R,R_{\mathrm{\nu}}(E))}\Big)^2}\Bigg),
    \label{eq:fluxf}
\end{equation}
\end{ceqn}
where $R_{\rm{\nu}}(E)$ is the neutrino surface 
radius for energy $E$. Far from the neutrino surface 
($R \gg R_{\nu}$) the neutrino flux points 
toward the observer direction and the propagation  
angle is 0, i.e. $\mu_{\rm{\nu}}(E,R)= 1$.
Close to the neutrino surface ($R \sim R_{\nu}$)
and assuming isotropic neutrino emission
above the plane 
tangential to it
$\mu_{\rm{\nu}}(E,R) \sim 1/2$.

Given the spectral, specific rates 
$r_{\nu}(E,\textbf{x})$ and 
$h_{\nu}(E,\textbf{x})$ at each point from
Eqs.~(\ref{eq:ser1D}) and (\ref{eq:heating}),
the energy-integrated emission and 
absorption specific rates are
\begin{ceqn}
\begin{align}
    R_{\mathrm{\nu}}^{k}(\textbf{x})&= \int_0^{+\infty} r_{\mathrm{\nu}}(E,\textbf{x})\:E^{2+k}\:{\rm d}E,
    \label{eq:emission}\\
    H_{\mathrm{\nu}}^{k}(\textbf{x})&= \int_0^{+\infty} h_{\mathrm{\nu}}(E,\textbf{x})\:E^{2+k}\:{\rm d}E,
    \label{eq:absorption}
\end{align}
\end{ceqn}
respectively, where $k = 0$ 
specifies the number rate 
($\rm{g^{-1}s^{-1}}$) and $k = 1$ 
the energy rate (erg $\rm{g^{-1}s^{-1}}$). Eqs.~(\ref{eq:emission}) and
(\ref{eq:absorption}) define 
the specific number and energy net rates $\Dot{Q}_{\mathrm{\nu}}^{k = 0}(\textbf{x})$ and  $\Dot{Q}_{\mathrm{\nu}}^{k = 1}(\textbf{x})$
\begin{ceqn}
\begin{align}
    \Dot{Q}_{\mathrm{\nu}}^{k = 0}(\textbf{x})= R_{\mathrm{\nu}}^{k = 0}(\textbf{x}) - H_{\mathrm{\nu}}^{k = 0}(\textbf{x}),\\
    \Dot{Q}_{\mathrm{\nu}}^{k = 1}(\textbf{x})= R_{\mathrm{\nu}}^{k = 1}(\textbf{x}) - H_{\mathrm{\nu}}^{k = 1}(\textbf{x}),
    \label{eq:ennetrate}
\end{align}
\end{ceqn}
from which the total neutrino number net rate $L_{\nu}^{k = 0}$ 
and the neutrino luminosity $L_{\nu}^{k = 1}$ 
can be derived by integrating over the volume $V$
of the fluid
\begin{ceqn}
\begin{align}
    L_{\mathrm{\nu}}^{k = 0}&= \int_V\Dot{Q}_{\mathrm{\nu}}^{k = 0}(\textbf{x})\rho(\textbf{x}){\rm d}V,\\
    L_{\mathrm{\nu}}^{k = 1}&= \int_V\Dot{Q}_{\mathrm{\nu}}^{k = 1}(\textbf{x})\rho(\textbf{x}){\rm d}V.
    \label{eq:Len}
\end{align}
\end{ceqn}
From the last two equations
the neutrino average energy is
calculated as
\begin{ceqn}
\begin{equation}
    \langle{}E_{\nu}\rangle{}= \frac{L_{\mathrm{\nu}}^{k = 1}}{L_{\mathrm{\nu}}^{k = 0}}
    \label{eq:meanen}.
\end{equation}
\end{ceqn}
The root-mean squared (rms) energy 
can be defined too as
\begin{ceqn}
\begin{equation}
    E_{\mathrm{rms}}= \sqrt{\frac{L_{\mathrm{\nu}}^{k = 2}}
    {L_{\mathrm{\nu}}^{k = 0}}}
    \label{eq:rms}.
\end{equation}
\end{ceqn}
In Sec~(\ref{sec:results}) we will mainly
refer to Eq.~(\ref{eq:meanen}) to describe
the neutrino energy, but we 
additionally provide values for the rms
energies for completeness.
From Eqs.~(\ref{eq:emission}) and (\ref{eq:absorption}) 
we can also recover the local net 
change in the total lepton number fraction $\Dot{Y}_{\rm{l}}(\textbf{x})$
\begin{ceqn}
\begin{equation}
\label{eq:Yldot}
    \Dot{Y}_{\mathrm{l}}(\textbf{x})= m_{\mathrm{b}}\:\big(H_{\mathrm{\nu_e}}^{k = 0}(\textbf{x}) - H_{\mathrm{\bar{\nu}_e}}^{k = 0}(\textbf{x}) + R_{\mathrm{\bar{\nu}_e}}^{k = 0}(\textbf{x}) - R_{\mathrm{\nu_e}}^{k = 0}(\textbf{x})\big)
\end{equation}
\end{ceqn}
where $m_{\rm{b}}$ is the baryon mass, 
and the change of
the total specific 
matter internal energy $\Dot{u}(\textbf{x})$
\begin{ceqn}
\begin{equation}
\label{eq:udot}
\begin{split}
    \Dot{u}(\textbf{x})= &
    - \big(R_{\mathrm{\nu_e}}^{k = 1}(\textbf{x}) +  R_{\mathrm{\bar{\nu}_e}}^{k = 1}(\textbf{x}) +  4R_{\mathrm{\bar{\nu}_{\mu,\tau}}}^{k = 1}(\textbf{x})\big)  + \\
    & \quad 
    + \big(H_{\mathrm{\nu_e}}^{k = 1}(\textbf{x}) + H_{\mathrm{\bar{\nu}_e}}^{k = 1}(\textbf{x})\big).
\end{split}
\end{equation}
\end{ceqn}
Both $\Dot{Y}_l(\textbf{x})$ and $\Dot{u}(\textbf{x})$
contain variations in the lepton number fraction and
specific internal energy by local emission and 
absorption of neutrinos and by neutrino diffusion
that would dominate in the optically thick regime. 
Denoting the variation of the number 
and energy of trapped neutrinos per baryon
as $\Dot{Y}_{\rm{\nu}}(\textbf{x})$ and 
$\Dot{Z}_{\rm{\nu}}(\textbf{x})$ driven 
by diffusion, we can recover the change 
in the electron 
fraction $\Dot{Y}_{\rm{e}}(\textbf{x})$ 
\begin{ceqn}
\begin{equation} 
\label{eq:yedot}
    \Dot{Y}_{\mathrm{e}}(\textbf{x})= \Dot{Y}_{\mathrm{l}}(\textbf{x}) - \Dot{Y}_{\mathrm{\nu_e}}(\textbf{x}) + \Dot{Y}_{\mathrm{\bar{\nu}_e}}(\textbf{x})
\end{equation}
\end{ceqn}
and the rate of change of 
the specific internal energy
due to local neutrino 
emission and absorption 
$\Dot{e}(\textbf{x})$
\begin{ceqn}
\begin{equation}
    \Dot{e}(\textbf{x})= 
    \Dot{u}(\textbf{x}) - \frac{1}{m_{\mathrm{b}}}\:\big(\Dot{Z}_{\mathrm{\nu_e}}(\textbf{x}) 
    + \Dot{Z}_{\mathrm{\bar{\nu}_e}}(\textbf{x}) + 4\Dot{Z}_{\mathrm{\nu_{\mu,\tau}}}(\textbf{x})\big).
\end{equation}
\end{ceqn}
$\Dot{Y}_{\rm{\nu}}(\textbf{x})$ and 
$\Dot{Z}_{\rm{\nu}}(\textbf{x})$ 
are evaluated at first order in time as
\begin{ceqn}
\begin{align}
\Dot{Y}_{\nu}(\textbf{x})&= \frac{Y_{\nu,t+\Delta t}(\textbf{x})-Y_{\nu}(\textbf{x})}{\Delta t},\\
\Dot{Z}_{\nu}(\textbf{x})&= \frac{Z_{\nu,t+\Delta t}(\textbf{x})-Z_{\nu}(\textbf{x})}{\Delta t},
\end{align}
\end{ceqn}
where $\Delta t$ is the current time step. 
An adequate time step should ensure that the 
variation of all the  variables for which 
the ASL provide a source term in the 
hydrodynamic equations is less than a given, 
small percentage (namely $\lesssim 1\%$).
The number 
and energy of trapped neutrinos per baryon
at time $t$ and location $\textbf{x}$ are related to
the neutrino-trapped distribution function
$f_{\nu,t}^{\rm{tr}}(E,\textbf{x})$ by 
\begin{ceqn}
\begin{align}
\label{eq:trapped1}
    Y_{\nu,t}(\textbf{x})&=
    \frac{4\pi}{(hc)^3}\frac{m_{\rm{b}}}{\rho(\textbf{x})}
    \int f_{\nu,t}^{\rm{tr}}(E,\textbf{x})
    E^2{\rm d}E,\\
\label{eq:trapped2}
    Z_{\nu,t}(\textbf{x})&=
    \frac{4\pi}{(hc)^3}\frac{m_{\rm{b}}}{\rho(\textbf{x})}
    \int f_{\nu,t}^{\rm{tr}}(E,\textbf{x})
    E^3{\rm d}E,
\end{align}
\end{ceqn}
where $h$ is the Planck constant.
Starting from $Y_{\nu,t}(\textbf{x})$ and $Z_{\nu,t}(\textbf{x})$, 
$f_{\nu,t}^{\rm{tr}}(E,\textbf{x})$ is first recovered
on the basis of equilibrium arguments. In particular, we assume a distribution of the form
\begin{ceqn}
\begin{equation}
     f_{\nu}^{\rm{tr}}(E,\textbf{x})= 
     f_{\nu}^{\rm{eq}}(E,\textbf{x})\big(1-e^{ 
     -\tau_{\nu,\rm{en}}(E,\textbf{x})}\big),
\end{equation}
\end{ceqn}
at time $t$, where $f_{\nu}^{\rm{eq}}(E,\textbf{x})$ is a
Fermi-Dirac distribution
\begin{ceqn}
\begin{equation}
     f_{\nu}^{\rm{eq}}(E,\textbf{x})=
     \frac{1}{e^{(E/ \left( k_{\rm B} T_{\nu}(\textbf{x}) \right)-\eta_{\nu}(\textbf{x}))}+1},
\end{equation}
\end{ceqn}
with $T_{\nu}(\textbf{x})$ being the neutrino
temperature, which is assumed to be equal to the matter
temperature, and 
$\eta_{\nu}(\textbf{x})$ the 
degeneracy parameter, evaluated by assuming
weak equilibrium.
Second, $f_{\nu,t}^{\rm{tr}}(E,\textbf{x})$ is 
evolved between $t$ and $t+\Delta t$
considering production and diffusion of neutrinos
as two competing processes.
Namely, we integrate the equation
\begin{ceqn}
\begin{equation}
  \frac{d f_{\nu}^{\mathrm{tr}}}{dt}=
  \dot{f}_{\nu,\mathrm{prod}}^{\mathrm{tr}} + 
  \dot{f}_{\nu,\mathrm{diff}}^{\mathrm{tr}},
\end{equation}
\end{ceqn}
where
\begin{ceqn}
\begin{equation}
\dot{f}_{\nu,\mathrm{prod}}^{\mathrm{tr}}=
\frac{f_{\nu}^{\mathrm{eq}} - f_{\nu}^{\mathrm{tr}}}{\mathrm{max}(t_{\nu,\mathrm{prod}},\Delta t)}\mathrm{exp}\bigg(-\frac{t_{\nu,\mathrm{prod}}}{t_{\nu,\mathrm{diff}}}\bigg)
\end{equation}
\end{ceqn}
and 
\begin{ceqn}
\begin{equation}
\dot{f}_{\nu,\mathrm{diff}}^{\mathrm{tr}}=
-\frac{f_{\nu}^{\mathrm{tr}}}{\mathrm{max}(t_{\nu,\mathrm{diff}},\Delta t)}\mathrm{exp}\bigg(-\frac{t_{\nu,\mathrm{diff}}}{t_{\nu,\mathrm{prod}}}\bigg).
\end{equation}
\end{ceqn}
At last, $Y_{\nu,t + \Delta t}$ and $Z_{\nu,t + \Delta t}$
are recovered by using $f_{\nu,t + \Delta t}^{\rm{tr}}$ in Eqs.~(\ref{eq:trapped1}) and (\ref{eq:trapped2}).
For more details, see \cite{Perego2016}.\\
Note that in Eqs.~(\ref{eq:Yldot}) 
and (\ref{eq:udot})
we have neglected absorption by 
heavy-lepton neutrinos in 
the semi-transparent regime,
because at the decoupling surfaces 
they do not have
enough energy to produce muons and taus
by charged-current interactions.
Moreover, heating by heavy-lepton 
neutrino annihilation and 
inverse nucleon-nucleon bremsstrahlung
provides negligible contributions 
in the semi-transparent regime, 
since the opacities
are small
\citep{Endrizzi2019}.
Neglecting heating
by heavy-lepton neutrinos is therefore a
reasonable assumption.

\subsection{Multi-D implementation}
\label{sec:3Dmeth}
All physical quantities shown 
in Sec.~\ref{sec:1Dmeth}
are local and independent 
of the system geometry, except for Eqs.~(\ref{eq:opdep}),(\ref{eq:nudensity}) and 
(\ref{eq:fluxf}).
In particular, 
Eq.~(\ref{eq:fluxf}) is straightforward 
to use only in cases where the
neutrino surface is easy to reconstruct. 
While this argument is certainly 
valid for a spherically 
symmetric configuration, for a more 
complex geometry 
like a neutron star merger remnant is not. 
Indeed, given
the presence of a torus
around the central compact object
the neutrino decoupling surface has 
larger radii on the equatorial 
plane than along the polar axis \citep{Dessart2009,Perego2014}. 
In the following, we describe
our implementation of 
Eqs.~(\ref{eq:opdep}),(\ref{eq:nudensity})
and (\ref{eq:fluxf}) to a multi-D 
configuration.

\subsubsection{Optical depth}
The computation of the 
optical depth is performed by taking the minimum among 
values of the optical depth calculated by integrating 
the neutrino mean free path over a set of predefined 
radii. In particular, given a point (x,y,z) 
we consider the following outgoing paths:
\begin{itemize}
    \item fixed (y,z), path along x
    \item fixed (x,z), path along y
    \item fixed (x,y), path along z
    \item fixed x, diagonal path along y,z
    \item fixed y, diagonal path along x,z
    \item fixed z, diagonal path along x,y
    \item diagonal path along x,y,z.
\end{itemize}
The likelihood of being close to
the true minimum optical depth 
at a point increases by increasing number of paths.
For this reason, the choice of diagonal paths ensures 
a more accurate calculation of the optical depth by
avoiding local
overestimates that would arise otherwise.
However, it is important to
stress that this algorithm
leads inevitably to an overestimation
of the local optical depth, 
as a consequence of 
the limited number of paths
that can be practically chosen for calculations.
\subsubsection{Flux factor}
To construct a more general form for 
the flux factor that still 
resembles the general properties of 
Eq.~(\ref{eq:fluxf}) we borrow the linear dependence of the inverse
flux factor from the optical depth from equation~(31) 
of \cite{Oconnor2010}. Although their equation~(31) does 
not consider a spectral distribution of energies, we 
take that form to get an approximate 
expression of the flux factor 
at any energy by just extending the grey interpolation 
formula to a spectral form by adding the energy dependence.
Therefore, we use
\begin{equation}
\label{eq:finalfluxf}
\frac{1}{\mu_{\mathrm{\nu}}}(E,\textbf{x})= \begin{cases} 1.5\:\tau_{\mathrm{\nu,tot}}(E,\textbf{x}) + 1 &\text{if  $\tau_{\mathrm{\nu,tot}}(E,\textbf{x}) \leq 2/3$}\\
2 &\text{$\mathrm{otherwise}$}
\end{cases}.
\end{equation}
Using this expression we mimic Eq.~(\ref{eq:fluxf}) 
with a flux factor $\mu_{\nu}(E,\textbf{x})$ 
tending to 1 for small optical depths 
and having its minimum value
at $\tau_{\rm{\nu,tot}}(E,\textbf{x})= 2/3$ equal to 
1/2. 
Moreover, we enforce a similar
constraint as in Eq.~(\ref{eq:heating}) 
by setting the value of the flux 
factor to be 1/2 for any optical depth larger than 2/3.
Eq.~(\ref{eq:finalfluxf}) 
is more suitable than Eq.~(\ref{eq:fluxf}) for 
a general geometry since it only depends on
the local optical depth, and no previous 
knowledge of the radius of the neutrino surface 
is required. However, a different choice of the 
flux factor might lead to noticeable variations of 
the amount of heating. 
We will quantify the effects of this choice of flux 
factor in the next section.

\subsubsection{Flux anisotropy}
Modelling the neutrino density distribution in space 
accurately requires the 
knowledge of the path along which neutrinos 
propagate from the optically thick to 
the optically thin region. While this is trivial 
in spherical symmetry, it is not so 
for a general geometry
such as a merger remnant. 
Sophisticated 
algorithms have been 
designed \citep{Perego2014}, 
but come at large 
computational expense.
Here we take a simpler approach.
We modify Eq.~(\ref{eq:nudensity}) 
according to the 
equation~(3) of \cite{Martin2018}.
In their work the total,
axially symmetric
neutrino flux an observer 
receives far from the 
neutrino emitting region 
is approximated by
\begin{ceqn}
\begin{equation}
    J_{\nu}(\theta,R)= \frac{3(1+\beta_{\nu}\:\cos^2\theta)}{3+\beta_{\nu}} \frac{L_{\mathrm{\nu}}^{k = 1}(\theta,R)}
    {4\pi R^2\langle{}E_{\nu}\rangle{}},
    \label{eq:nuflux}
\end{equation}
\end{ceqn}
where $\theta$ is the angle of the observer with
respect to the source pole and $\beta_{\nu}= 
\frac{J_{\nu}(\theta=0,R)}{J_{\nu}(\theta=\pi/2,R)} - 1$
measures the degree of flux anisotropy
which depends on the geometry of the source.
In particular, a spherically symmetric source
would have $\beta_{\nu}= 0$. On the other hand,
in the context of mergers
the presence of a torus implies
more escaping neutrinos along 
the polar region rather 
than along the optically thick 
equatorial region, and 
therefore $\beta_{\nu} > 0$
\citep{Rosswog2003,Dessart2009,Perego2014,Foucart2016a}.
In order to adapt Eq.~(\ref{eq:nuflux}) to our
problem of reconstructing the neutrino density
distribution in our domain, we assume a 
similar modulation $\sim \cos^b(\theta)$,
but we gauge the exponent $b$ by 
comparison with an M1 neutrino
transport approach.
Moreover, we make the assumption that
the value of $\beta_{\nu}$ that would in
principle be a function of the distance to the 
source of neutrino emission 
is constant and equal to 
the value measured at large distances.
Furthermore, at distances close to the neutrino
surface the effects of the neutrino angular 
distribution arise and the hypothesis of pure
radial fluxes breaks down. We thus 
calculate the neutrino
density $n_{\nu,\tau \lesssim 1}(E,\textbf{x})$
at each radius $R=|\textbf{x}|$ 
as in Eq.~(\ref{eq:nudensity}) by retaining 
a spherically symmetric neutrino spectral
number rate $l_{\nu}(E,|\textbf{x}|)$, but 
on one side we add the flux factor 
$\mu_{\nu}(E,\textbf{x})$ of 
Eq.~(\ref{eq:finalfluxf}) and on the other
side we include 
the modulation factor $\sim \rm{cos}^\textit{b}(\theta)$
to keep track of
the effect of the geometry of the system
on the neutrino fluxes.
Therefore we replace Eq.~(\ref{eq:nudensity}) with
\begin{ceqn}
\begin{equation}
    n_{\mathrm{\nu,\tau \lesssim 1}}(E,\textbf{x})= \frac{(1+b)(1+\beta_{\nu}\cos^b(\theta))}{1+b+\beta_{\nu}} \frac{l_{\mathrm{\nu}}(E,|\textbf{x}|)}{4\pi |\textbf{x}|^2c\mu_{\mathrm{\nu}}(E,\textbf{x})},
    \label{eq:new_nudensity}
\end{equation}
\end{ceqn}
where $\theta \equiv \theta(\textbf{x})$
and we rewrite the angular
dependent pre-factor such
that the integral over the solid
angle is 4$\pi$.
Note that unlike Eq.~(\ref{eq:fluxf})
we use $\mu_{\nu}(E,\textbf{x})$ from 
Eq.~(\ref{eq:finalfluxf}) and therefore
points in space at same distance $R=|\textbf{x}|$
to the centre of the system can 
potentially have different values of 
the flux factor.
Only for spherically symmetric systems
they are the same and 
we recover Eq.~(\ref{eq:nudensity}) 
by setting $\beta_{\nu}=0$. 
Note that in Eq.~(\ref{eq:nudensity})
we keep the assumption of axial symmetry when
calculating the flux modulation. 
A high degree of axisymmetry is 
generally expected after few tens of ms 
irrespective of the mass ratio
\citep{Perego2019}.
This timescale reduces even more 
if angular momentum
transport by turbulent magnetic viscosity
is effective. Therefore, except the first few ms
after the merger, our assumption of axis symmetry
is reasonable.
As last step we estimate 
$\beta_{\nu}$ considering
an approach similar 
to the one of \cite{Rosswog2003}. 
We divide neutrinos in two groups: 
diffusive and free streaming neutrinos. 
The former propagate outward following 
the direction of the
gradient $\hat{n}_{\rho}= 
-\nabla \rho/|\nabla \rho|$, while the latter
are emitted isotropically. The luminosity 
at a polar angle $\theta$ coming from 
the diffusive neutrinos is selected by choosing 
those neutrinos emitted within a ring of 
width $\Delta \theta$ around the $\theta$ direction. 
Therefore, the luminosity per solid angle 
$\Lambda_{\nu}(\theta)=\frac{\Delta L_{\nu}^{k = 1}}
{\Delta \Omega}$ is
\begin{ceqn}
\begin{align}
    \Lambda_{\nu}(\theta)= \frac{\sum_i R_{i,\nu,\rm diff}
    ^{k = 1}(\theta)m_i}{2 \pi \sin(\theta) \Delta \theta} +
    \frac{\sum_j R_{j,\nu,\mathrm{prod}}^{k = 1}m_j}{4\pi},
\end{align}
\label{eq:lambda}
\end{ceqn} 
where the index $i$ in the sum 
is limited to those fluid points 
for which $\theta - \Delta \theta/2 <
\theta_i < \theta + \Delta \theta/2$ 
and $\mathrm{cos}(\theta_{i})= \hat{n}_{\rho}
\cdot \hat{e}_{z}$, $\hat{e}_{z}$ being
the unit vector along the z-axis,
and the index $j$ 
extends over the whole volume.
$m_i$ and $m_j$ are the masses of the fluid 
elements $i$ and $j$.
The diffusive and free 
streaming contributions to
the luminosity
are calculated as
\begin{ceqn}
\begin{align}
R_{i,\nu,\rm diff}^{k = 1}(\theta)&= \int_0^{+\infty} 
f_{i,\nu,\rm diff}(E)r_{i,\nu}(E)E^3{\rm d}E \\
R_{j,\nu,\rm prod}^{k = 1}&= \int_0^{+\infty} 
f_{j,\nu,\rm prod}(E)r_{j,\nu}(E)E^3{\rm d}E
\end{align}
\end{ceqn}
with $f_{i,\nu,\rm{diff}}(E)$ and $f_{j,\nu,\rm{prod}}(E)$ 
being the fractions of diffusive and 
free streaming contribution to the emission
rates $r_{i,\nu}(E)$ and $r_{j,\nu}(E)$ at $i$ 
and $j$ points respectively,
which can be approximated by 
\footnote{We neglect the Pauli blocking and
the thermalization correction for this
calculation.}
\begin{ceqn}
\begin{align}
f_{i,\nu,\rm{diff}}(E) &\approx \frac{\tilde{r}_{i,\nu}(E)}{r_{i,\nu,\rm{diff}}(E)}, \\
f_{j,\nu,\rm{prod}}(E) &\approx \frac{\tilde{r}_{j,\nu}(E)}{r_{j,\nu,\rm{prod}}(E)}.
\end{align}
\end{ceqn}
The value of $\beta_{\nu}$ is then estimated from
the ratio of the fluxes $\sim \frac{\Lambda_{\nu}}{R^2}$ 
an observer close to pole and equator would see
at a fixed distance $R$ to the source
\footnote{We do not exactly choose the angles 
$\theta=0 \degree$ and $\theta= 90 \degree$ for two reasons:
first, the solid angle corresponding to $\theta=0 \degree$
is small and the value of $\Lambda_{\nu}(\theta=0 \degree)$
would be associated to a small region that would
not be representative of the flux at 
the pole. Second, the cosine dependence 
is only meant as an approximate trend 
of the flux between pole and equator.
We therefore set our fiducial
angles close to pole and
equator to $\theta=10\degree$
and $\theta=80\degree$ 
respectively.}

\begin{ceqn}
\begin{equation}
\label{eq:beta}
\beta_{\nu}= \frac{\Lambda_{\nu}(\theta \approx 0 \degree)}
{\Lambda_{\nu}(\theta \approx 90 \degree)} - 1.
\end{equation}
\end{ceqn}
It is important to note that in 
the computation of $\beta_{\nu}$ we distinguish 
between neutrino species as the relevant 
neutrino-matter interactions differ between them
and they also have different decoupling surfaces
(see Figure \ref{fig:BNS_nusphere} 
for example), but for each species we do not  
consider different $\beta_{\nu}$ for 
different neutrino energies. Nevertheless,
our algorithm accounts for the different
contributions to $\beta_{\nu}$ from different
energies and therefore we assume Eq.~(\ref{eq:beta})
as indicative for a suitable estimate of 
the spectral neutrino density of Eq.~(\ref{eq:new_nudensity}).

\section{Results}
\label{sec:results}
In this section we summarise
our results. We begin by applying the ASL to 
a 1D core-collapse supernova profile.
Subsequently we map this profile on a 3D
grid and apply our multi-D implementation of
the ASL, see Sec.~\ref{sec:3Dmeth}.
Finally, we use the ASL to extract 
the neutrino physics from a neutron 
star merger remnant. Our tests 
are performed by taking snapshots of density,
temperature and electron fraction from
dynamical simulations \citep{Rosswog17a} 
as a background on which we evolve the neutrino
quantities until they achieve a steady state.
For the core collapse supernovae 
tests, we use the Lattimer-Swesty Equation 
of State (EoS) \citep{Lattimer1991} with nuclear
incompressibility $K= 220\:\rm{MeV}$, 
a standard choice widely used 
throughout the literature.
In light of the recent constraints
on the EoS from 
gravitational and electromagnetic
observations of 
GW170817 \citep{Coughlin2018,Radice2018,Most2018,
Abbott2018} 
for the binary merger remnant we 
consider the SFHo EoS, which provides masses in 
agreement with the highest neutron star masses  
measured until very recently \citep{Demorest2010,Antoniadis2013},
but in tension with the most recent 
measurement of a $2.17^{+0.11}_{-0.10}\; \Msun$ neutron 
star \citep{cromartie19}.
The neutrino transport is run with a 
spectrum of 20 geometrically increasing energy 
groups from 3 MeV to 300 MeV. We also explore 
for fixed energy interval the convergence 
of our results by changing the number 
of energy bins, and for fixed number of bins
we also increase and decrease 
the energy interval to test the 
dependence of our results 
on the energy spectrum.
Our ASL results are 
compared with a two moment scheme 
(M1) \citep{Thorne1981}.
In particular, for the core-collapse 
supernovae tests we compare with 
the M1 scheme of the spherically 
symmetric Eulerian hydrodynamics
code GR1D \citep{Oconnor2010,OConnor2015,OConnor2013}.
Since we take the 1D core-collapse 
supernova snapshot 
from the dynamical simulations of
\cite{Perego2016}, which uses the same ASL
described in Sec.~\ref{sec:1Dmeth}, 
we perform a dynamical simulation
of the same progenitor with GR1D and 
take the outcome from the simulation at
the same time post-bounce to make 
a consistent comparison.
For the binary neutron star merger case
we compare with the 
M1 scheme implemented in the
Eulerian hydrodynamics code FLASH \citep{Fryxell2000,Oconnor2018}.
Unlike the core-collapse case, 
we take a snapshot of a post-merger remnant from the 
simulations of \cite{Rosswog2003},
which uses a grey leakage scheme,
and apply both the ASL implementation
described in Sec.~\ref{sec:3Dmeth}
and the M1 scheme in FLASH to it.
Applying a different transport approach
to the one adopted for obtaining
the snapshot introduces inconsistencies
that will be quantified.

\subsection{ASL in 1D core-collapse
supernovae}
\label{sec:ASLvsGR1D}
\begin{figure*}
\begin{minipage}{0.33 \linewidth}
\centering
\includegraphics[width = 1 \linewidth]{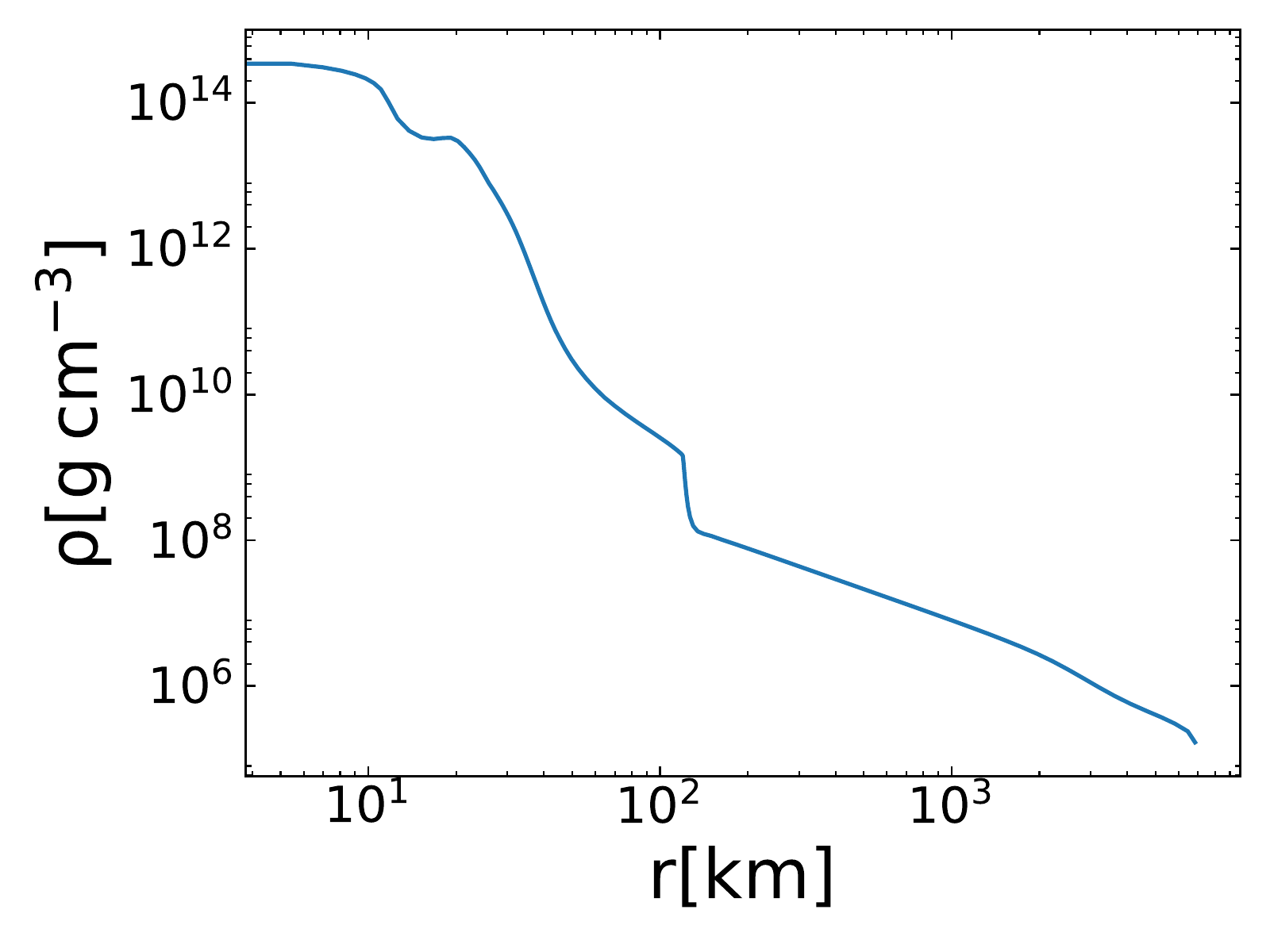}
\end{minipage}
\begin{minipage}{0.33 \linewidth}
\centering
\includegraphics[width = 1 \linewidth]{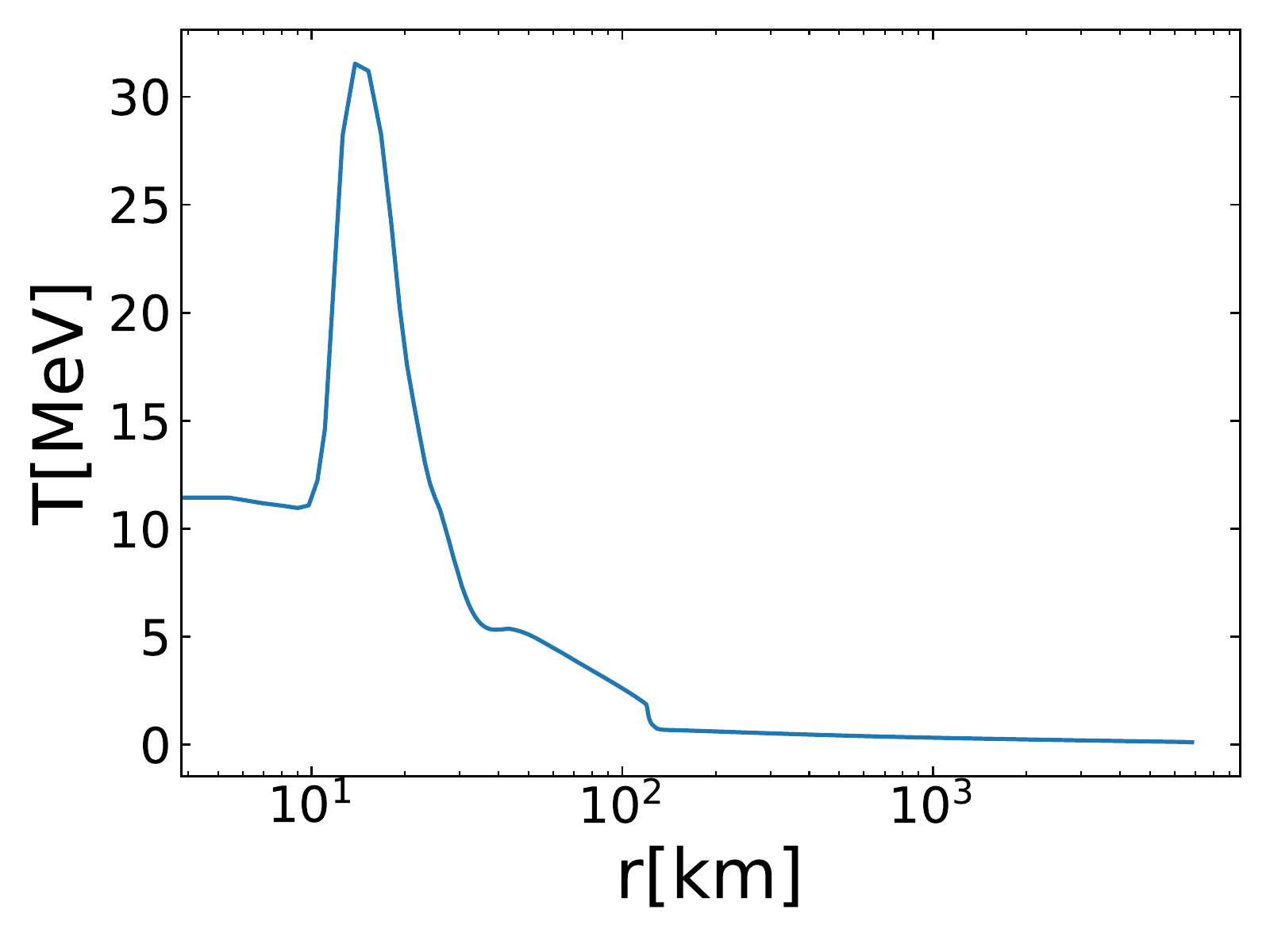}
\end{minipage}
\begin{minipage}{0.33 \linewidth}
\centering
\includegraphics[width = 1 \linewidth]{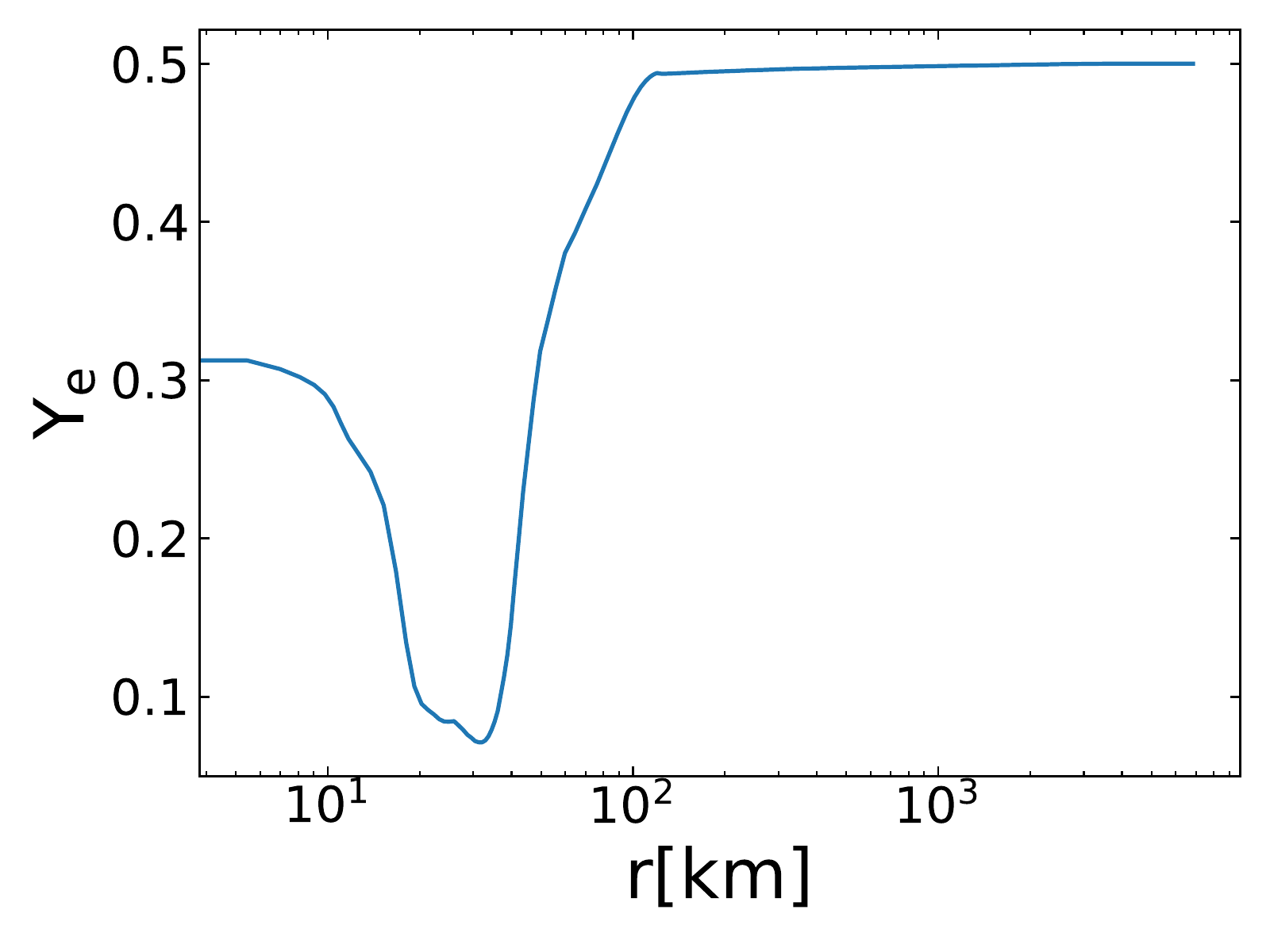}
\end{minipage}\\
\begin{minipage}{0.33 \linewidth}
\centering
\includegraphics[width = 1 \linewidth]{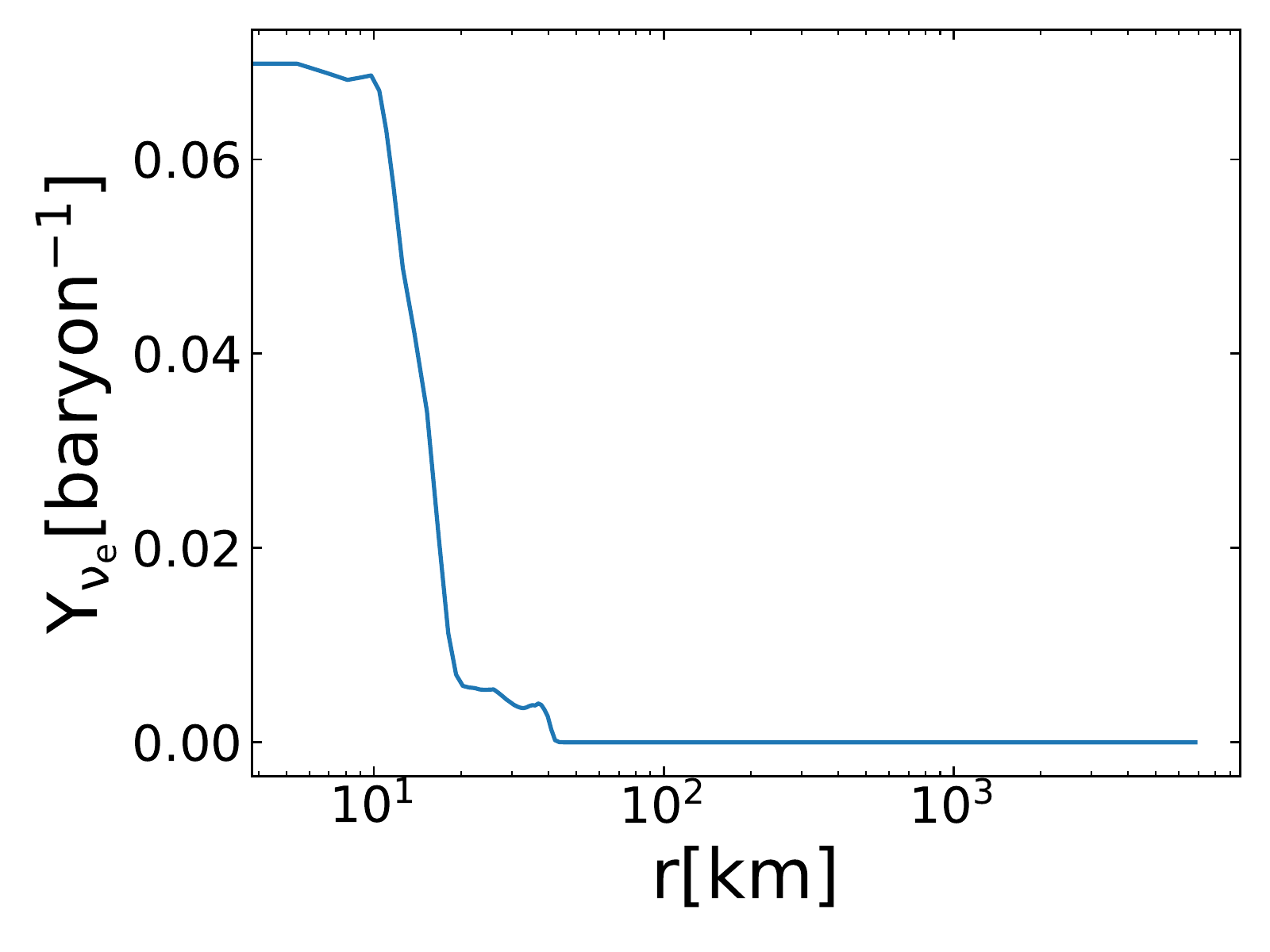}
\end{minipage}
\begin{minipage}{0.33 \linewidth}
\centering
\includegraphics[width = 1 \linewidth]{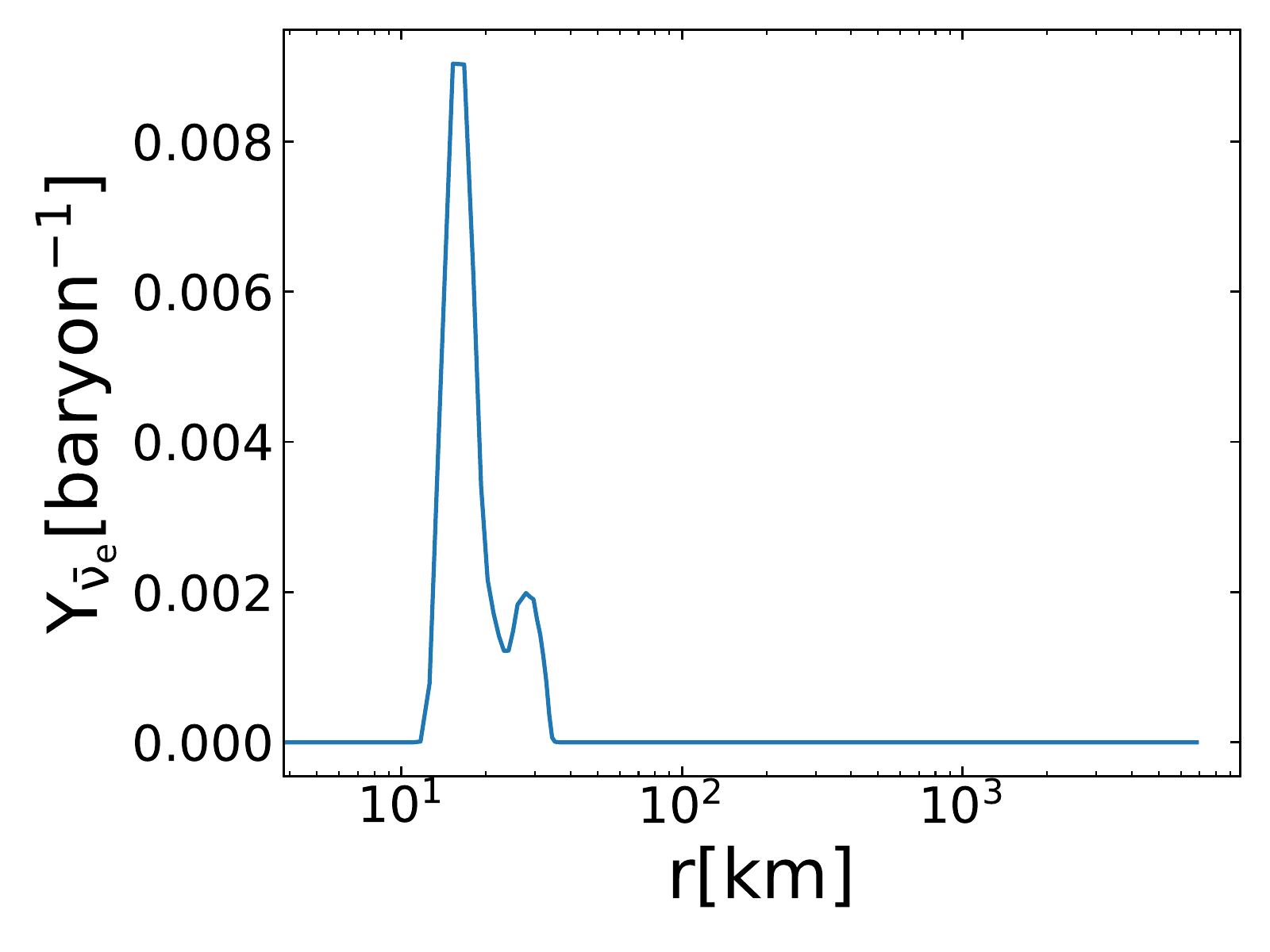}
\end{minipage}
\begin{minipage}{0.33 \linewidth}
\centering
\includegraphics[width = 1 \linewidth]{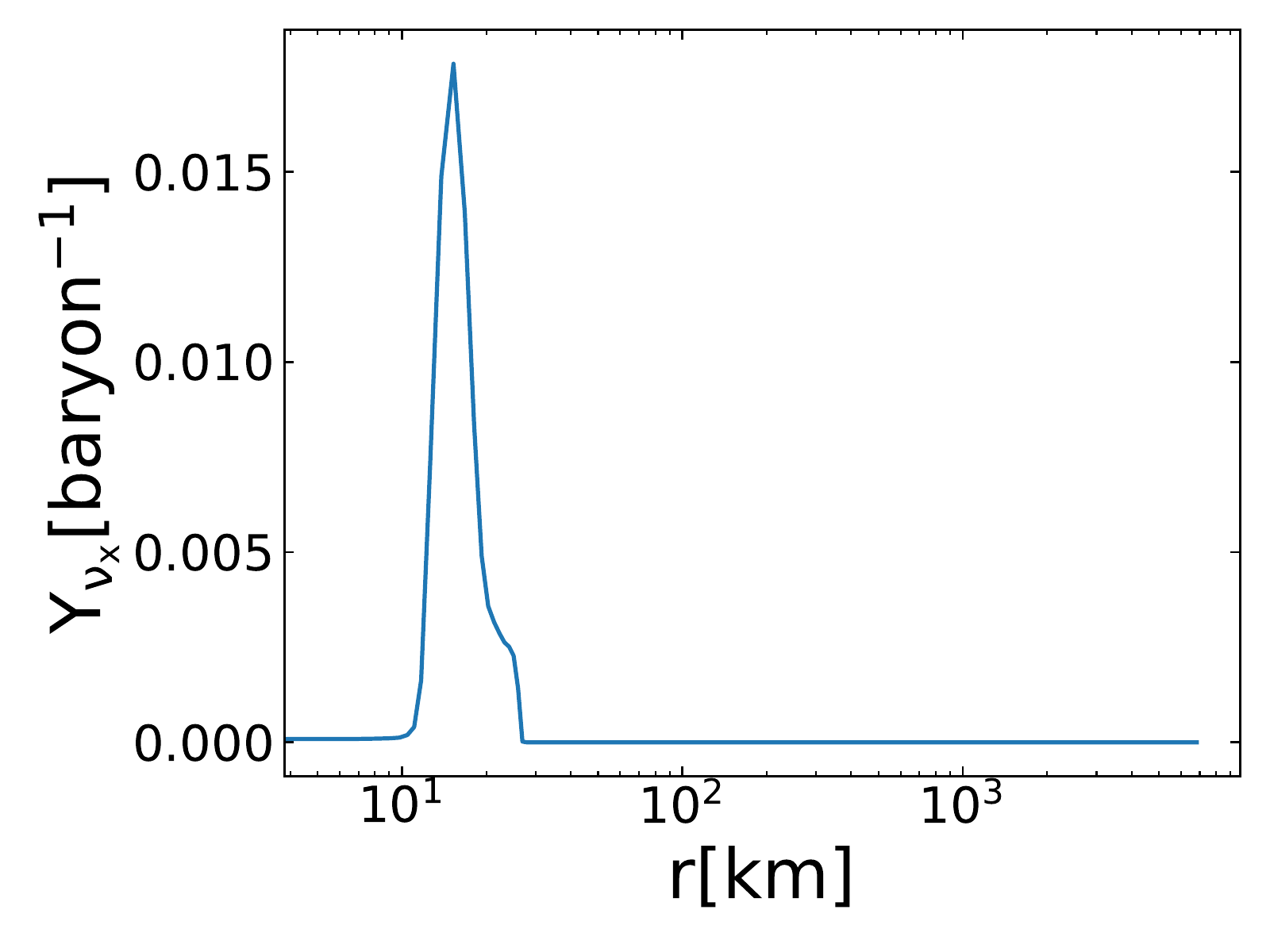}
\end{minipage}\\
\begin{minipage}{0.33 \linewidth}
\centering
\includegraphics[width = 1 \linewidth]{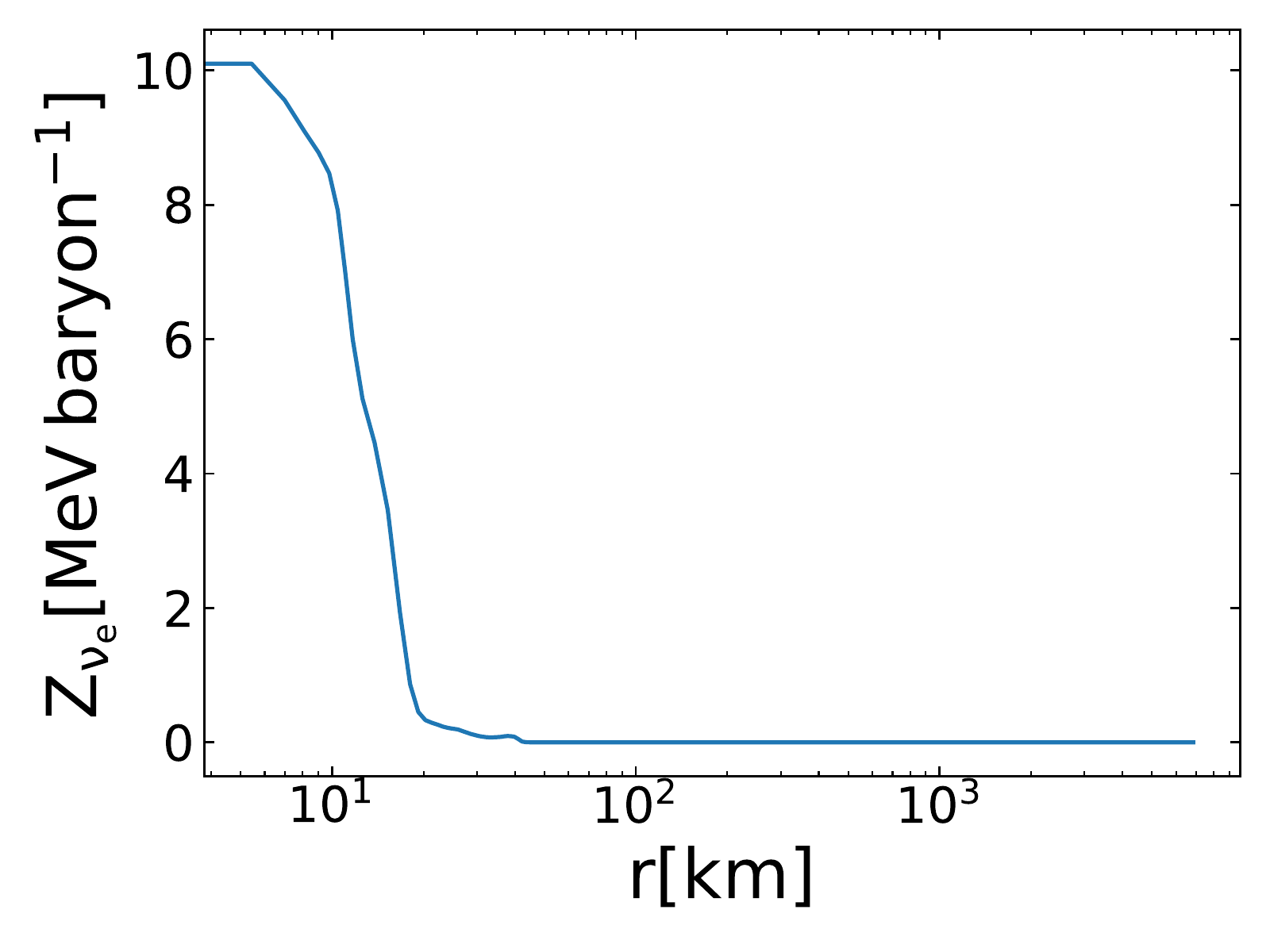}
\end{minipage}
\begin{minipage}{0.33 \linewidth}
\centering
\includegraphics[width = 1 \linewidth]{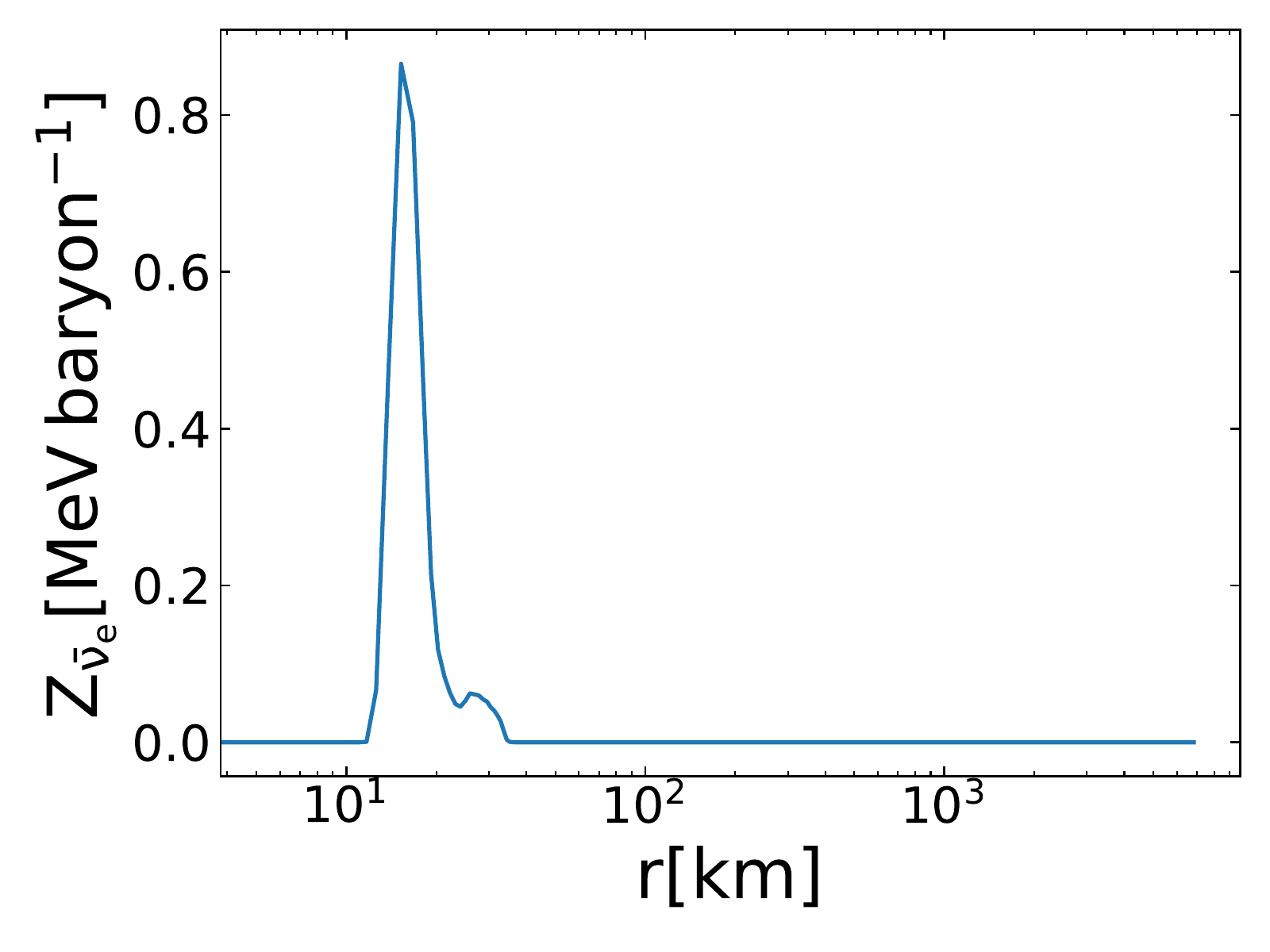}
\end{minipage}
\begin{minipage}{0.33 \linewidth}
\begin{center}
\includegraphics[width = 1 \linewidth]{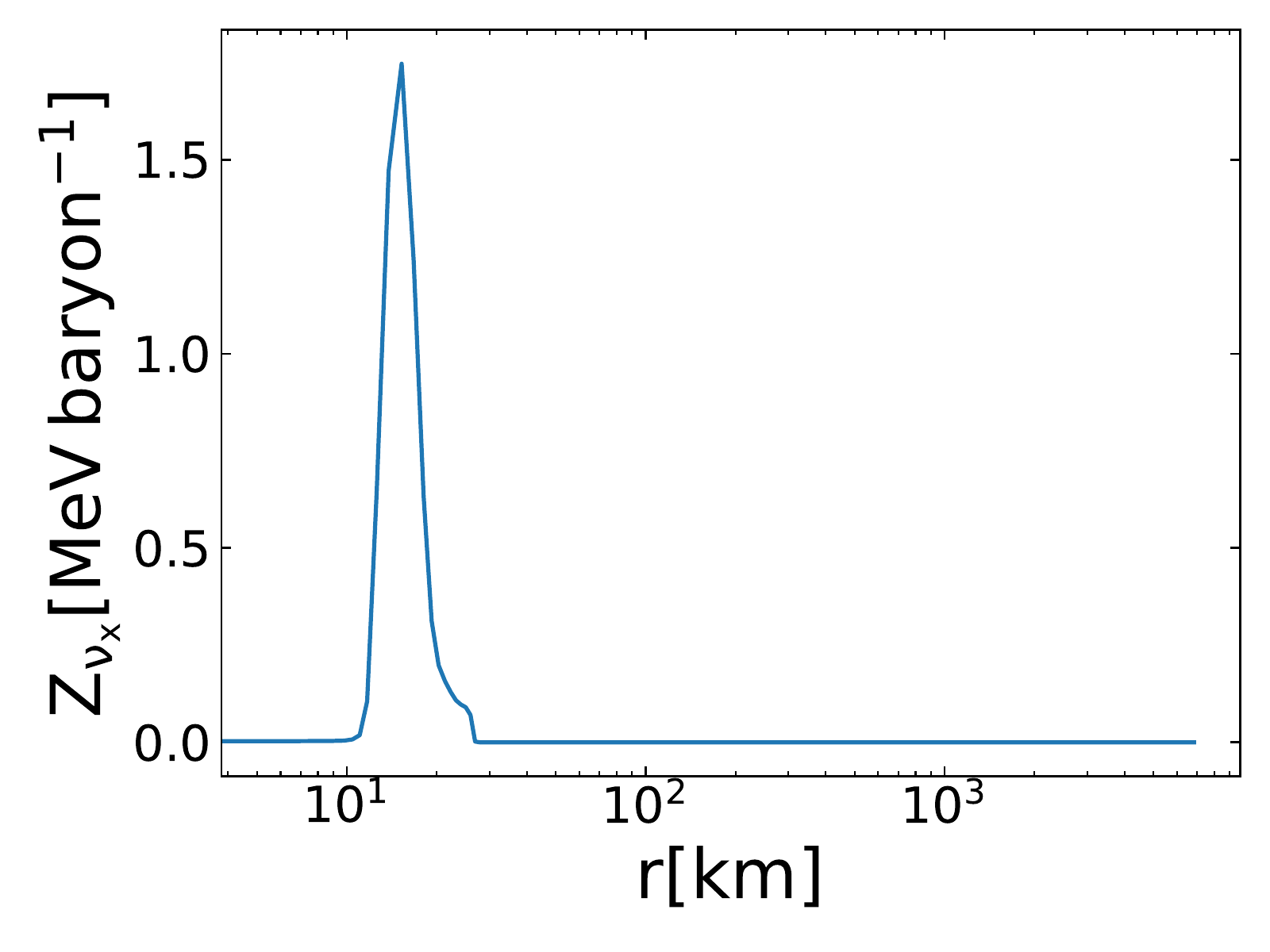}
\end{center}
\end{minipage}\\
\begin{minipage}{0.33 \linewidth}
\centering
\includegraphics[width = 1 \linewidth]{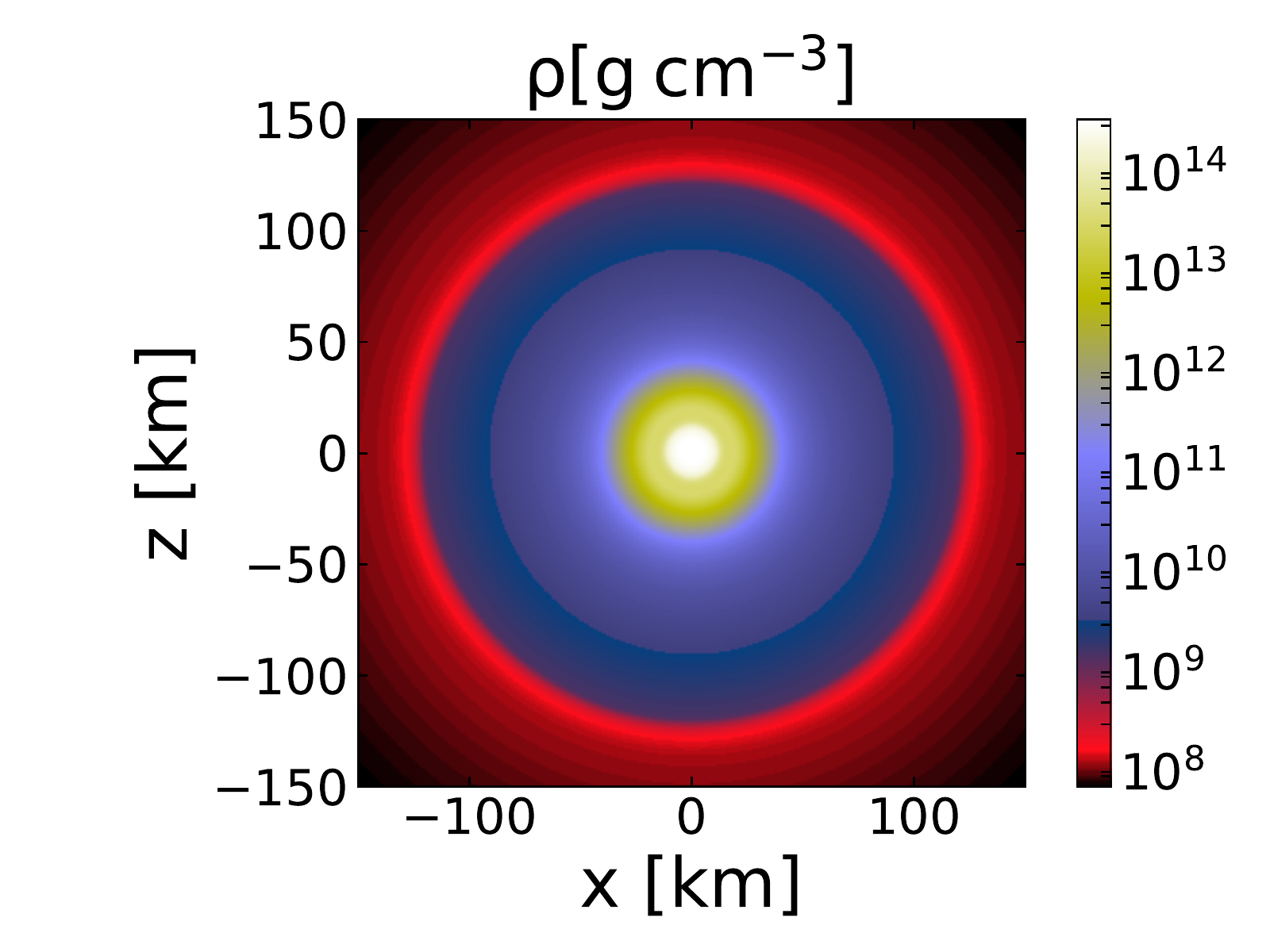}
\end{minipage}
\begin{minipage}{0.33 \linewidth}
\centering
\includegraphics[width = 1 \linewidth]{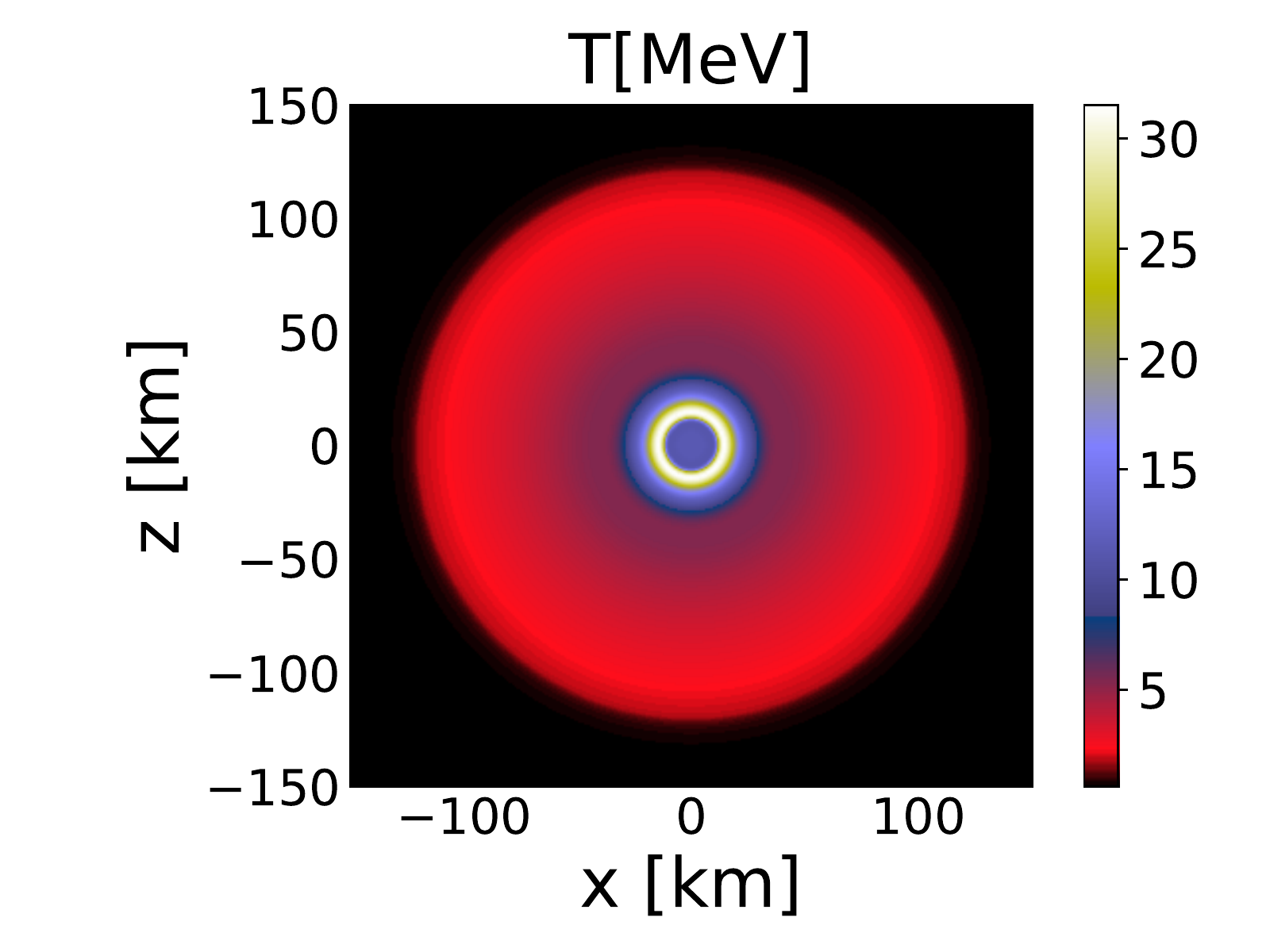}
\end{minipage}
\begin{minipage}{0.33 \linewidth}
\centering
\includegraphics[width = 1 \linewidth]{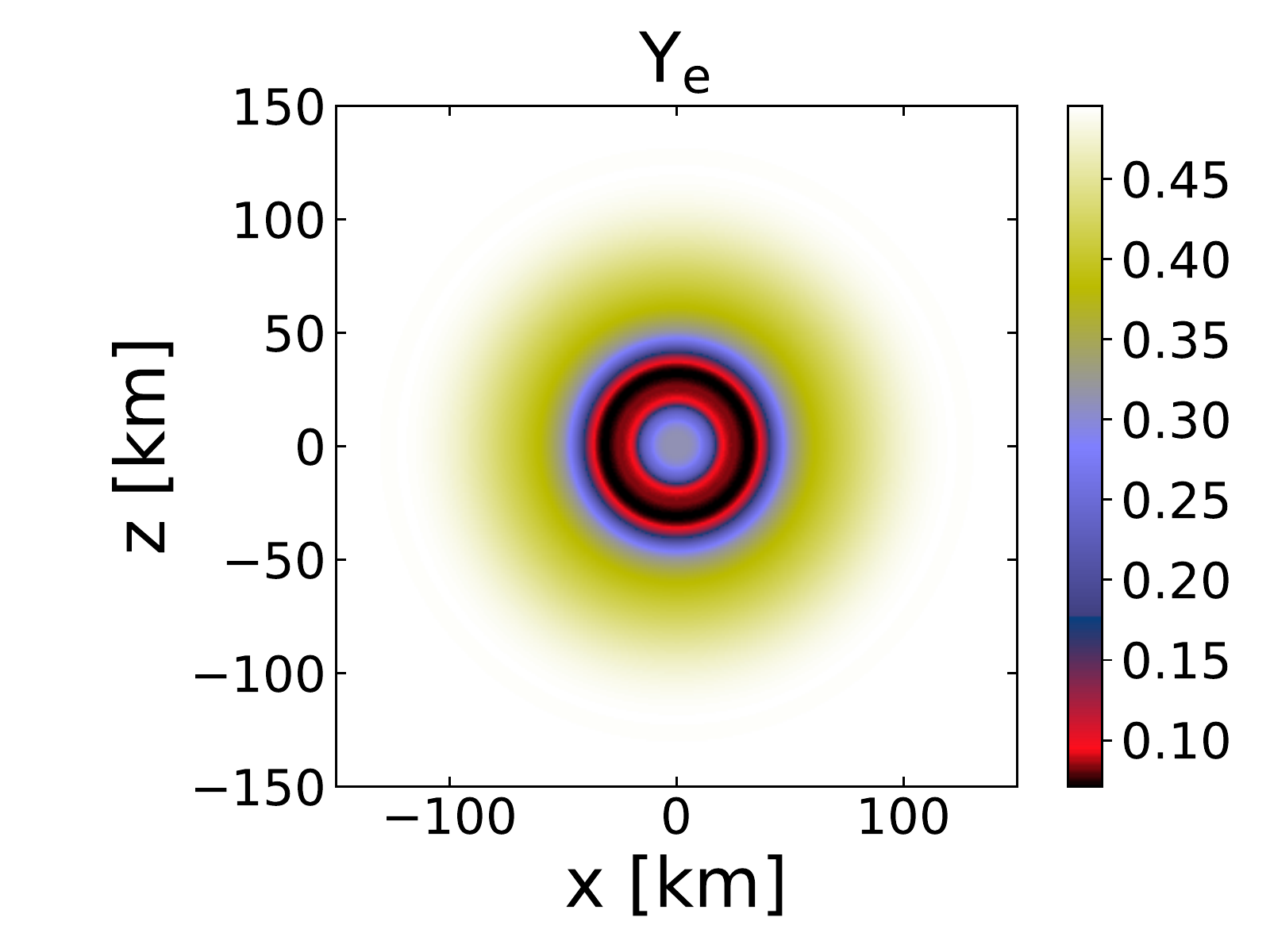}
\end{minipage}\\
\caption{\textbf{Top row:} 1D profile of density (left panel), temperature (middle panel) and electron fraction (right panel) at 275 ms after core bounce. The shock is located 
at $\sim 120$ km from the core, clearly visible
from the jump of density and temperature. 
Their values in the core are about 
$\rho \sim 3 \cdot 10^{14} \: \rm{g\:cm^{-3}}$ and 
$T \sim 12$ MeV respectively. The effect of the 
deleptonization until neutrino trapping leads to
a inner electron fraction of $Y_e \sim 0.3$. 
The passage of the shock during its propagation
affects both $T$ and $Y_e$ sensitively. 
In particular, neutrino-matter interactions 
after iron nuclei dissociation decrease the
electron fraction down to $Y_e \sim 0.1$. 
Conversely, temperatures reach their maximum
at $T \sim 30 \: \rm{MeV}$. Matter properties
in the outermost layers are untouched by the 
shock and therefore have their original
$Y_e$ and low values of $T$.
\textbf{Second and third rows:} number and energy of trapped neutrinos per baryon along the 1D profile. From left to right, electron neutrinos, electron anti-neutrinos and heavy-lepton neutrinos. Electron neutrinos are confined within a region
of $\sim 20$ km from the core, while electron 
anti-neutrinos and heavy-lepton neutrinos are 
within $10-30$ km.
\textbf{Bottom row:} density (left panel),
temperature (middle panel) and electron fraction 
(right panel) of the core-collapse snapshot at 275 
ms after bounce on the y=0 plane of the 3D grid.
The peak in the temperature that is visible at
$T \sim 20$ km in the 1D profile is shown here 
in yellow. In the same 
way, the drop in the electron fraction down 
to $Y_e \sim 0.1$ corresponds to the black-red
region.}
\label{fig:initialdata}
\end{figure*}
\begin{figure*}
\begin{minipage}{0.4 \linewidth}
\centering
\includegraphics[width = 1 \linewidth]{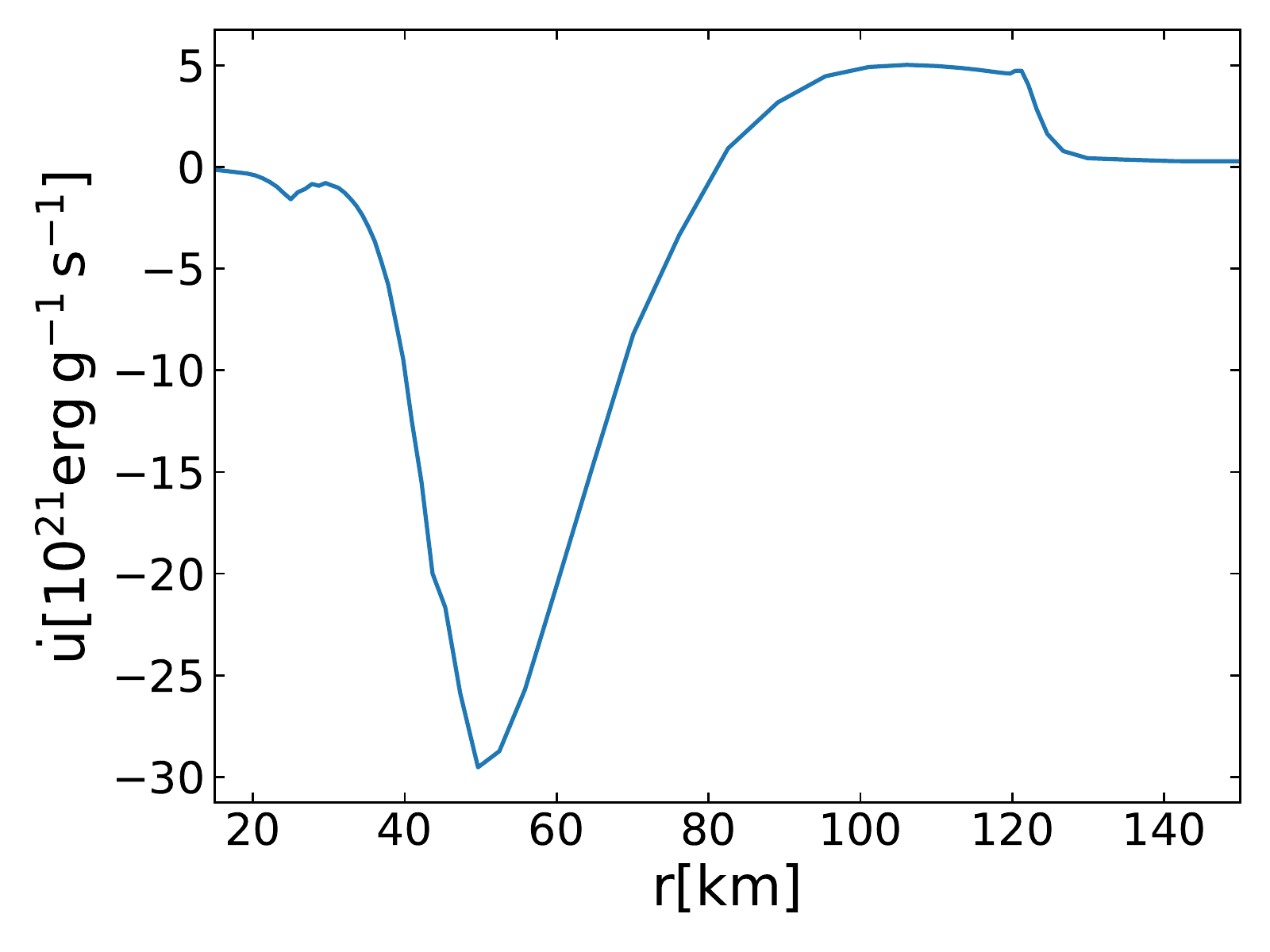}
\end{minipage}
\begin{minipage}{0.4 \linewidth}
\centering
\includegraphics[width = 1 \linewidth]{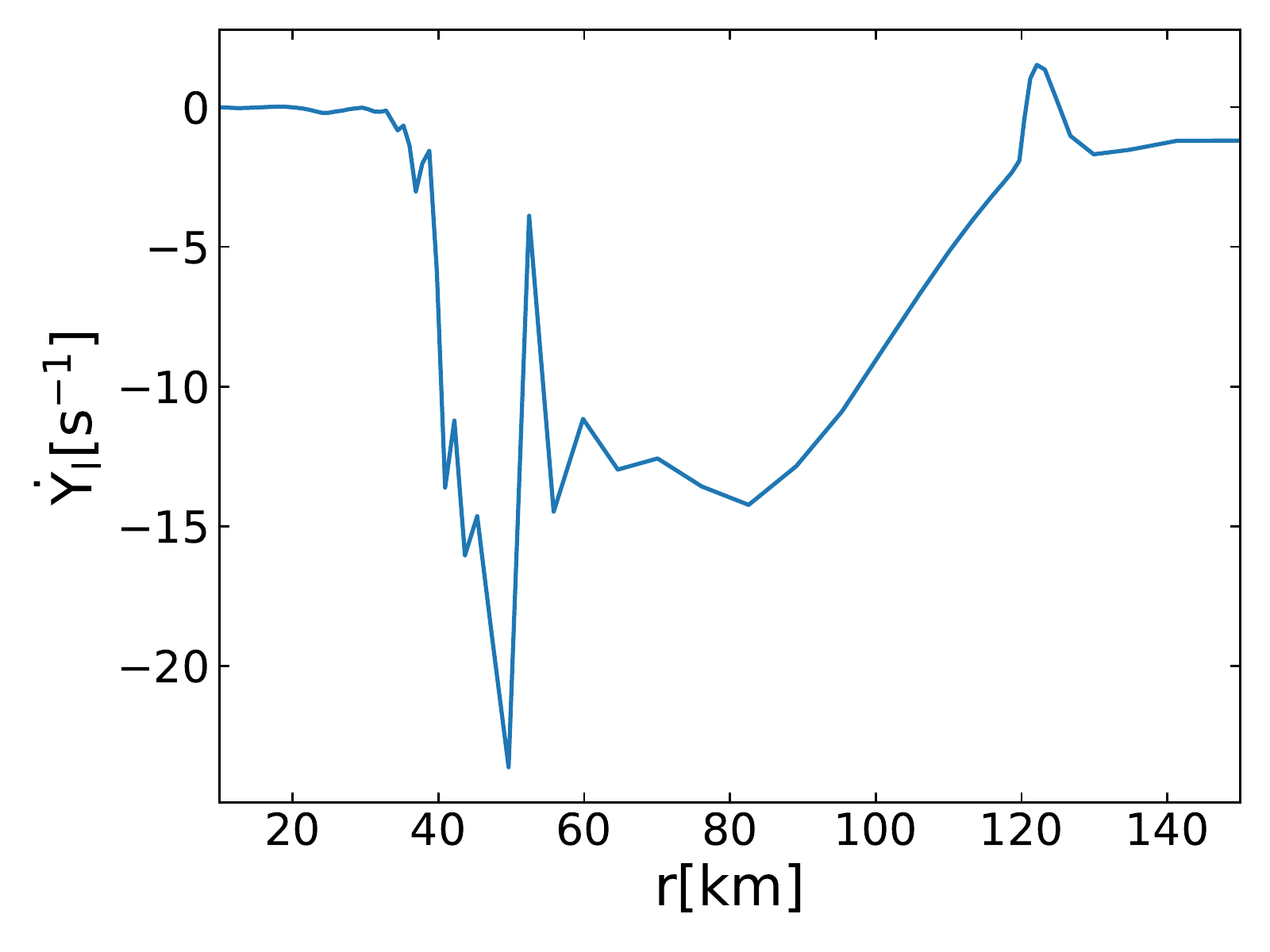}
\end{minipage}\\
\caption{Specific total internal energy rate (left panel) and lepton number fraction rate (right panel) 
once steady state is reached. 
Negative values
of $\Dot{u}$ describe the loss of internal 
energy due to neutrino emission at equilibrium,
while positive values mark the region where the 
heating dominates. The negative electron 
fraction rate is a result of the conversion
of electrons into neutrinos.}
\label{fig:relaxation}
\end{figure*}
We take a snapshot at 275 ms post-bounce 
of the 15 $\Msun$ progenitor of \cite{Perego2016},
whose dynamical evolution has 
been simulated with
the spherically symmetric hydrodynamics
code \textit{Agile} \citep{Liebendorfer2002}.
The radial profile has a variable resolution 
with radius ranging from $\lesssim 1$ km to 
tens of km moving from 
the inner to the outer regions,
with a maximum radial extension of 6832 km.
In order to describe the profile properties, 
it is worth summarising the previous dynamics.
During the collapse phase the deleptonization 
of the iron core reduces the fraction of 
electrons $Y_e$ in the core by producing 
electron neutrinos. 
Neutrinos freely stream out initially and 
therefore the total lepton number decreases.
Once neutrinos get trapped, 
the decrease in $Y_e$ is compensated by the 
fraction of trapped neutrinos. At
core bounce a shock 
forms and propagates outward.
During the shock propagation, 
iron nuclei falling into the
shock are photo-dissociated into neutrons and 
protons. At this stage, neutrinos of all flavours
are largely produced by charged current interactions 
and pair processes. Dissociation of iron 
nuclei into nucleons, the ram pressure of the still 
infalling outer layers, and the energy losses due to
the neutrino burst when it surpasses the neutrinosphere,
cause the shock to stall. Neutrino
absorption behind the shock in the so called gain
region helps the shock to revive and in some cases
leads to the final explosion.
In the top row 
panels of Figure~\ref{fig:initialdata} 
we show density, temperature and electron fraction as 
a function of the radius, while in
the second and third rows we show the 
number and energy of trapped neutrinos
per baryon, denoted
as $Y_{\nu}$ and $Z_{\nu}$ respectively.\\
We evolve such fractions until steady state is
reached while keeping density, 
temperature and electron fraction of the fluid
fixed. Once in equilibrium, the  rates of the
trapped neutrino  components vanish, which 
translates into $\Dot{u}=\Dot{e}$ 
and $\Dot{Y}_l= \Dot{Y}_e$, 
see Eqs.~(\ref{eq:udot}) and (\ref{eq:yedot}). 
The distribution of  $\Dot{u}$ and $\Dot{Y}_l$
(from a simulation using the flux factor of Eq.~(\ref{eq:finalfluxf}))
is shown in Figure \ref{fig:relaxation}.
All the results we are going to show 
are referred to this equilibrium state.
\subsubsection{Neutrino rates}
\begin{figure*}
\begin{minipage}{0.33 \linewidth}
\centering
\includegraphics[width = 1 \linewidth]{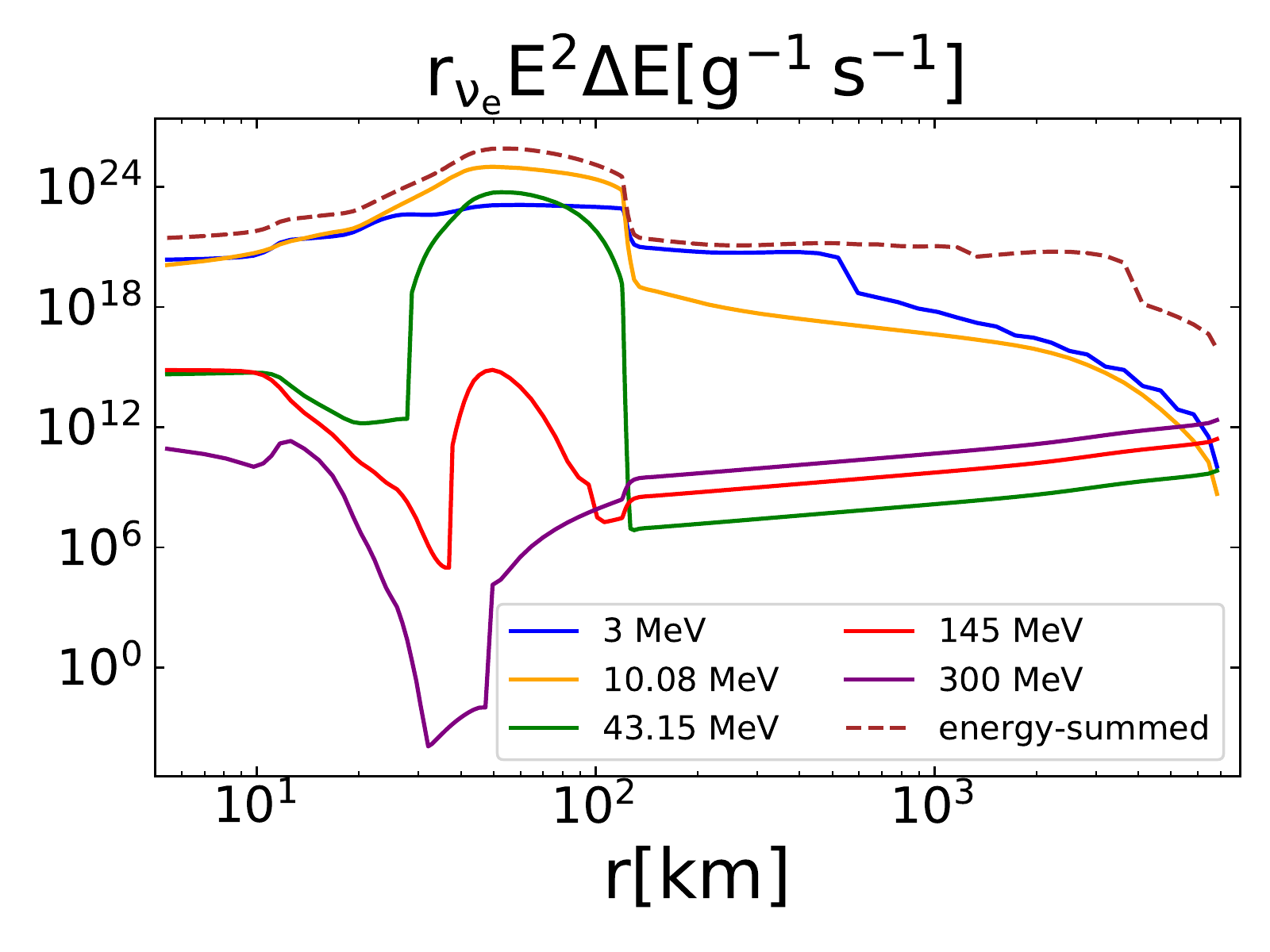}
\end{minipage}
\begin{minipage}{0.33 \linewidth}
\centering
\includegraphics[width = 1 \linewidth]{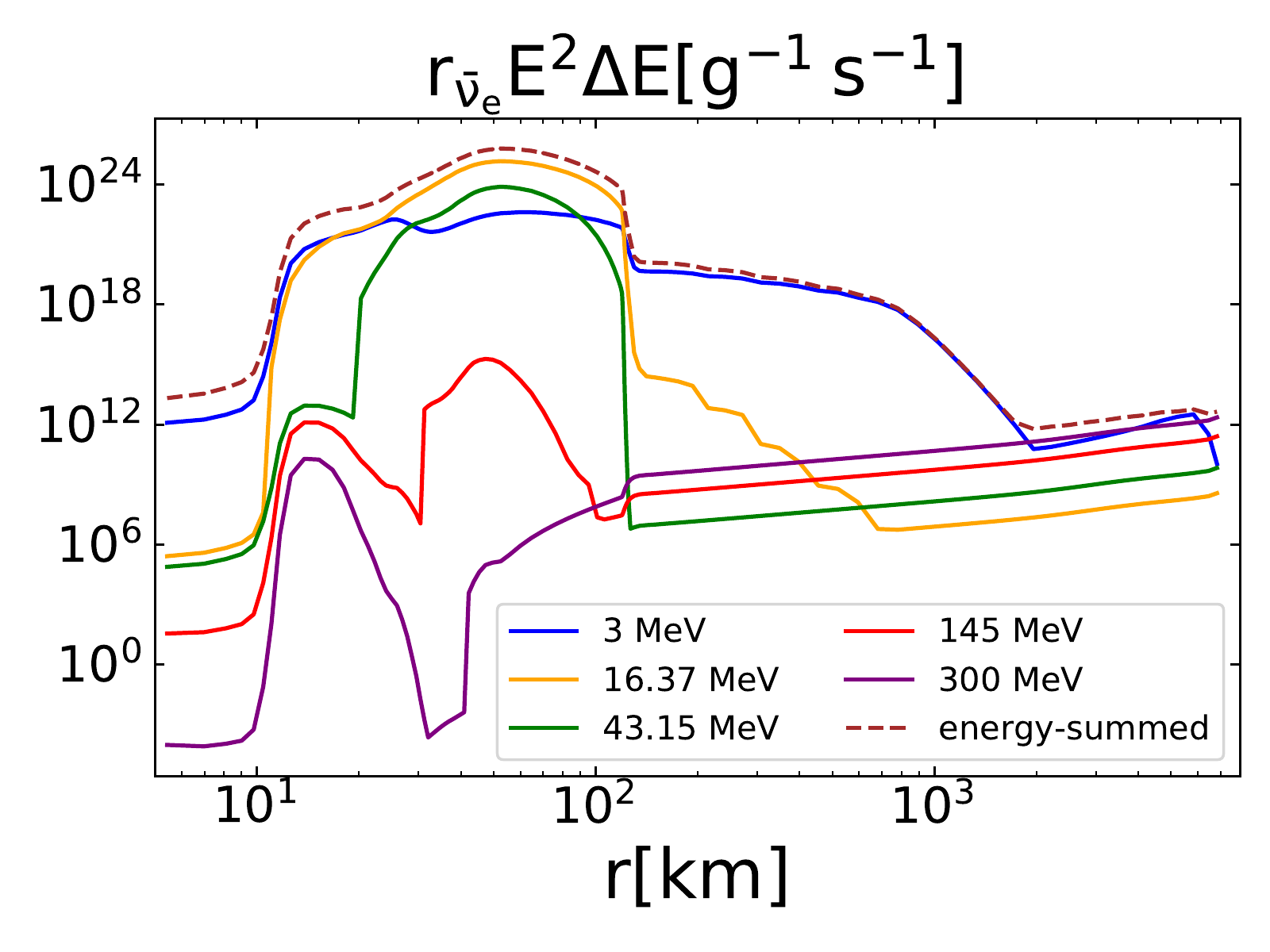}
\end{minipage}
\begin{minipage}{0.33 \linewidth}
\centering
\includegraphics[width = 1 \linewidth]{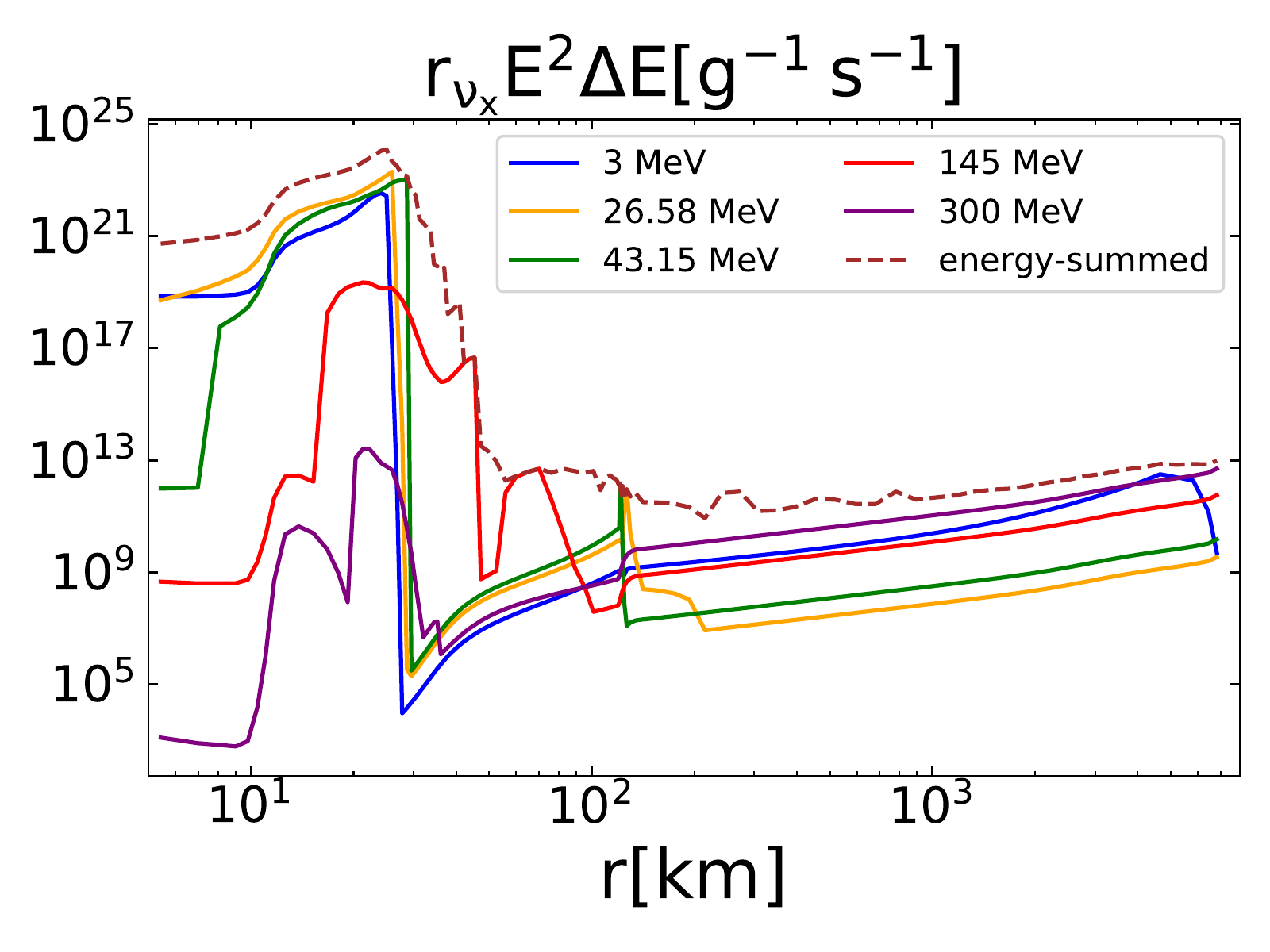}
\end{minipage}\\
\caption{Specific emission rate at different energies between 3 MeV to 300 MeV for electron neutrinos (left panel), electron anti-neutrinos (middle panel) and heavy-lepton neutrinos (right panel). The dashed-line
corresponds to the energy-summed rate.
In contrast to electron neutrinos and anti-neutrinos,
heavy-lepton neutrinos can still provide  
an important contribution to the 
emission at energies 
$\gtrsim 40$ MeV. Low-energy neutrinos
contribute the most to the local emission.}
\label{fig:ser1D}
\end{figure*}
\begin{figure*}
\begin{minipage}{0.49 \linewidth}
\centering
\includegraphics[width = 1 \linewidth]{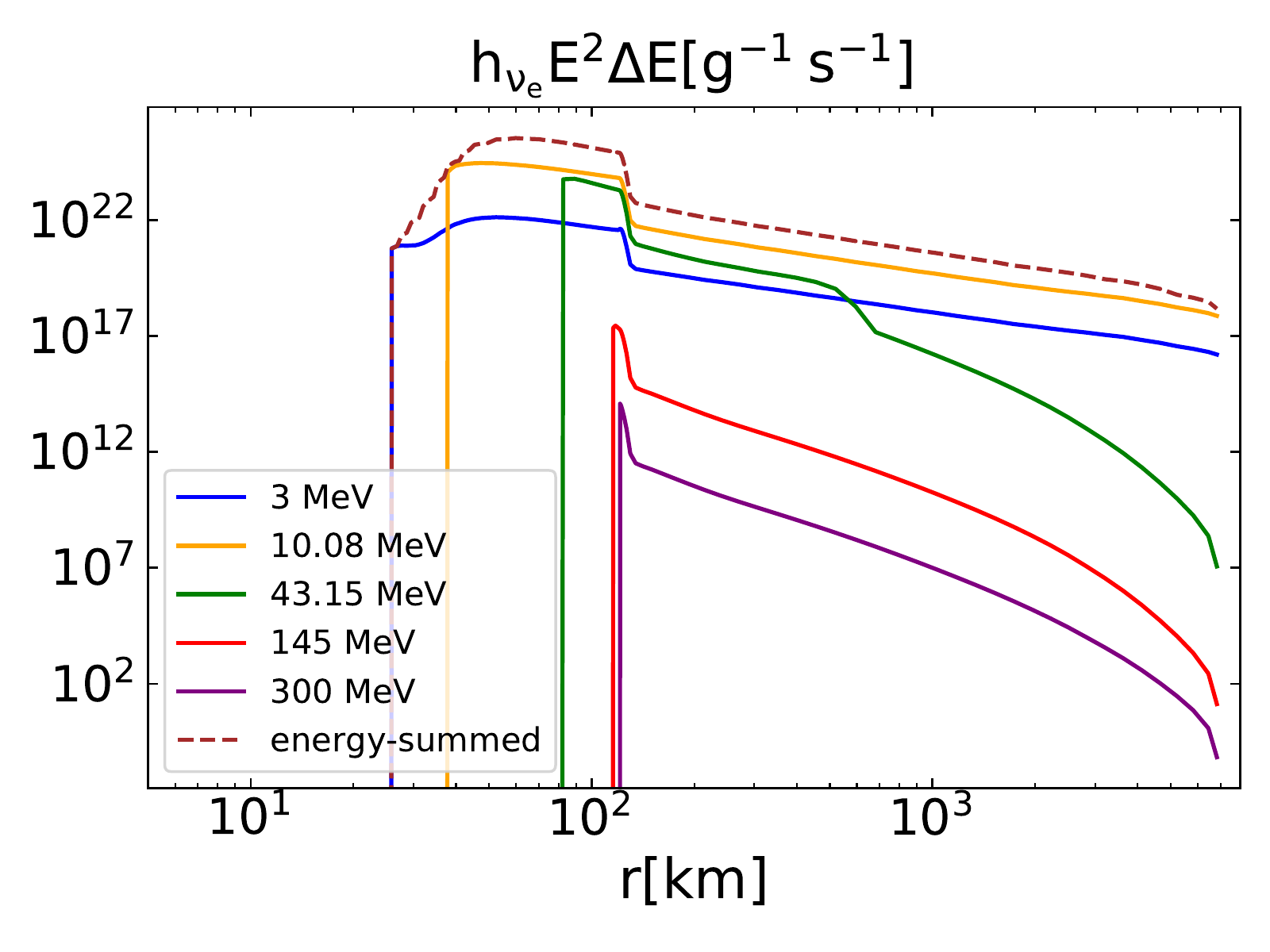}
\end{minipage}
\begin{minipage}{0.49 \linewidth}
\centering
\includegraphics[width = 1 \linewidth]{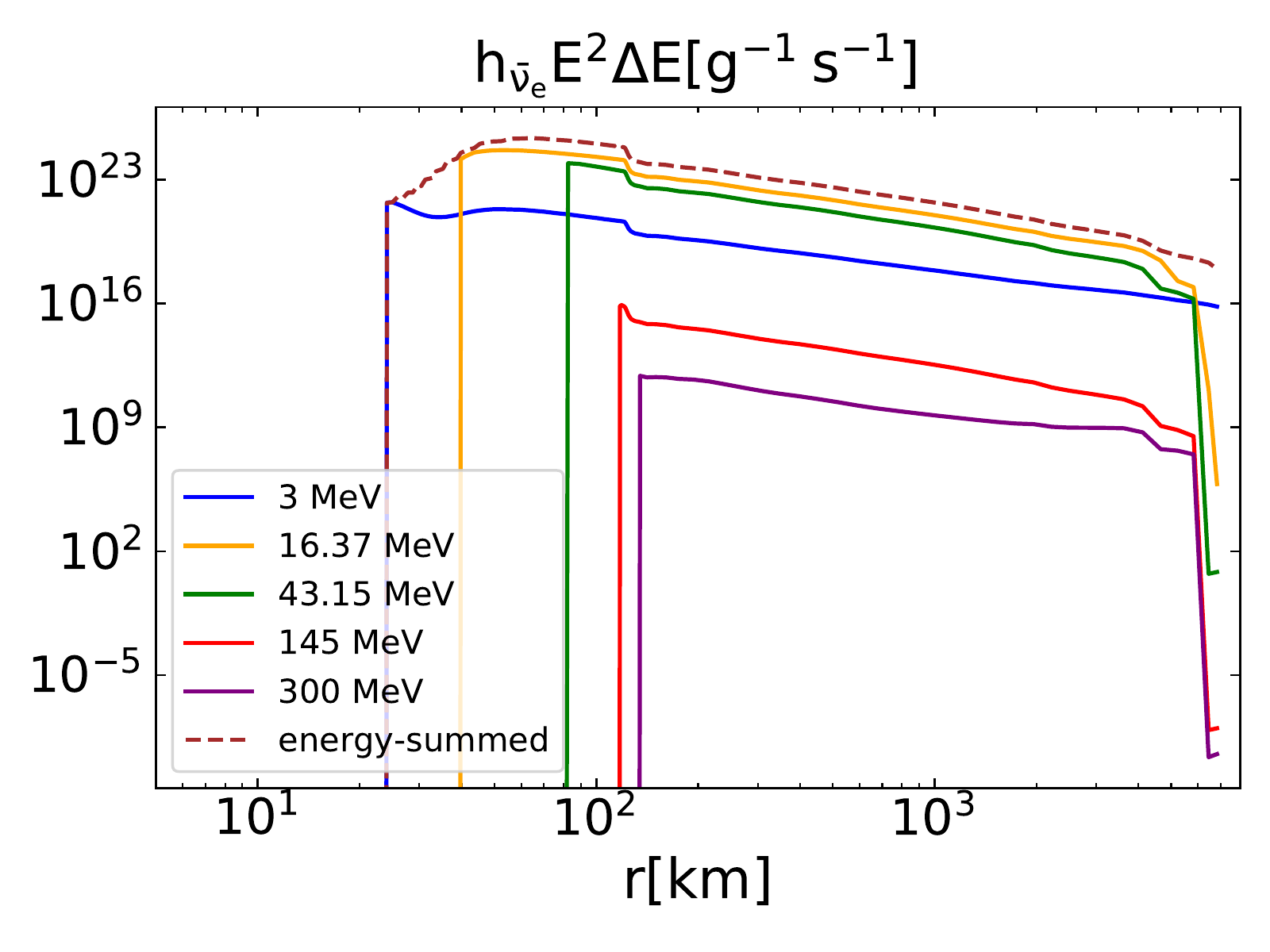}
\end{minipage}\\
\caption{Specific absorption rate at different energies between 3 MeV to 300 MeV for electron neutrinos (left panel) and electron anti-neutrinos (right panel). The dashed-line
corresponds to the energy-summed rate. 
The largest contribution to the local absorption is 
provided by low energy neutrinos.}
\label{fig:shr1D}
\end{figure*}
In Figure 
\ref{fig:ser1D} we show the specific
emission rate along the radial profile 
for different energies,
obtained from Eq.~(\ref{eq:ser1D}) 
by multiplying for each energy
by $E^2\Delta E$ of the corresponding bin,
as well as the cumulative contribution from all energies. 
The main contribution to 
the neutrino emission comes from neutrino
energies below $\sim$ few tens of MeV for all 
neutrino species. In particular, 
all species show a decreasing 
contribution to the emission with 
increasing energy at low radii because of the 
$\sim \tau_{\nu,\rm{tot}}^{-2} \sim E_{\nu}^{-2}$
dependence of the diffusion rate 
(equation (31) of \cite{Perego2016}) 
that causes high energy neutrinos to have 
a lower diffusion rate and therefore 
to diffuse out less efficiently 
than low energy neutrinos.
The production rate dominates over 
the diffusion rate at large radii.
Given the emissivity dependence $\sim j_{\nu}$
(equation (30) of \cite{Perego2016}), neutrino 
emission is governed by the low temperatures 
and is therefore suppressed at all energies.
Similarly, in Figure \ref{fig:shr1D} we show 
the specific absorption rate
for electron neutrinos and anti-neutrinos
(heavy-lepton neutrinos are not included
in the heating). Neutrino absorption is 
calculated locally for each
neutrino energy if the condition $\tau_{\rm{en},\nu}(E,\textbf{x}) \leq 1$ is satisfied,
and depends on the local amount 
of available neutrinos, 
which in turn depends on both local
production and on neutrinos coming 
from innermost regions. 
Local neutrino production 
decreases as the radius increases,
as a consequence of the decrease
in temperature and density. 
In addition, the amount of 
neutrinos at a given radius 
coming from innermost 
regions is reduced both by absorption
occurring at smaller radii and
by the $\sim 1/R^2$ dependence
in Eq.~(\ref{eq:new_nudensity}).
Therefore, all energies show 
a decrease in the local neutrino
absorption with increasing distance
from the centre. We also notice 
in the same way of the emission rate 
that the larger contribution 
to the heating rate comes from neutrinos of $\sim$
few tens of MeV, as a result of
the dependence of the neutrino 
absorption on the emission via
the number rate $l_{\nu}(E,\textbf{x})$.
\begin{figure*}
\begin{minipage}{0.49 \linewidth}
\centering
\includegraphics[width = 1 \linewidth]{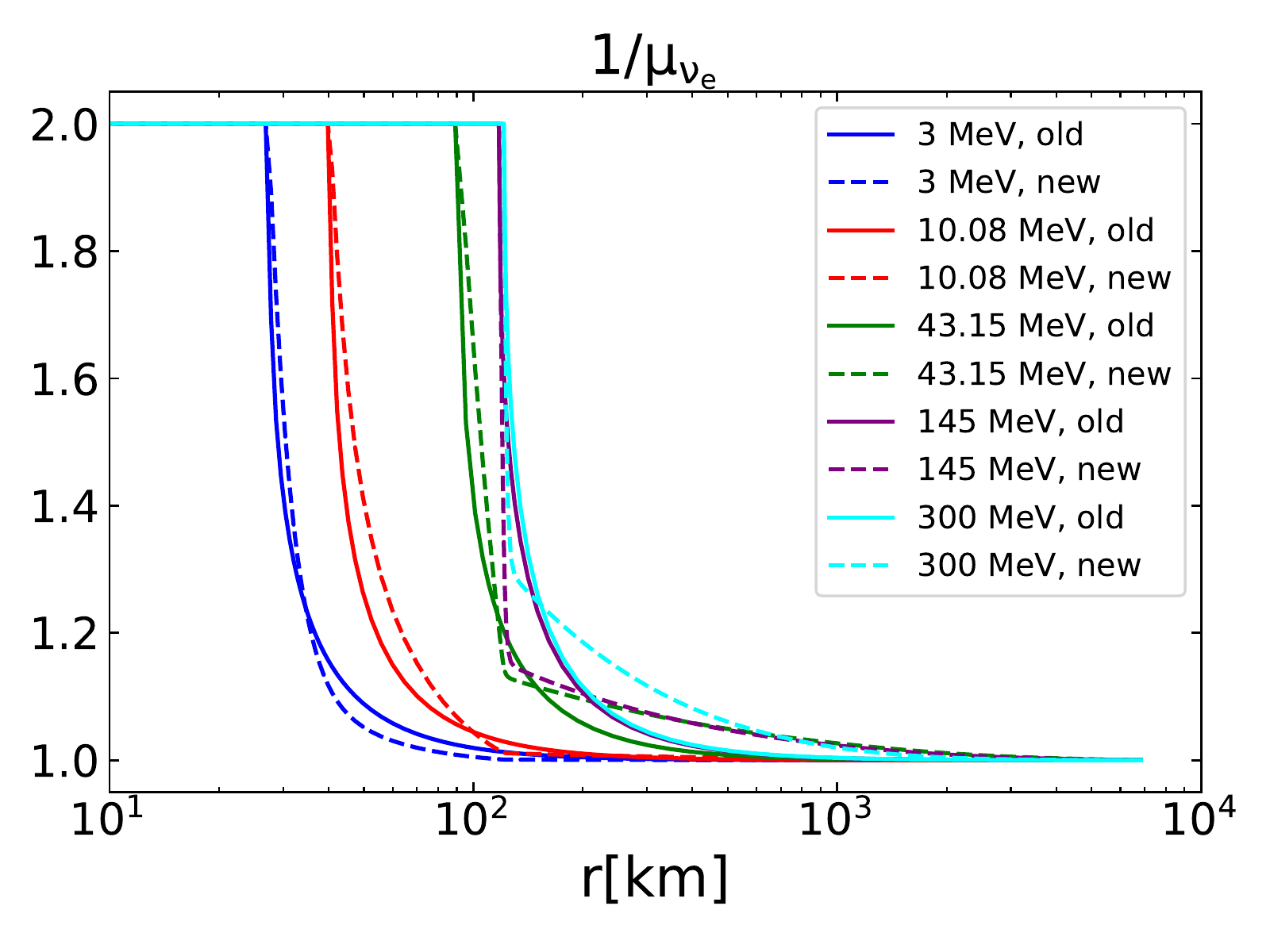}
\end{minipage}
\begin{minipage}{0.49 \linewidth}
\centering
\includegraphics[width = 1 \linewidth]{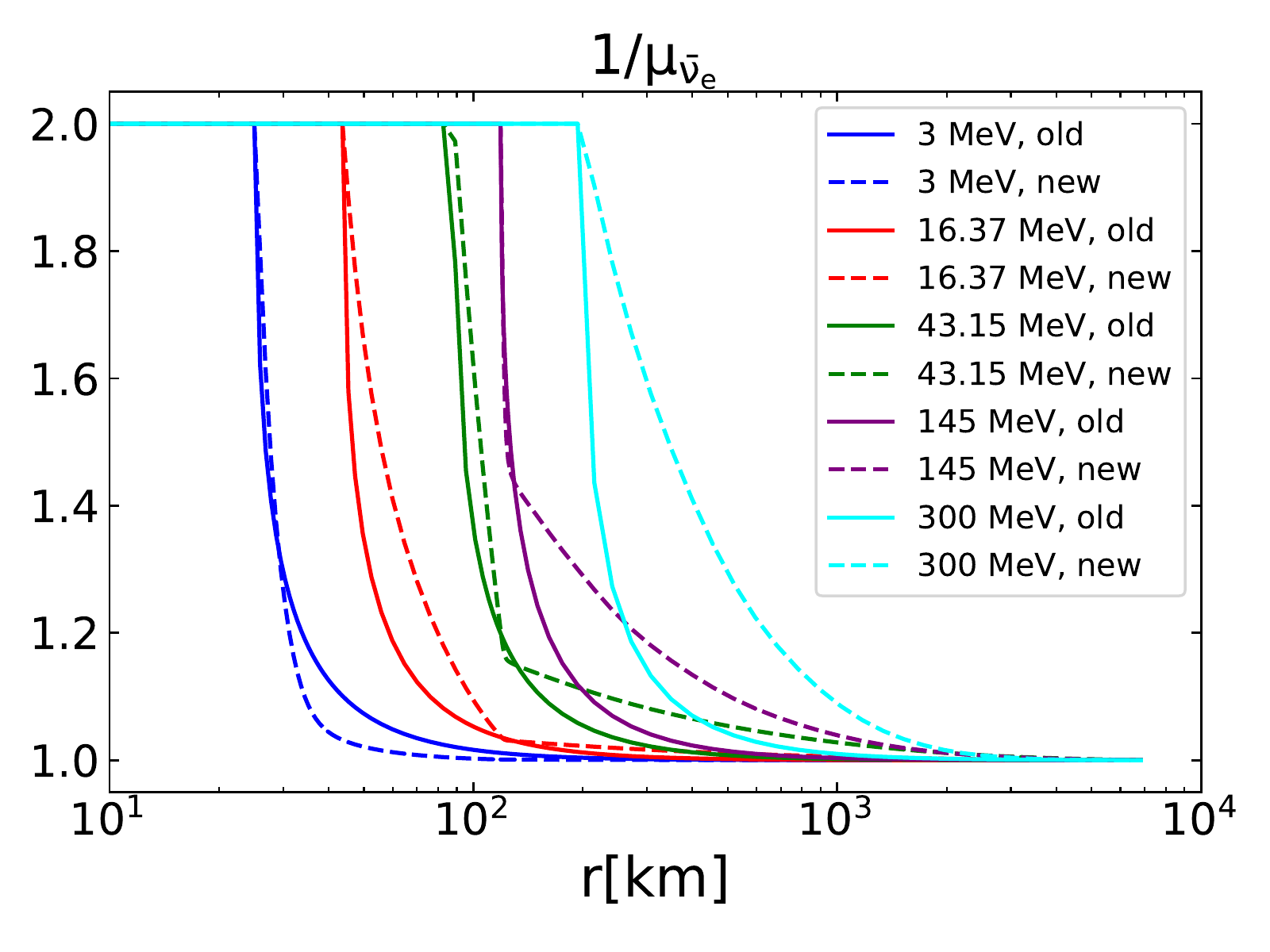}
\end{minipage}\\
\caption{Inverse of the electron neutrino (left panel) and anti-neutrino (right panel) flux factor for the prescriptions of Eqs.~(\ref{eq:fluxf}) and (\ref{eq:finalfluxf}), labelled as \textit{old} and \textit{new} respectively. Different neutrino energies of the global spectrum are chosen as reference. Note the larger deviations from the old prescription at high energies, due to the $\tau$-dependence of the
new prescription.} 
\label{fig:comparisonfluxf}
\end{figure*}
\subsubsection{Flux factor and heating}
In Figure \ref{fig:comparisonfluxf} 
we compare the inverse
of the flux factor for the two prescriptions of Eqs.~(\ref{eq:fluxf}) and 
(\ref{eq:finalfluxf}), 
labelled as \textit{old} and \textit{new} 
respectively, for electron neutrinos and anti-neutrinos
at different energies, 
as it enters Eq.~(\ref{eq:heating}) and it 
therefore affects the local heating.
The largest differences are at energies 
$\gtrsim 40$ MeV with discrepancies 
up to $\sim 25-30\%$.
Such differences are expected since the flux
factor from Eq.~(\ref{eq:finalfluxf})
depends on the optical depth rather than 
the radius as in Eq.~(\ref{eq:fluxf}), and
therefore the trend in Figure 
\ref{fig:comparisonfluxf}
resembles the trend of the optical depth.
Nevertheless, as we have seen in 
Figure \ref{fig:shr1D}
the largest contribution to the
heating comes from neutrinos 
of energies of few tens of MeV. 
We therefore do not expect such 
differences to contribute 
sensitively to the global neutrino 
luminosities for our snapshot calculations.
 In fact, in the bottom panels of 
Figure \ref{fig:netrate} we show 
the quantity $-\Dot{Q}_{\nu}^{k=1}$ in 
units of $10^{20}\:\rm{erg\:g^{-1}\:s^{-1}}$
calculated
from Eq.~(\ref{eq:ennetrate}),
which is a measure of the 
local heating rate. Overall, 
we see a very good agreement between  
the old (blue line) and new (red line) 
flux factor prescriptions, with 
differences at the order of a few percent
in the region where heating gets important 
($-\Dot{Q}_{\nu}^{k=1} > 0$).
More detailed quantification of the heating
with our new flux factor prescription
during full dynamical simulations will be
the subject of future work.\\
As an additional test, we show in Figure \ref{fig:netrate}
a comparison at the same time post-bounce of 
the same quantity $-\Dot{Q}_{\nu}^{k=1}$
provided by a dynamical 
evolution of the same progenitor 
starting at the onset of collapse
performed with GR1D (black-dashed line). 
Compared to
our test with the ASL, the net rate 
from GR1D provides less cooling 
(less negative $-\Dot{Q}_{\nu}^{k=1}$) in the 
ranges $\sim 30-50$ km and $\sim 65-80$ km,
with a larger one only in the range
$\sim 50-65$ km for electron neutrinos.
This is the result of a dynamical
evolution with a different neutrino transport, 
where at the same time after bounce the structure
of the star shows significant
differences with respect
to the ASL run, clearly visible in the density and 
temperature profiles on the top panels of Figure \ref{fig:netrate}. 
In particular, the GR1D profile has a colder and 
less compact layer within $\sim 30-50$ km where
most of the emission occurs.
Note also the different location of the shock, 
which is located at $\sim 105$ km for the GR1D case. 
On the other hand, we do not see a 
difference in the peak heating
rate near 100 km, with $-\Dot{Q}_{\nu}^{k=1} 
\sim 2.5\cdot10^{20}$ erg 
$\rm{g}^{-1} \rm{s}^{-1}$ for both runs.

\subsubsection{Neutrino luminosities and average-rms energies}
Given the specific net rates we calculate the
total neutrino luminosities and average 
energies given by Eqs.~(\ref{eq:Len})
and (\ref{eq:meanen}).
The results are shown in the first row of 
Table \ref{Table3} for both the old and 
the new flux factor prescriptions.
The latter reduces the luminosities 
compared to the former by $\lesssim 5\%$. 
The average energies are 
less affected
with differences $\lesssim 1\%$.
For completeness, we also 
calculate the rms neutrino energies 
from Eq.~(\ref{eq:rms}) and find 
$E_{\rm{rms},\nu_e}= 14.21$ MeV,
$E_{\rm{rms},\bar{\nu}_e}= 17.07$ MeV, 
$E_{\rm{rms},\nu_x}= 25.33$ MeV, 
in agreement with \cite{Perego2016}. We
further show the results obtained 
by performing the same calculations with GR1D in the
last row of Table \ref{Table3}. 
The electron neutrino and anti-neutrino
luminosities are 
lower by $\lesssim 10\%$ compared 
to the ASL runs due to the combination of an 
overall weaker neutrino cooling 
and comparable heating. 
Heavy lepton neutrinos have instead 
$\sim 7\%$ lower luminosity. 
By looking at the average energies,
differences are at the level of
$\lesssim 5\%$ for electron neutrinos 
and anti-neutrinos, and of $\sim 7\%$
for heavy-lepton neutrinos.
A similar trend is seen in 
the rms energies.
Overall, beside the different grid
setups, these discrepancies
are a consequence of the usage of
different neutrino transport schemes.

\subsubsection{Number of energy bins}
We finally determine the 
number of energy bins that 
are needed for trustworthy results for the luminosities and
average energies. We first vary the number of 
energy bins in the fixed energy 
interval [3, 300] MeV. 
Table \ref{Table1} summarises our 
findings. A number of energy bins $\lesssim 10$ causes 
sensitive deviation from a regime of convergence
that is visible at larger numbers
(which includes our preliminary
choice of 20 energy groups in [3, 300] MeV). In particular,
we notice a decrease in the luminosities and to 
a minor extent in the average energies, as a result 
of poorly resolved energy-integrated emission and 
absorption rates. We therefore choose 20 energy bins 
and vary the energy interval. In this way, variations
in the simulation outcome  by a reduction 
of the spectrum size would provide information 
on those energy ranges that 
are too small regardless
of the number of energy bins. Moreover,
we can assess whether the regime of convergence 
with 20 bins is satisfied for wider intervals of
energies or if it is strictly bound to certain 
energy ranges. Results are shown in Table 
\ref{Table2}. 
We see that cutting the energy 
spectrum [3, 300] MeV at high energies leads to
a significant decrease 
of $L_{\nu_x}^1$ that starts
appearing from spectra with upper energies below
$\sim 75$ MeV and that worsen by up to about 
$50\%$ decrease for the smallest
range [3, 30] MeV.
The average energy $\langle{}E_{\nu_x}\rangle{}$ is
also reduced to 14.37 MeV. This reduction is due 
to the lack of contribution coming from energies 
$\gtrsim 40$ MeV, that can still be relevant 
to the emission from heavy-leptons 
(see left panel of Figure \ref{fig:ser1D}).
In contrast, electron neutrino 
and anti-neutrino values remain 
almost stationary. In the same way, cutting the 
energy spectrum at low energies from 3 MeV has a
strong impact on the luminosities, with a reduction
of up to $70-80\%$ for electron neutrinos and 
anti-neutrinos in the case of the [25, 300] MeV
range. The average energies show the 
opposite trend, 
increasing as a result
of the high energies 
giving contribution to the
luminosity $L_{\nu}^1$
via Eqs.~(\ref{eq:emission}),
(\ref{eq:absorption}) and
(\ref{eq:ennetrate}) and thus affecting the
mean of Eq.~(\ref{eq:meanen}). 
On the other hand, we notice a convergence in
the values of $L_{\nu}^1$ and  
$\langle{}E_{\nu}\rangle{}$ 
for spectra 
spanning a range from 3 
MeV to hundreds of MeV.
In particular, no sensitive 
variations are seen
by extending the interval 
of energies above 300 MeV.
The smallest interval above which we start 
seeing convergence is [3, 75] MeV. 
However, it is important to specify 
that we are here basing our 
convergence tests by just looking at
global neutrino luminosities
and neglecting the convergence
of the fractions of neutrino 
trapped components $Y_{\nu}$ and 
$Z_{\nu}$ which are 
crucial in the modelling of the
diffusion while performing
dynamical simulations.
Such convergence requires
neutrinos of energies 
at least equal to
the neutrino chemical potential
in the core (set by the 
beta-equilibrium condition),
i.e. $\gtrsim 100$ MeV.
Moreover, we want
to stress that our
convergence tests are performed 
over a snapshot, but the way the 
selected energy interval affects 
the emission can change for different
stages of a dynamical evolution.
Summarising,
a number of 20 energy bins for the neutrino
spectrum is a reasonable choice 
despite the assumed interval of energies, provided that 
there are neither cuts in the spectrum at low 
energies nor at energies that would exclude 
neutrinos of $\gtrsim 100$ MeV. 
\begin{figure*}
\begin{minipage}{0.4 \linewidth}
\centering
\includegraphics[width = 1 \linewidth]{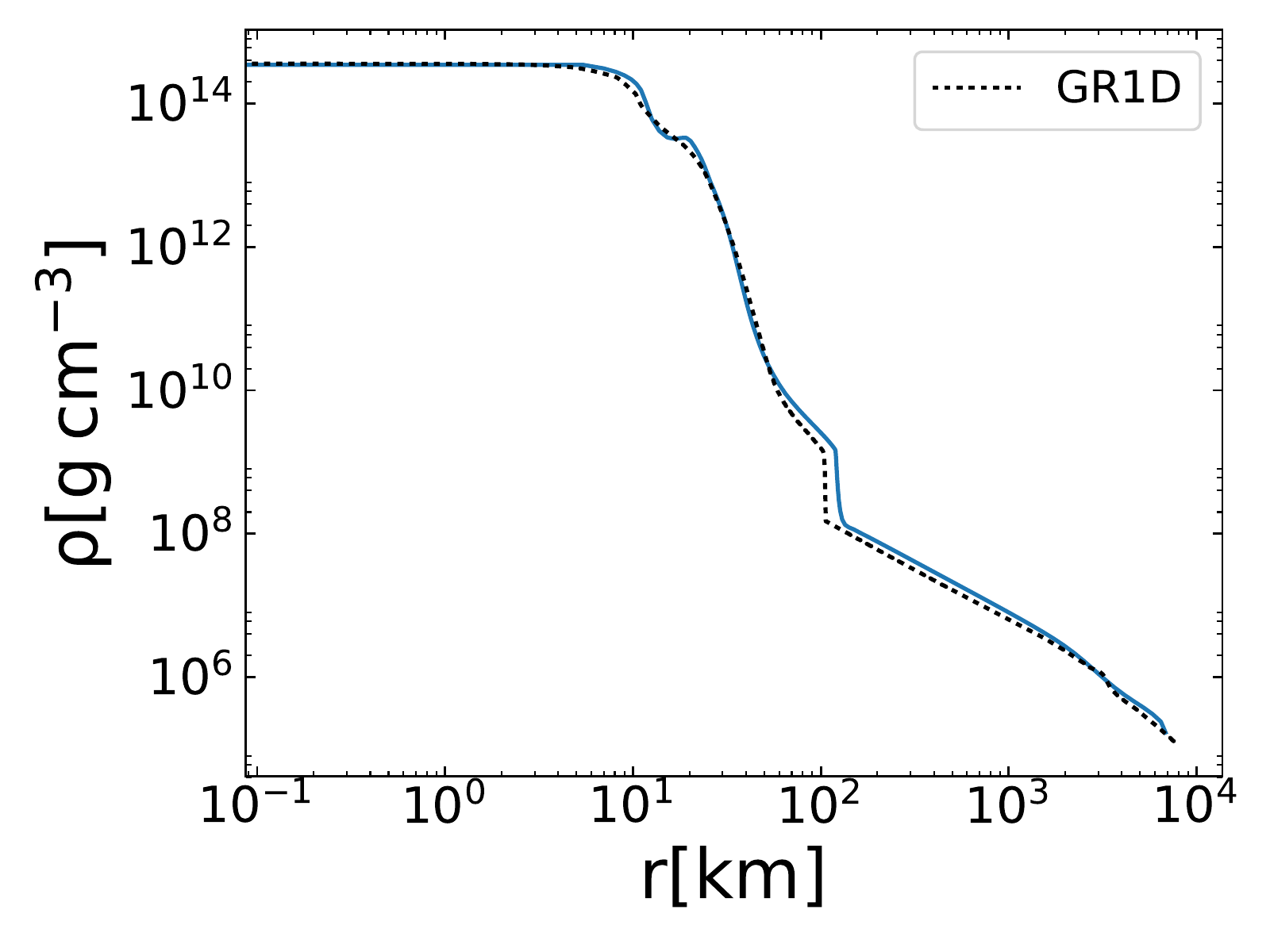}
\end{minipage}
\begin{minipage}{0.4 \linewidth}
\centering
\includegraphics[width = 1 \linewidth]{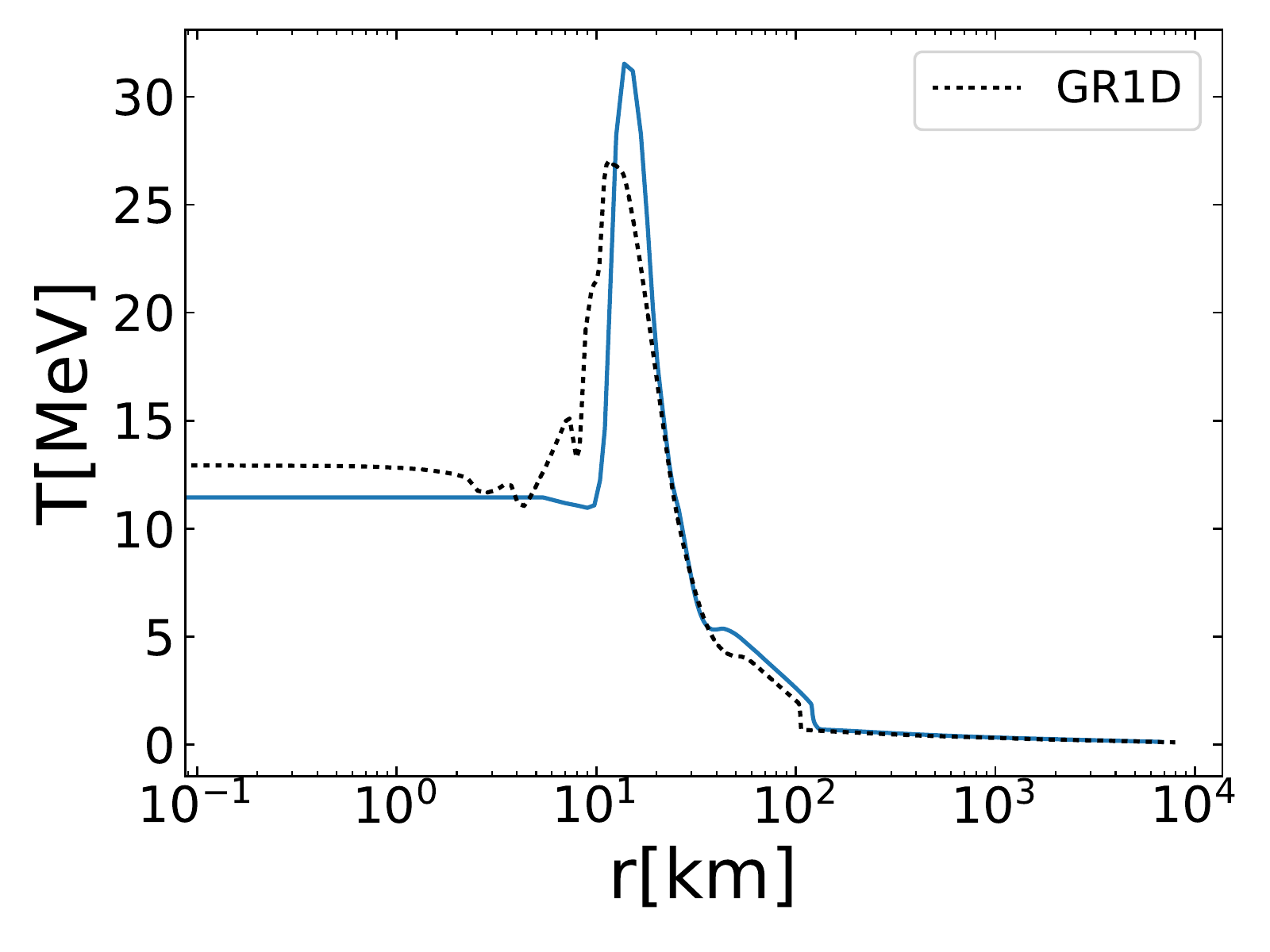}
\end{minipage}\\
\begin{minipage}{0.4 \linewidth}
\centering
\includegraphics[width = 1 \linewidth]{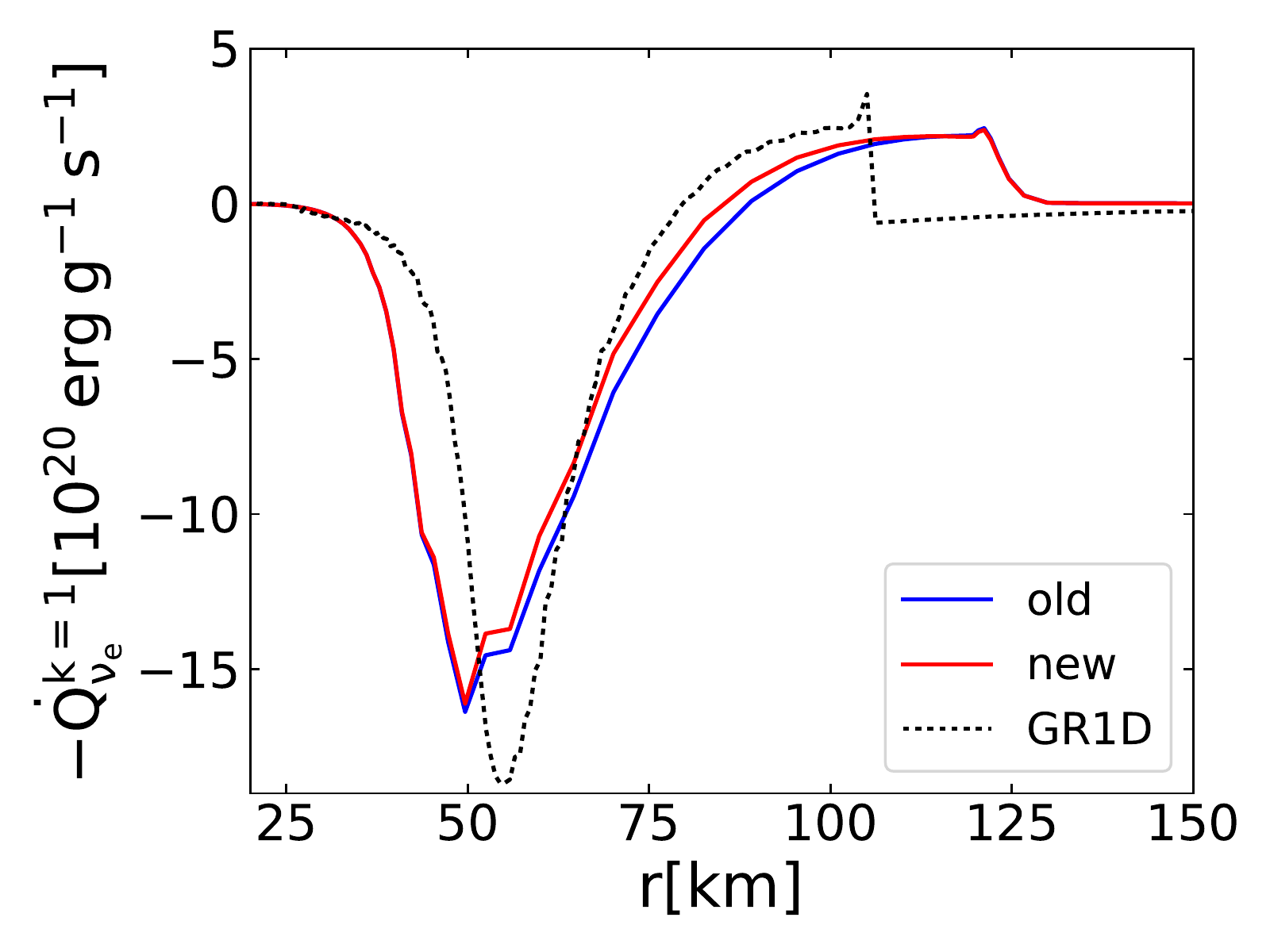}
\end{minipage}
\begin{minipage}{0.4 \linewidth}
\centering
\includegraphics[width = 1 \linewidth]{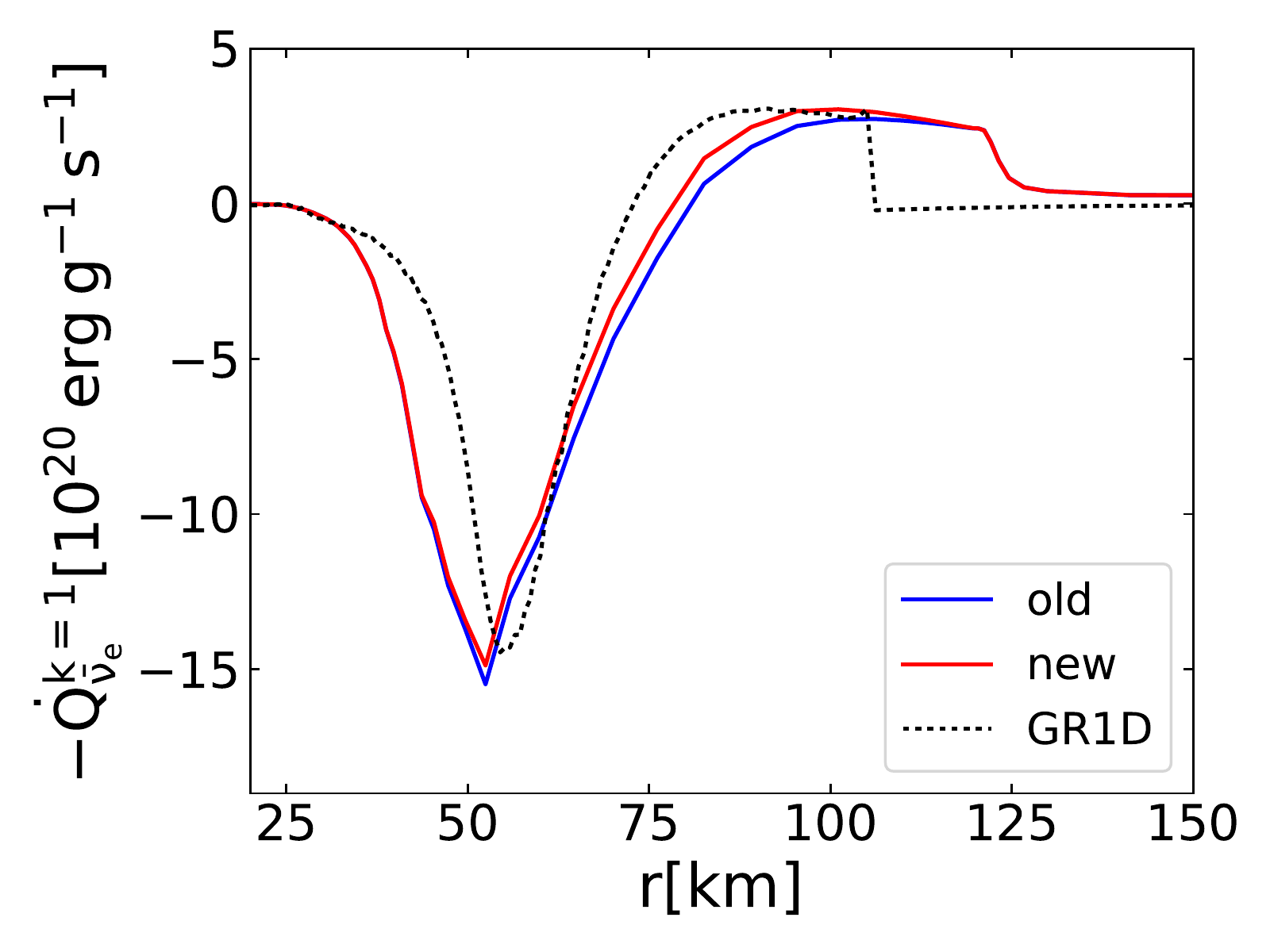}
\end{minipage}\\
\caption{\textbf{Top row:} density (left panel) and temperature (right panel) profiles as in Figure \ref{fig:initialdata} but with the addition of the data from the GR1D run. \textbf{Bottom row:} Measure of the heating rate in units of $10^{20}\:\rm{erg\:g^{-1}\:s^{-1}}$ along the radial profile for electron neutrinos (left panel) and anti-neutrinos (right panel) with the old (blue) and new (red) prescription for the flux factor. An overall good agreement between the two choices of flux factor is visible, with
differences at the level of few percents. 
Positive values mark the 
dominance of neutrino absorption over emission
which can be seen behind the shock location at
$\sim 120$ km.
The dashed line shows the result 
from a dynamical 
simulation with GR1D.}
\label{fig:netrate}
\end{figure*}
\begin{table*}
    \setlength{\tabcolsep}{10pt}
    \renewcommand{\arraystretch}{1.5}
	\centering
	\caption{Summary of the values of neutrino luminosities and average energies with ASL for both prescriptions of flux factors and for both 1D and 3D implementations.  Considering the comparison with the 1D implementation of Sec.~\ref{sec:ASLvsGR1D}, the 3D implementation provides electron neutrino and anti-neutrino luminosities larger by $\sim 7-8\%$ with the new choice of flux factor. Heavy lepton neutrinos are less affected with a discrepancy of $\sim 1\%$ instead. Similar trend is observed with the old flux factor prescription, with deviations reaching
	$\sim 10\%$ for electron neutrinos and anti-neutrinos. Such discrepancies reduce to $\sim 3-4\%$ and to $\sim 1-2\%$ with old and new prescription respectively when comparing with the same 1D radial profile but with uniform resolution of 1 km. Comparing the two flux factor prescriptions in 3D, the new one provides electron neutrino and anti-neutrino luminosities $\sim 6-7\%$ smaller than the old one. The 1D comparison between the two flux factor choices provides a slightly lower percentage of $\sim 4-5\%$ instead. Overall, discrepancies in the average energies are of the order of $\lesssim 1-2\%$. We additionally show the values obtained by performing the same calculations 
	with the M1 scheme of GR1D.}
	\begin{tabular}{c|c|c}
		\hline
		\multicolumn{1}{c}{Implementation} &
		\multicolumn{1}{c}{flux factor, old choice} &
		\multicolumn{1}{c}{flux factor, new choice}\\
		\hline
		\multirow{6}{*}{1D, variable resolution, ASL} & $L_{\nu_e}^1= 4.70\cdot10^{52} \rm{erg\:s^{-1}}$ & $L_{\nu_e}^1= 4.50\cdot10^{52}\:\rm{erg\:s^{-1}}$\\
		& $L_{\bar{\nu}_e}^1= 4.61\cdot10^{52}\:\rm{erg\:s^{-1}}$ & $L_{\bar{\nu}_e}^1= 4.43\cdot10^{52}\:\rm{erg\:s^{-1}}$ \\
		& $L_{\nu_x}^1= 2.04\cdot10^{52}\:\rm{erg\:s^{-1}}$ & $L_{\nu_x}^1= 2.04\cdot10^{52}\:\rm{erg\:s^{-1}}$\\ 
	    & $\langle{}E_{\nu_e}\rangle{}= 12.86\:\rm{MeV}$ &  $\langle{}E_{\nu_e}\rangle{}= 12.77\:\rm{MeV}$ \\
	    & $\langle{}E_{\bar{\nu}_e}\rangle{}= 15.69\:\rm{MeV}$ &  $\langle{}E_{\bar{\nu}_e}\rangle{}= 15.52\:\rm{MeV}$ \\
	    & $\langle{}E_{\nu_x}\rangle{}= 20.35\:\rm{MeV}$ &  $\langle{}E_{\nu_x}\rangle{}= 20.35\:\rm{MeV}$ \\ \hline
	    \multirow{6}{*}{1D, uniform resolution of 1 km, ASL} & $L_{\nu_e}^1= 5.00\cdot10^{52} \rm{erg\:s^{-1}}$ & $L_{\nu_e}^1= 4.76\cdot10^{52}\:\rm{erg\:s^{-1}}$\\
		& $L_{\bar{\nu}_e}^1= 4.88\cdot10^{52}\:\rm{erg\:s^{-1}}$ & $L_{\bar{\nu}_e}^1= 4.67\cdot10^{52}\:\rm{erg\:s^{-1}}$ \\
		& $L_{\nu_x}^1= 2.01\cdot10^{52}\:\rm{erg\:s^{-1}}$ & $L_{\nu_x}^1= 2.01\cdot10^{52}\:\rm{erg\:s^{-1}}$\\ 
	    & $\langle{}E_{\nu_e}\rangle{}= 12.98\:\rm{MeV}$ &  $\langle{}E_{\nu_e}\rangle{}= 12.82\:\rm{MeV}$ \\
	    & $\langle{}E_{\bar{\nu}_e}\rangle{}= 15.72\:\rm{MeV}$ &  $\langle{}E_{\bar{\nu}_e}\rangle{}= 15.56\:\rm{MeV}$ \\
	    & $\langle{}E_{\nu_x}\rangle{}= 20.33\:\rm{MeV}$ &  $\langle{}E_{\nu_x}\rangle{}= 20.33\:\rm{MeV}$ \\ \hline
	    \multirow{6}{*}{3D, uniform resolution of 1 km, ASL} & $L_{\nu_e}^1= 5.17\cdot10^{52}\: \rm{erg\:s^{-1}}$ & $L_{\nu_e}^1= 4.82\cdot10^{52}\:\rm{erg\:s^{-1}}$\\
		& $L_{\bar{\nu}_e}^1= 5.11\cdot10^{52}\:\rm{erg\:s^{-1}}$ & $L_{\bar{\nu}_e}^1= 4.79\cdot10^{52}\:\rm{erg\:s^{-1}}$ \\
		& $L_{\nu_x}^1= 2.01\cdot10^{52}\:\rm{erg\:s^{-1}}$ & $L_{\nu_x}^1= 2.01\cdot10^{52}\:\rm{erg\:s^{-1}}$\\ 
	    & $\langle{}E_{\nu_e}\rangle{}= 12.89\:\rm{MeV}$ &  $\langle{}E_{\nu_e}\rangle{}= 12.63\:\rm{MeV}$ \\
	    & $\langle{}E_{\bar{\nu}_e}\rangle{}= 15.77\:\rm{MeV}$ &  $\langle{}E_{\bar{\nu}_e}\rangle{}= 15.50\:\rm{MeV}$ \\
	    & $\langle{}E_{\nu_x}\rangle{}= 20.17\:\rm{MeV}$ &  $\langle{}E_{\nu_x}\rangle{}= 20.17\:\rm{MeV}$ \\ \hline
	    \multirow{6}{*}{GR1D, M1} &
	    \multicolumn{2}{c}{$L_{\nu_e}^1= 4.20\cdot10^{52}\: \rm{erg\:s^{-1}}$} \\
	    \multirow{6}{*}{} &
	    \multicolumn{2}{c}{$L_{\bar{\nu}_e}^1= 4.10\cdot10^{52}\: \rm{erg\:s^{-1}}$} \\
	    \multirow{6}{*}{} &
	    \multicolumn{2}{c}{$L_{\nu_x}^1= 1.90\cdot10^{52}\: \rm{erg\:s^{-1}}$}\\
	    \multirow{6}{*}{} &
	    \multicolumn{2}{c}
	    {$\langle{}E_{\nu_e}\rangle{}= 13.30\:\rm{MeV}$} \\
	    \multirow{6}{*}{} &
	    \multicolumn{2}{c}
	    {$\langle{}E_{\bar{\nu}_e}\rangle{}= 16.40\:\rm{MeV}$}\\
	    \multirow{6}{*}{} &
	    \multicolumn{2}{c}
	    {$\langle{}E_{\nu_x}\rangle{}= 18.90\:\rm{MeV}$} \\
	    \hline
	\end{tabular}
	\label{Table3}
\end{table*}
\begin{table*}
	\centering
	\caption{Variation of the neutrino luminosities and average energies by changing the number of energy bins in the interval [3, 300] MeV. Choosing a number of energy bins larger than $\sim 10$ leads to stable values, whereas fewer bins cause notable deviations.}
	\begin{tabular}{ccccccc} % four columns, alignment for each
		\hline
		Energy bins & $L_{\nu_e}^1 (\rm{erg\:s^{-1}})$ & $L_{\bar{\nu}_e}^1 (\rm{erg\:s^{-1}})$ & $L_{\nu_x}^1 (\rm{erg\:s^{-1}})$ & $\langle{}E_{\nu_e}\rangle{}$ (MeV) &  $\langle{}E_{\bar{\nu}_e}\rangle{}$ (MeV)& 
		$\langle{}E_{\nu_x}\rangle{}$ (MeV)\\
		\hline
		5 & $3.43\cdot10^{52}$ & $3.36\cdot10^{52}$ & $1.67\cdot10^{52}$ & 11.66 & 15.45 & 18.88\\
		10 & $4.45\cdot10^{52}$ & $4.43\cdot10^{52}$ & $2.00\cdot10^{52}$ & 12.71 & 15.59 & 20.13\\
		15 & $4.59\cdot10^{52}$ & $4.48\cdot10^{52}$ & $2.02\cdot10^{52}$ & 12.78 & 15.57 & 20.25\\
		20 & $4.54\cdot10^{52}$ & $4.48\cdot10^{52}$ & $2.04\cdot10^{52}$ & 12.77 & 15.52 & 20.35\\
		25 & $4.65\cdot10^{52}$ & $4.51\cdot10^{52}$ & $2.01\cdot10^{52}$ & 12.88 & 15.57 & 20.22\\
		30 & $4.66\cdot10^{52}$ & $4.55\cdot10^{52}$ & $2.05\cdot10^{52}$ & 12.88 & 15.62 & 20.41\\
		35 & $4.57\cdot10^{52}$ & $4.53\cdot10^{52}$ & $2.03\cdot10^{52}$ & 12.77 & 15.59 & 20.34\\
		40 & $4.66\cdot10^{52}$ & $4.56\cdot10^{52}$ & $2.03\cdot10^{52}$ & 12.89 & 15.64 & 20.34\\
		45 & $4.62\cdot10^{52}$ & $4.55\cdot10^{52}$ & $2.04\cdot10^{52}$ & 12.82 & 15.61 & 20.38\\
		50 & $4.62\cdot10^{52}$ & $4.55\cdot10^{52}$ & $2.03\cdot10^{52}$ & 12.83 & 15.61 & 20.37\\
		55 & $4.58\cdot10^{52}$ & $4.57\cdot10^{52}$ & $2.04\cdot10^{52}$ & 12.82 & 15.63 & 20.40\\
		60 & $4.66\cdot10^{52}$ & $4.53\cdot10^{52}$ & $2.04\cdot10^{52}$ & 12.86 & 15.58 & 20.41\\
		\hline
	\end{tabular}
	\label{Table1}
\end{table*}
\begin{table*}
	\centering
	\caption{Variation of the luminosities and average energies for all neutrino species considered, with 20 energy bins in total but varying the energy interval. A stability is seen as long as the energy interval is between few MeV to few hundreds MeV. Either a restriction of the interval to energies below $\sim 75$ MeV or above $\sim 10$ MeV causes sensitive variations in either the luminosities or the average energies, or both. 
	Note that although a convergence in the luminosities
	is already seen for energies above $\sim 75$ MeV,
	dynamical simulations would also require 
	neutrino energies above $\sim 100$ MeV 
	to properly model the neutrino
	trapped components.}
	\begin{tabular}{ccccccc} % four columns, alignment for each
		\hline
		Energy range (MeV) & $L_{\nu_e}^1 (\rm{erg\:s^{-1}})$ & $L_{\bar{\nu}_e}^1 (\rm{erg\:s^{-1}})$ & $L_{\nu_x}^1 (\rm{erg\:s^{-1}})$ & $\langle{}E_{\nu_e}\rangle{}$ (MeV) &  $\langle{}E_{\bar{\nu}_e}\rangle{}$ (MeV)&  $\langle{}E_{\nu_x}\rangle{}$ (MeV)\\
		\hline
		$[3, 30]$ & $4.35\cdot10^{52}$ & $4.13\cdot10^{52}$ & $1.08\cdot10^{52}$ & 12.58 & 15.07 & 14.37\\
		$[3, 50]$ & $4.53\cdot10^{52}$ & $4.44\cdot10^{52}$ & $1.78\cdot10^{52}$ & 12.78 & 15.54 & 18.86\\
		$[3, 75]$ & $4.54\cdot10^{52}$ & $4.48\cdot10^{52}$ & $2.01\cdot10^{52}$ & 12.81 & 15.58 & 20.22\\
		$[3, 100]$ & $4.59\cdot10^{52}$ & $4.48\cdot10^{52}$ & $2.03\cdot10^{52}$ & 12.83 & 15.55 & 20.35\\
		$[3, 150]$ & $4.56\cdot10^{52}$ & $4.44\cdot10^{52}$ & $2.02\cdot10^{52}$ & 12.77 & 15.53 & 20.27\\
		$[3, 200]$ & $4.60\cdot10^{52}$ & $4.44\cdot10^{52}$ & $2.04\cdot10^{52}$ & 12.81 & 15.51 & 20.35\\
		$[3, 250]$ & $4.64\cdot10^{52}$ & $4.43\cdot10^{52}$ & $2.01\cdot10^{52}$ & 12.85 & 15.49 & 20.24\\
		$[3, 300]$ & $4.50\cdot10^{52}$ & $4.43\cdot10^{52}$ & $2.04\cdot10^{52}$ & 12.71 & 15.54 & 20.35\\
		$[3, 350]$ & $4.53\cdot10^{52}$ & $4.44\cdot10^{52}$ & $2.05\cdot10^{52}$ & 12.76 & 15.50 & 20.38\\
		$[3, 400]$ & $4.59\cdot10^{52}$ & $4.42\cdot10^{52}$ & $2.01\cdot10^{52}$ & 12.84 & 15.50 & 20.18\\
		$[3, 450]$ & $4.56\cdot10^{52}$ & $4.45\cdot10^{52}$ & $2.05\cdot10^{52}$ & 12.76 & 15.54 & 20.38\\
		$[3, 500]$ & $4.44\cdot10^{52}$ & $4.40\cdot10^{52}$ & $2.02\cdot10^{52}$ & 12.66 & 15.46 & 20.26\\
		$[3, 550]$ & $4.55\cdot10^{52}$ & $4.43\cdot10^{52}$ & $2.04\cdot10^{52}$ & 12.77 & 15.52 & 20.04\\
		$[3, 600]$ & $4.54\cdot10^{52}$ & $4.39\cdot10^{52}$ & $2.05\cdot10^{52}$ & 12.74 & 15.45 & 20.36\\
		$[10, 300]$ & $4.03\cdot10^{52}$ & $4.22\cdot10^{52}$ & $1.96\cdot10^{52}$ & 15.62 & 17.33 & 25.39\\
		$[15, 300]$ & $2.84\cdot10^{52}$ & $3.31\cdot10^{52}$ & $1.84\cdot10^{52}$ & 19.54 & 20.30 & 29.65\\
		$[25, 300]$ & $9.38\cdot10^{51}$ & $1.23\cdot10^{52}$ & $1.52\cdot10^{52}$ & 29.38 & 28.65 & 37.78\\
		\hline
	\end{tabular}
	\label{Table2}
\end{table*}
\begin{figure*}
\begin{minipage}{0.4 \linewidth}
\centering
\includegraphics[width = 1 \linewidth]{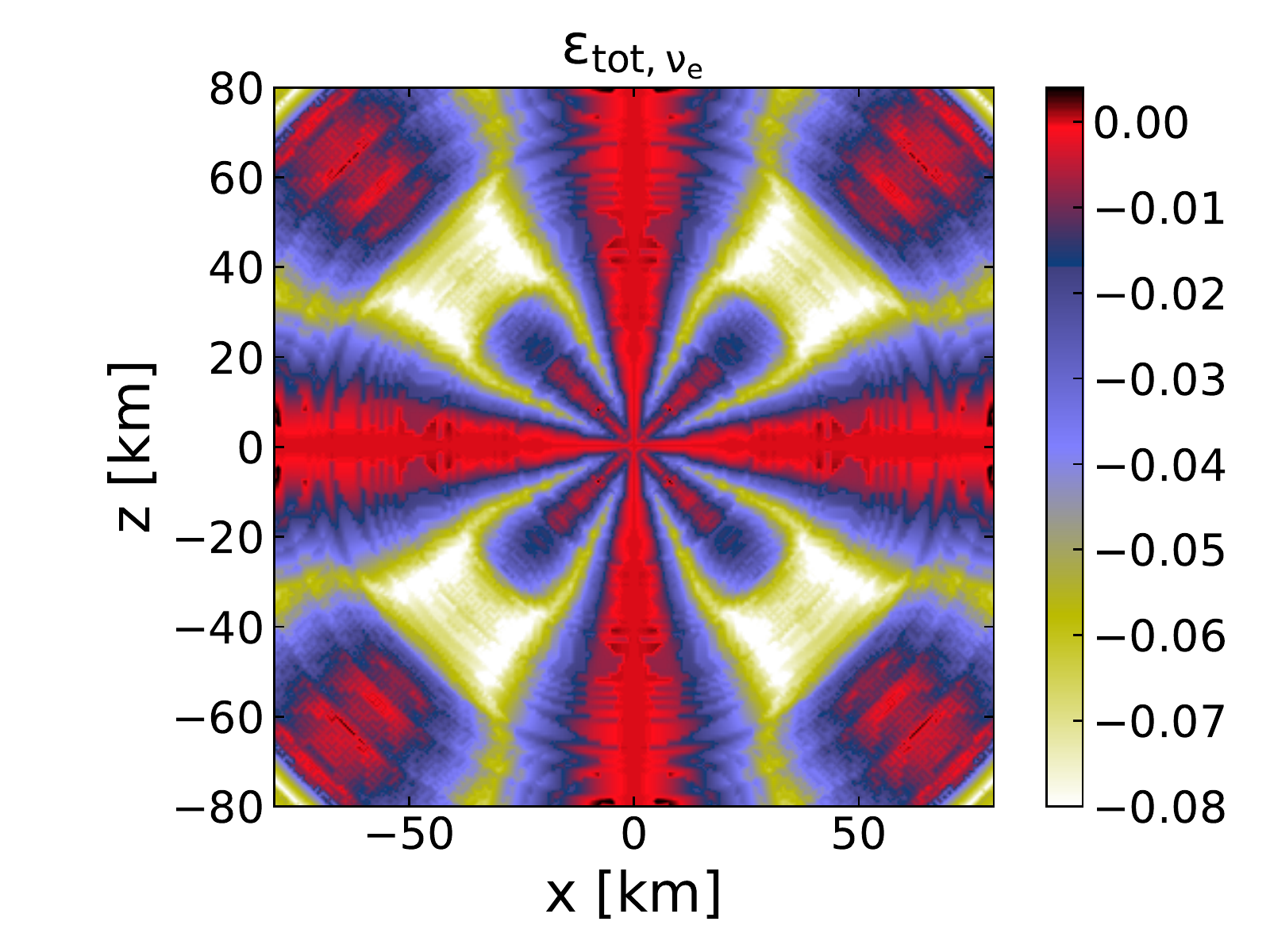}
\end{minipage}
\begin{minipage}{0.4 \linewidth}
\centering
\includegraphics[width = 1 \linewidth]{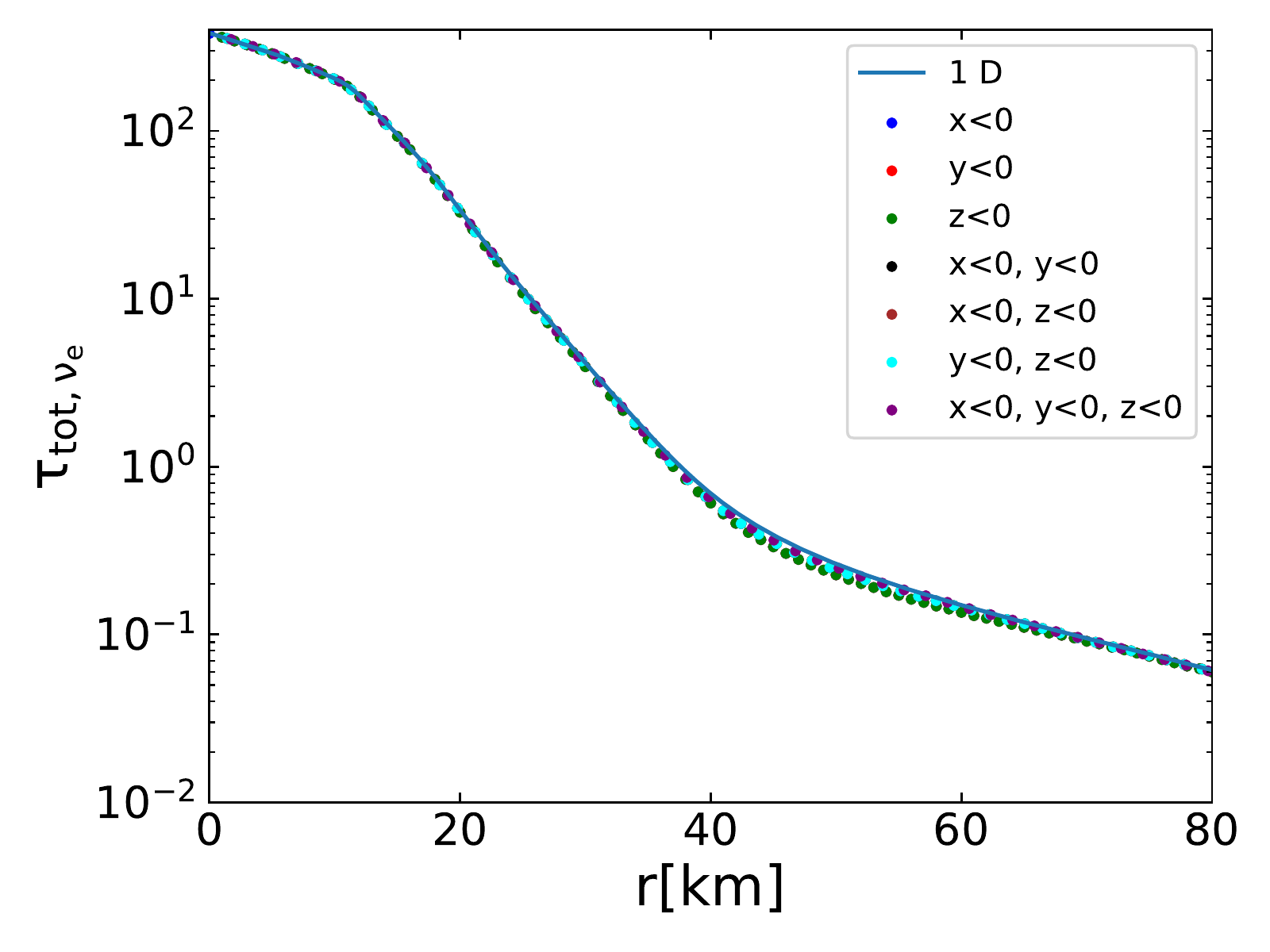}
\end{minipage}\\
\caption{\textbf{Left panel:} 
Relative error $\varepsilon_{\nu}=
\frac{\tau_{1D}-\tau_{3D}}{\tau_{1D}}$ of the 
total optical depth for electron neutrinos of
energy 10.08 MeV.
The 1D values of the 
optical depth on the plane are
obtained by creating a
similar 1D profile 
to the one of Sec.~\ref{sec:ASLvsGR1D}
but with uniform resolution of 1 km,
by integrating over this path,
and by mapping the resulting
values on the plane itself.
For the 3D calculations 
we use the approach explained in
Sec.~\ref{sec:3Dmeth}.
The 3D implementation 
generally overestimates the
optical depths by $\sim 8\%$ at the most.
\textbf{Right panel:} Total optical 
depth along the radius from the centre
for electron neutrinos of 10.08 MeV. 
Unlike the 1D profile 
of uniform resolution  
adopted in the left panel
to obtain the relative error, 
the blue line
corresponds to an integration
performed over the 1D profile
of variable resolution
of Sec.~\ref{sec:ASLvsGR1D}, 
while the dots 
correspond to a selection of radial
paths on the 3D grid. 'x<0','y<0' and
'z<0' are the paths along the x<0,y<0
and z<0 axis. 'x<0,y<0' is the diagonal
path on the z=0 plane with negative x
and y coordinates and x=y,'x<0,z<0' is the
diagonal path on the y=0 plane with negative
x and z coordinates and x=z,'y<0,z<0' 
is the diagonal path on the x=0 plane with negative
y and z coordinates and y=z,'x<0,y<0,z<0' is
the diagonal path with negative x,y and z
coordinates and x=y, x=z and y=z. 
The calculation of the 
optical depth along the 1D profile 
of variable resolution leads to a 
$\lesssim 10\%$ larger optical depth at
$\sim 40-60$ km from the centre
compared to the 3D implementation.}
\label{fig:opdep3D}
\end{figure*}
\begin{figure*}
\centering
\includegraphics[width=13 cm]{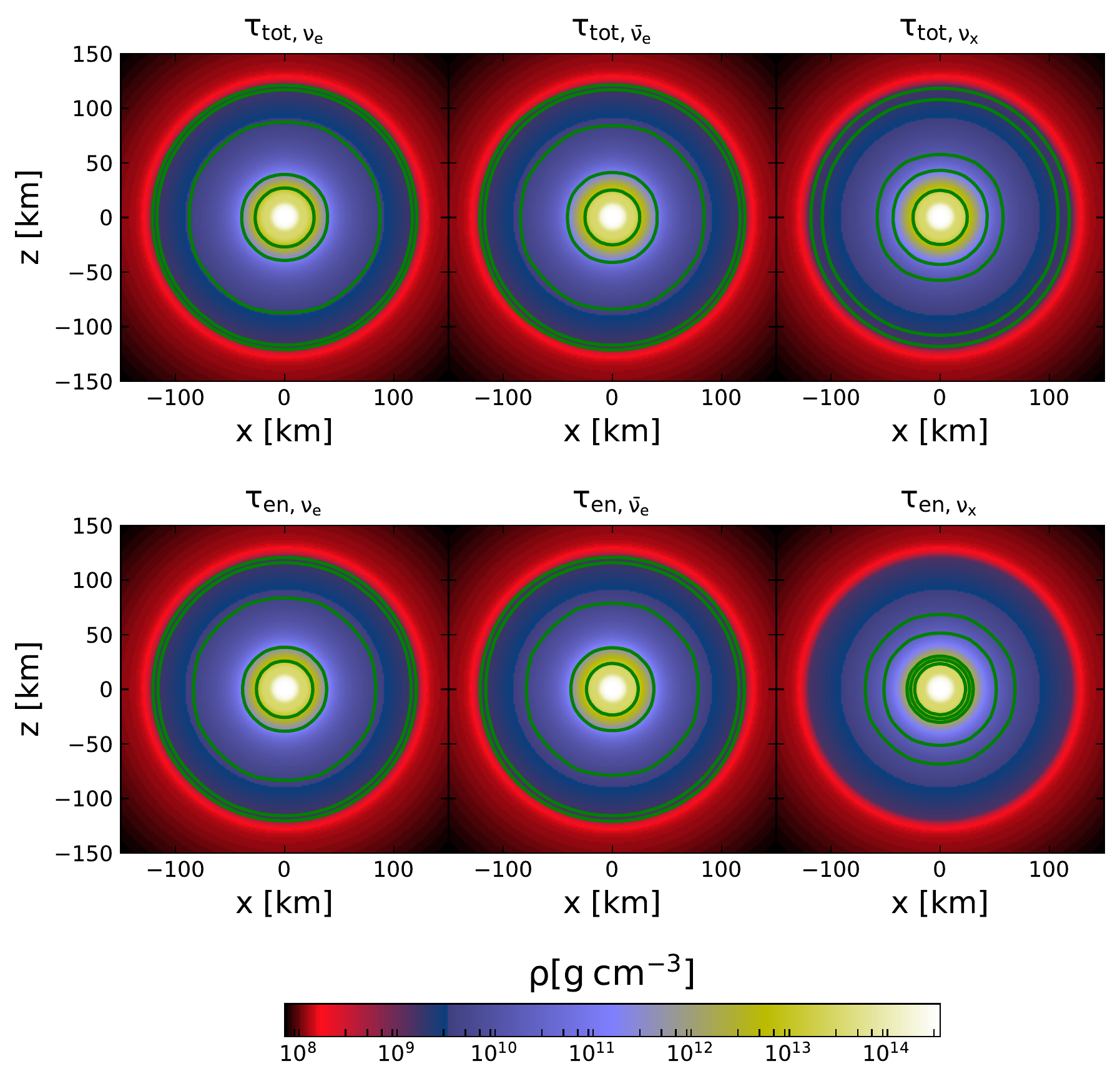}
\caption{Location of the neutrino surfaces
on the density map at y=0
for the same sets of neutrino energies 
used in Figures \ref{fig:ser1D} and
\ref{fig:shr1D}.
\textbf{Top row:} total optical depth 
and \textbf{Bottom row:} energy optical depth. 
For each row, we show electron neutrino 
(left panel), electron anti-neutrino 
(middle panel) and heavy-lepton neutrino 
(right panel). The location of the 
neutrinosphere is shown in green circles, 
starting from the inner one at E= 3 MeV 
to the outer one at E= 300 MeV, i.e. 
neutrino surface radii increase with neutrino energy.
While electron neutrinos and anti-neutrinos 
have similar neutrinosphere radii at all 
energies both in the total and in the energy 
optical depths, heavy-lepton neutrinos 
show smaller energy neutrinospheres than 
the other species because the only inelastic 
contribution comes from pair processes 
and bremsstrahlung. On the other hand, 
elastic scattering on nuclei and nucleons 
extends the heavy lepton total surfaces to
radii comparable with the other species. 
Comparable radii of total and energy neutrino
surfaces for electron neutrinos and 
anti-neutrinos indicate that 
neutrino emission and absorption reactions, 
efficient in thermalizing neutrinos,
provide also an important opacity contribution
to the total optical depth.}
\label{fig:neutrinosphere}
\end{figure*}
\subsection{ASL in 3D applications}
\label{sec:ASL3D}
\subsubsection{Comparison in spherical symmetry between 1D and 3D}
To scrutinise our multi-D implementation,
we start by taking the snapshot analyzed in 
Sec.~\ref{sec:ASLvsGR1D}, map the initial 
data on a 3D grid with uniform resolution
of 1 km and evolve the neutrino 
transport part until equilibrium. 
Instead of mapping the whole radial
profile which extends up to 6832 km at 
densities below $10^6\:\rm{g/cm^3}$ and 
temperatures of $\sim 0.1$ MeV, we take 
the profile information only up to 150 km
and neglect the remaining part to save 
computational time. Indeed, beyond such 
distances densities and temperatures are 
low enough that the neutrino contribution
to the outcome of the simulations is 
negligible, $\lesssim 1\%$ in the total luminosities 
and average energies. We choose the energy interval
[3, 300] MeV, see Sec.~\ref{sec:ASLvsGR1D}. 
We show the mapped initial conditions 
on the y=0 plane in the bottom 
panel of Figure \ref{fig:initialdata}.\\

\noindent{\em 3D optical depth}\\
To test our 3D implementation 
of the optical depths, we 
first calculate the optical
depth on the 3D grid
as explained in Sec.~\ref{sec:3Dmeth}.
We then create a 1D profile 
equivalent to the one 
used in Sec.~\ref{sec:ASLvsGR1D}
but with uniform resolution 
of 1 km, where we calculate
the optical depth by 
doing a simple integration
over the radial path. We finally map
such optical depth on the 
3D grid and calculate 
the relative error $\varepsilon_{\nu}=
\frac{\tau_{1D}-\tau_{3D}}{\tau_{1D}}$
on the y=0 plane.
In the left panel of 
Figure \ref{fig:opdep3D}
we show our result.
As reference case we
take the total optical 
depth for electron
neutrinos of energy 
$E_{\nu_e}= 10.08$ MeV, 
the other energies 
and neutrino species show
a similar behaviour. 
The differences between the
1D and the 3D calculations 
are at the level of 
$\sim 8\%$ at the most, with the
3D implementation providing
larger values overall. 
To get an idea of the distribution of the 
neutrino surfaces at different energies, we
show the location of the total and energy 
neutrinosphere radii on the density map at 
y=0 in Figure \ref{fig:neutrinosphere}. 
The selected set of energies are 
the same used to show the 
plots in Figures \ref{fig:ser1D} and
\ref{fig:shr1D}. 
The neutrino surfaces are almost
perfectly spherical, suggesting
that our optical depth algorithm 
overall captures the actual path that 
minimises the optical depth
and that neutrinos preferentially
cross, i.e. the radial path.
Obviously the larger the energy
the larger the radius of 
the neutrinosphere, since  
$\tau_{\nu} \sim E_{\nu}^2$. Comparing 
between species, heavy-lepton 
neutrinos have smaller energy neutrinospheres
because the only interactions where
they exchange energy with the
fluid are pair processes and bremsstrahlung.
On the contrary, elastic scattering on nucleons
and nuclei makes the total neutrino surfaces 
comparable with the other species. Electron neutrinos and anti-neutrinos show similar energy and total neutrinospheres 
as a result of the comparable amount 
of emission and absorption interactions involving both
species at this time of the post-bounce phase. Moreover, for each of these neutrino species we notice comparable 
radii of total and energy neutrino surface at 
each energy, indicating that 
neutrino emission and absorption reactions, 
efficient in thermalizing neutrinos,
provide also an important opacity contribution
to the total optical depth.
Overall, the total neutrino surfaces extend from 
$\sim 27$ km to $\sim 121$ km for electron
neutrinos, from $\sim 25$ km to $\sim 121$ 
km for electron anti-neutrinos, and from 
$\sim 25$ km to $\sim 118$ km for heavy- 
lepton neutrinos. Accordingly, the energy 
surfaces extend from $\sim 26$ km to $\sim 121$
km for electron neutrinos, from $\sim 24$ km 
to $\sim 121$ km for electron anti-neutrinos,
and from $\sim 23$ km to $\sim 69$ km for 
heavy-lepton neutrinos.\\

\noindent{\em Heating}\\
Calculation of the absorption rates on the
grid is done by applying Eq.~(\ref{eq:heating}).
Computation of the neutrino density is
performed by means of Eq.~(\ref{eq:new_nudensity}) 
with $\beta_{\nu}= 0$ (i.e. Eq.~(\ref{eq:nudensity})) 
and of Eq. (\ref{eq:spartlm}).
Given the spherical symmetry 
(see Figure \ref{fig:neutrinosphere}) 
the value of $l_{\nu}(E,R)$ over the
grid is approximately 
recovered by integrating 
Eq. (\ref{eq:spartlm}) over
a reference radial path from the origin,
and by mapping the obtained values on the path
over the rest of the grid.
\begin{figure*}
\begin{minipage}{0.4 \linewidth}
\centering
\includegraphics[width = 1 \linewidth]{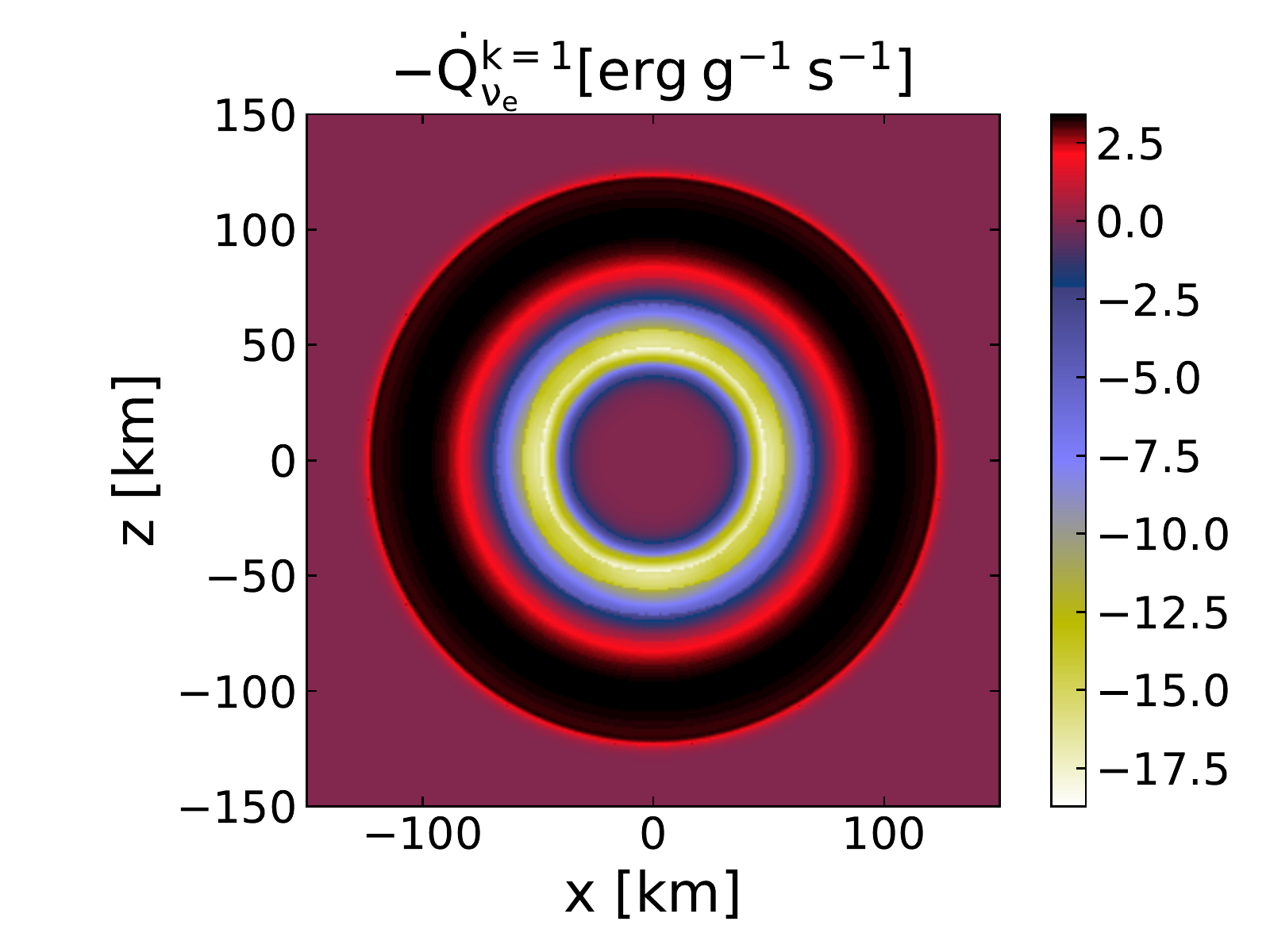}
\end{minipage}
\begin{minipage}{0.4 \linewidth}
\centering
\includegraphics[width = 1 \linewidth]{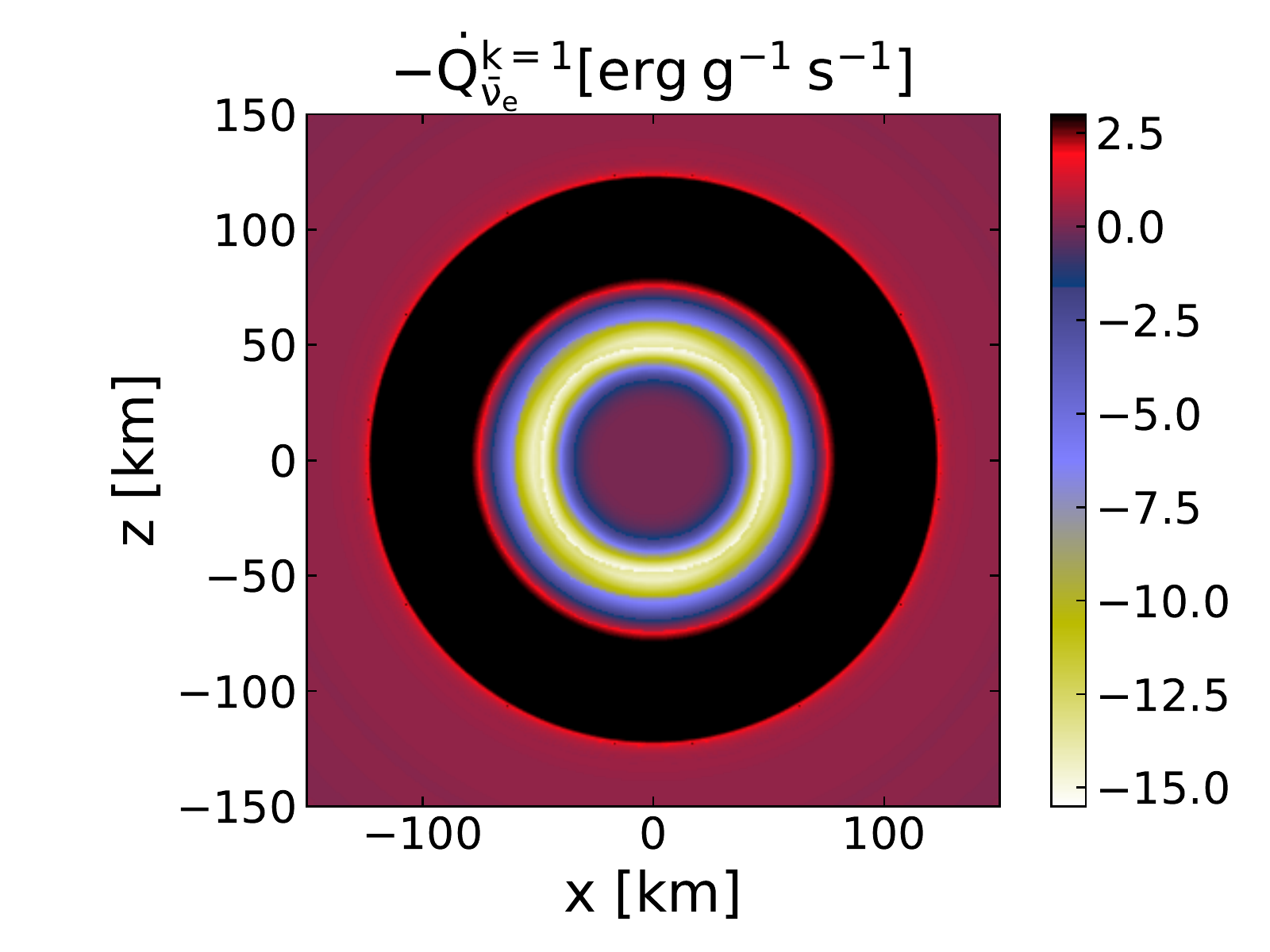}
\end{minipage}\\
\caption{Estimate of the heating rate on the y=0
in units of $10^{20}\:\rm{erg\:g^{-1}\:s^{-1}}$ 
with the new flux factor
prescription used for the modelling of the heating,
for electron neutrinos (left) and anti-neutrinos (right).
The values are in agreement with Figure 
\ref{fig:netrate}, with a positive rate between
$\sim 80$ and $\sim 120$ km from the origin. 
However, a stronger neutrino emission and 
absorption are observed with respect to the 1D 
implementation as a result of lower optical
depths leading to stronger diffusion rates 
in the transition from diffusion to free streaming.}
\label{fig:netrate3D}
\end{figure*}
Figure \ref{fig:netrate3D}
shows the heating rate 
for electron neutrinos and anti-neutrinos on 
the plane y=0 with the new flux factor prescription 
in units of $10^{20}\:\rm{erg\:g^{-1}\:s^{-1}}$, resembling the values 
from Figure \ref{fig:netrate}. Moving outward from 
the centre, the net rate decreases and reaches
its minimum at $\sim 50-60$ km from the centre,
then increasing and getting to positive values
where neutrino heating dominates between 80 km 
and 120 km. Unlike Figure \ref{fig:netrate} we 
notice a slightly lower minimum that goes below
$-17.5\cdot10^{20}\:\rm{erg\:g^{-1}\:s^{-1}}$
and $-15\cdot10^{20}\:\rm{erg\:g^{-1}\:s^{-1}}$ for electron neutrinos and 
anti-neutrinos respectively, and a 
larger maximum above $\sim 2.5\cdot10^{20} 
\:\rm{erg\:g^{-1}\:s^{-1}}$ for the heating.
This is expected because 
the original
profile of Sec.~\ref{sec:ASLvsGR1D}
has decreasing
resolution with increasing distance
from the centre, leading to larger
optical depths in the transition
between the optically thick and 
the optically thin regime. 
This is clearly visible in the right panel
of Figure \ref{fig:opdep3D}, where
we plot the total optical depth
for electron neutrinos of energy 10.08 MeV
(taken as reference)
calculated by integrating along the 
profile of Sec.~\ref{sec:ASLvsGR1D}
(blue line), together with values of
the optical depth calculated
along several paths
of the 3D grid. In the range
$\sim 40-60$ km the 1D 
optical depth is $\lesssim 10\%$ larger
than the 3D ones.
The lower values from the 3D
calculations lead to  
lower diffusion timescales
and therefore to 
stronger emission and
absorption rates.\\

\noindent{\em Neutrino luminosities and average-rms energies}\\
Given the heating rate, we
calculate the total neutrino luminosities 
and average energies. The results are shown 
in the third row of Table \ref{Table3}.
Differences with respect to the 1D case
(Sec.~\ref{sec:ASLvsGR1D}) 
are of the order of $\sim 7-8\%$ 
for the electron neutrino and anti-neutrino 
luminosities, and of $\sim 1\%$ for the
heavy-lepton neutrinos. 
Average energies all differ by 
$\lesssim 1\%$. Considering the results
with the old flux factor prescription,
differences in the luminosities compared to the 
1D case are of the order of $\sim 10\%$ for 
electron neutrinos and anti-neutrinos and 
$\sim 1\%$ for heavy lepton neutrinos. 
In addition, 
we repeat the calculation
for the case of a 1D radial 
profile with uniform 
resolution of 1 km. 
Results are shown in the second 
row of Table \ref{Table3}.
Errors in the luminosities between 
the 3D and the 1D implementation 
reduce to $\sim 3-4\%$ and to
$\sim 1-2\%$ with the old 
and new flux factor prescription 
respectively, confirming that the 
resolution contributes as 
source of variability to the 
outcome of the simulations. 
As we have similarly seen already 
in Sec.~\ref{sec:ASLvsGR1D}, 
the choice of the new flux 
factor in our 3D simulations provides 
luminosities $\lesssim 6-7\%$ lower respect 
to the old prescription, while average energies are 
less affected with discrepancies 
of $\lesssim 2\%$. 
Overall, we state that the
modification adopted to the standard ASL of 
\cite{Perego2016} perform well in the context
of core-collapse supernova, although more 
precise assessments require full dynamical evolutions. 

\subsubsection{Neutron star merger remnants in 3D}
\begin{figure*}
\begin{minipage}{0.33 \linewidth}
\centering
\includegraphics[width = 1 \linewidth]{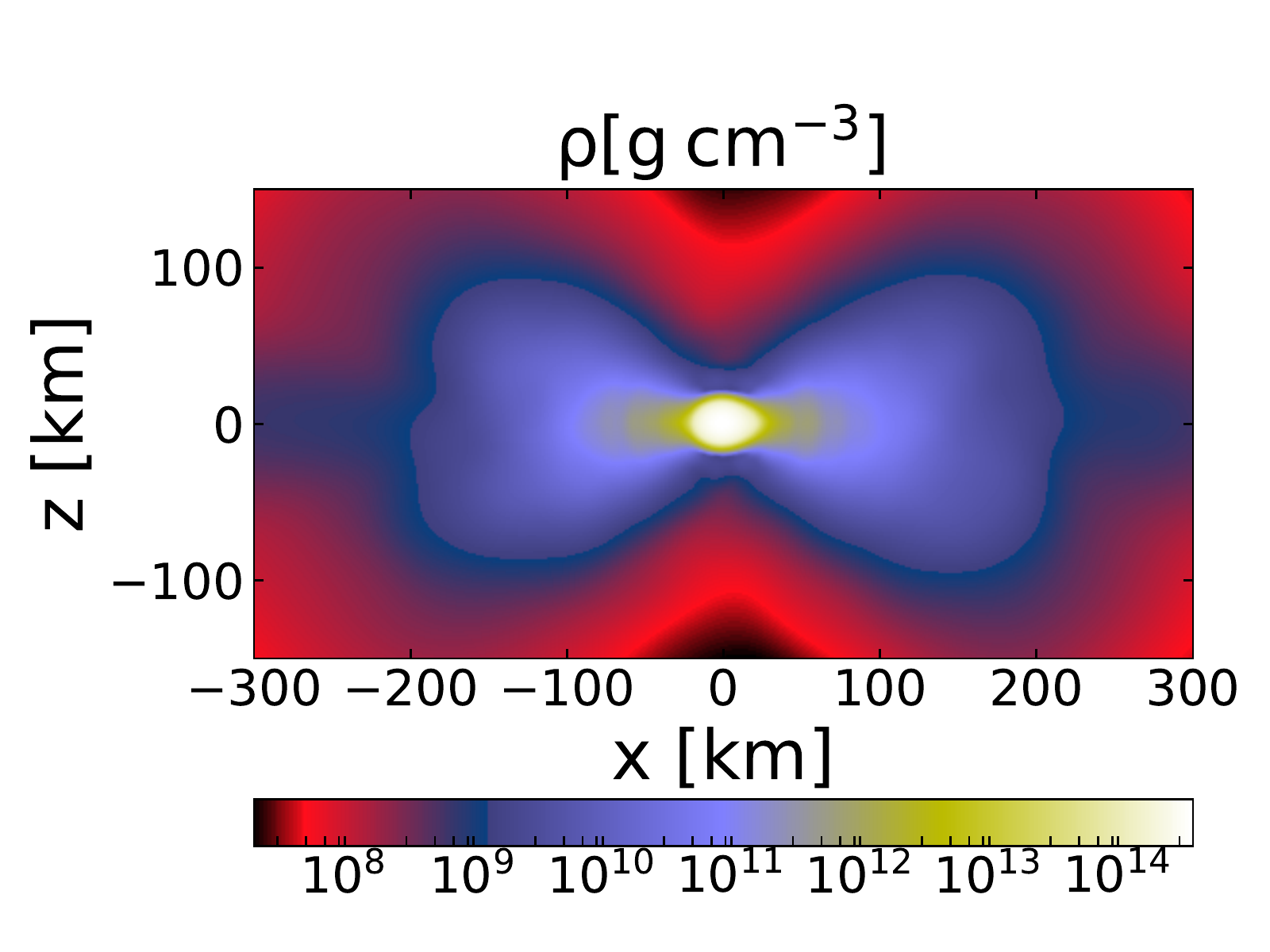}
\end{minipage}
\begin{minipage}{0.33 \linewidth}
\centering
\includegraphics[width = 1 \linewidth]{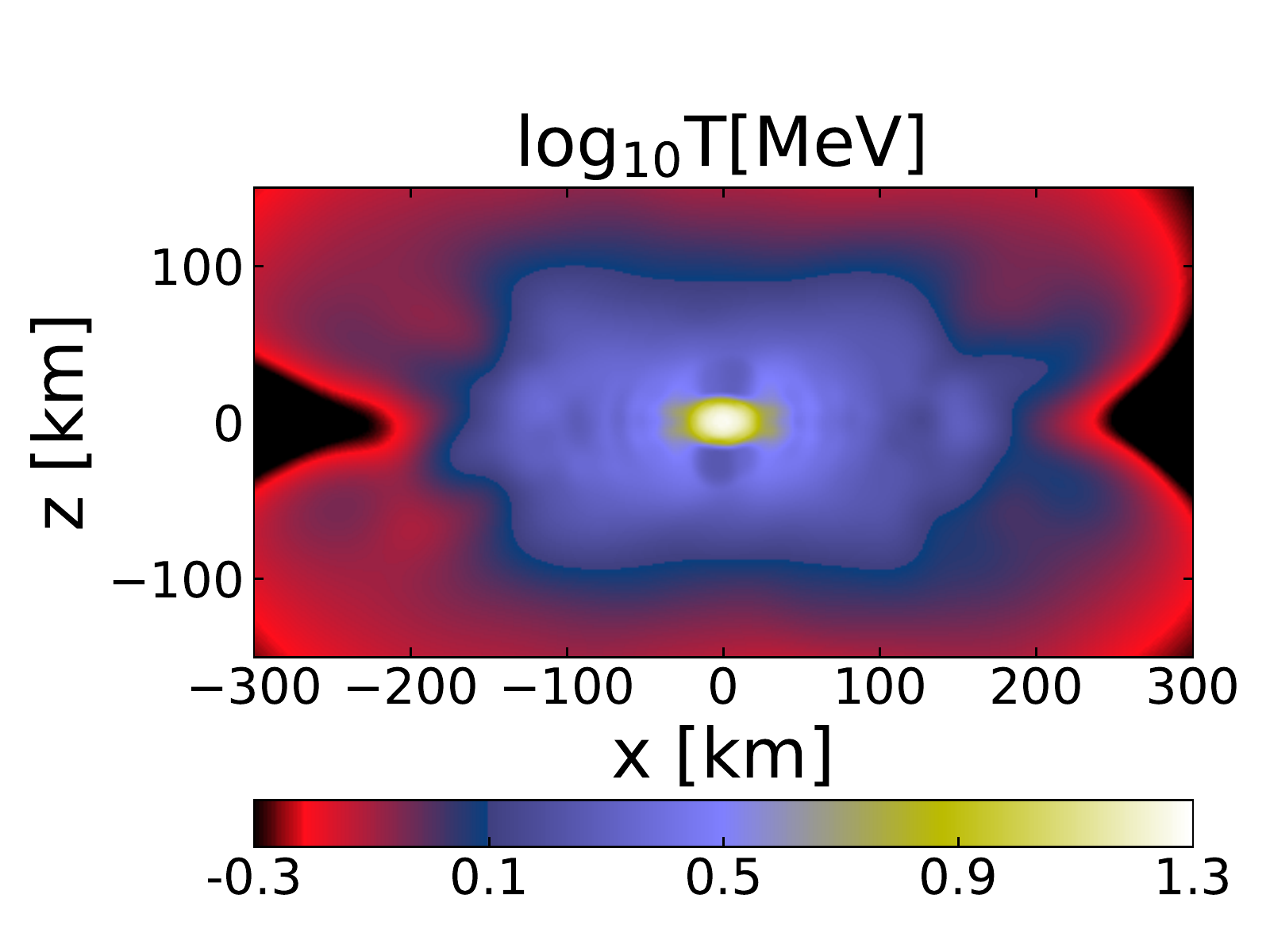}
\end{minipage}
\begin{minipage}{0.33 \linewidth}
\centering
\includegraphics[width = 1 \linewidth]{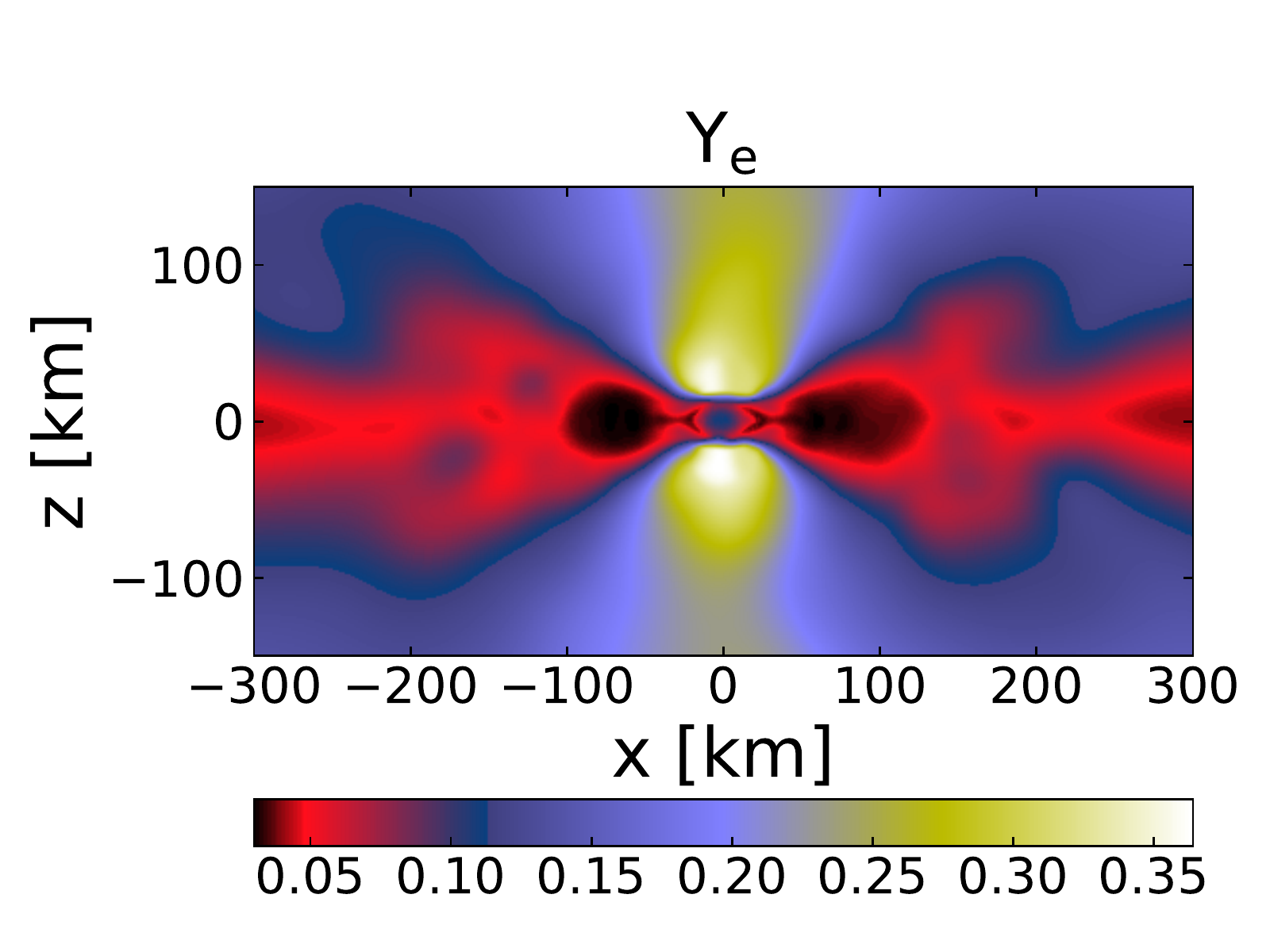}
\end{minipage}\\
\begin{minipage}{0.33 \linewidth}
\centering
\includegraphics[width = 1 \linewidth]{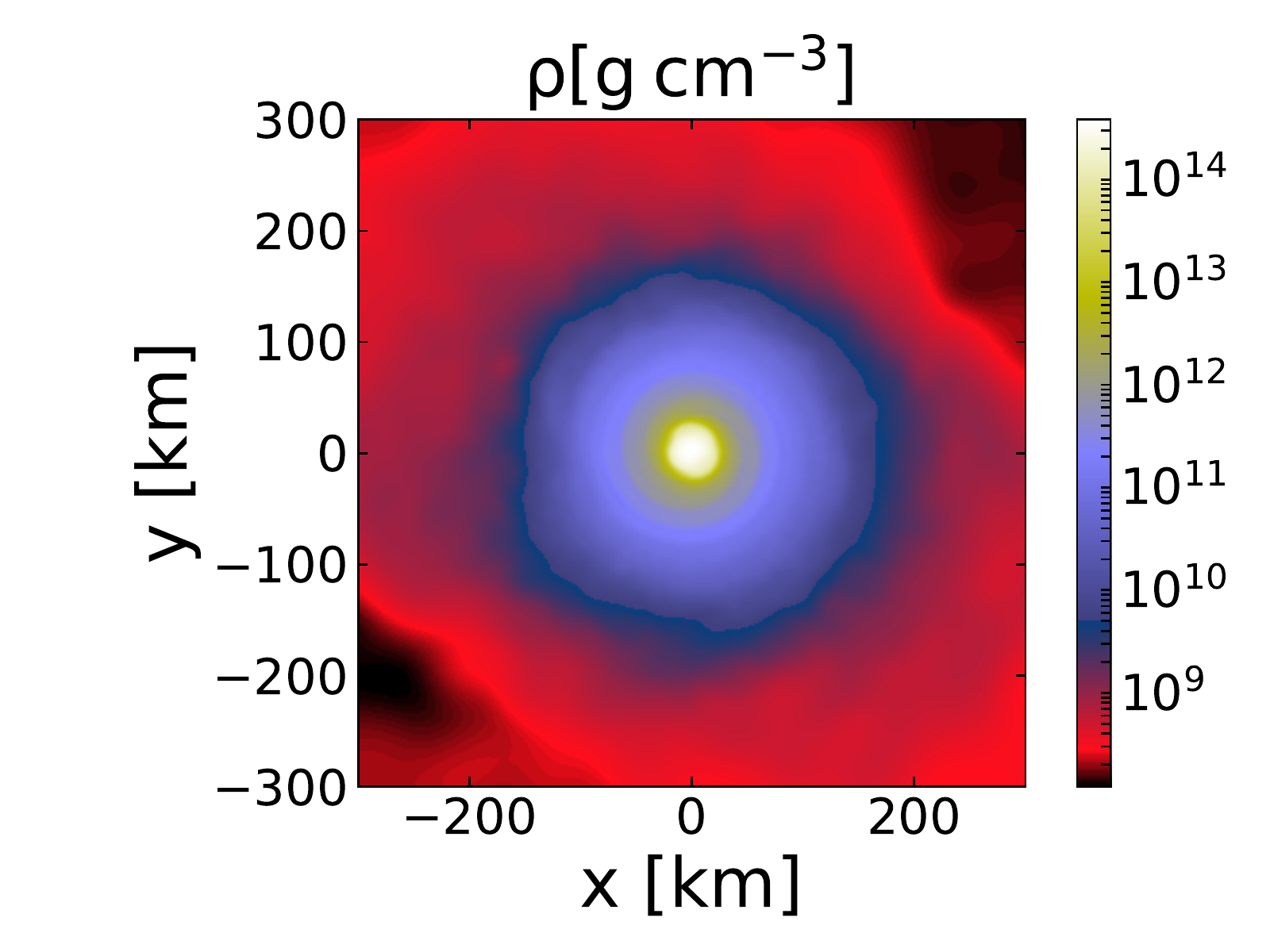}
\end{minipage}
\begin{minipage}{0.33 \linewidth}
\centering
\includegraphics[width = 1 \linewidth]{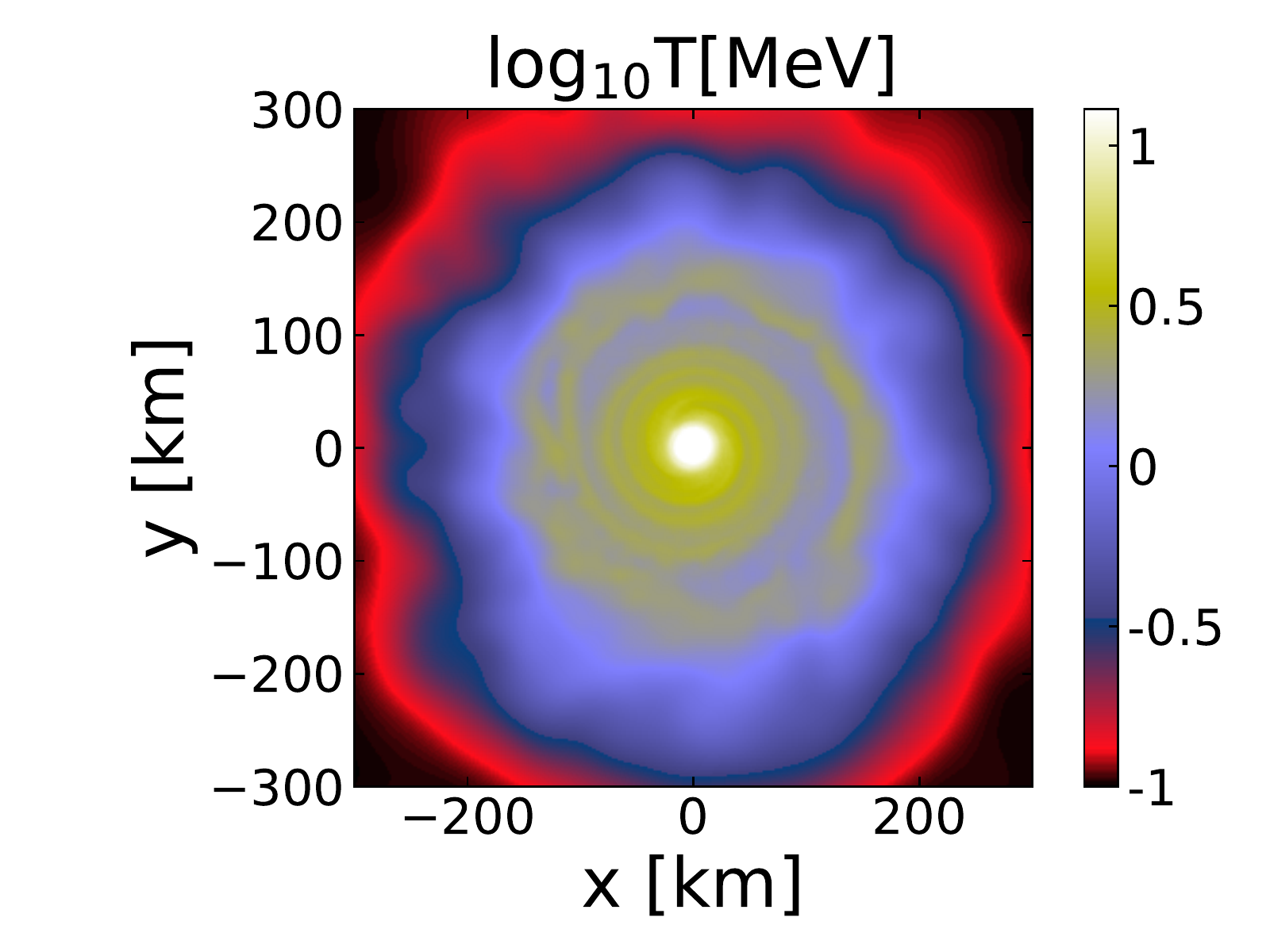}
\end{minipage}
\begin{minipage}{0.33 \linewidth}
\centering
\includegraphics[width = 1 \linewidth]{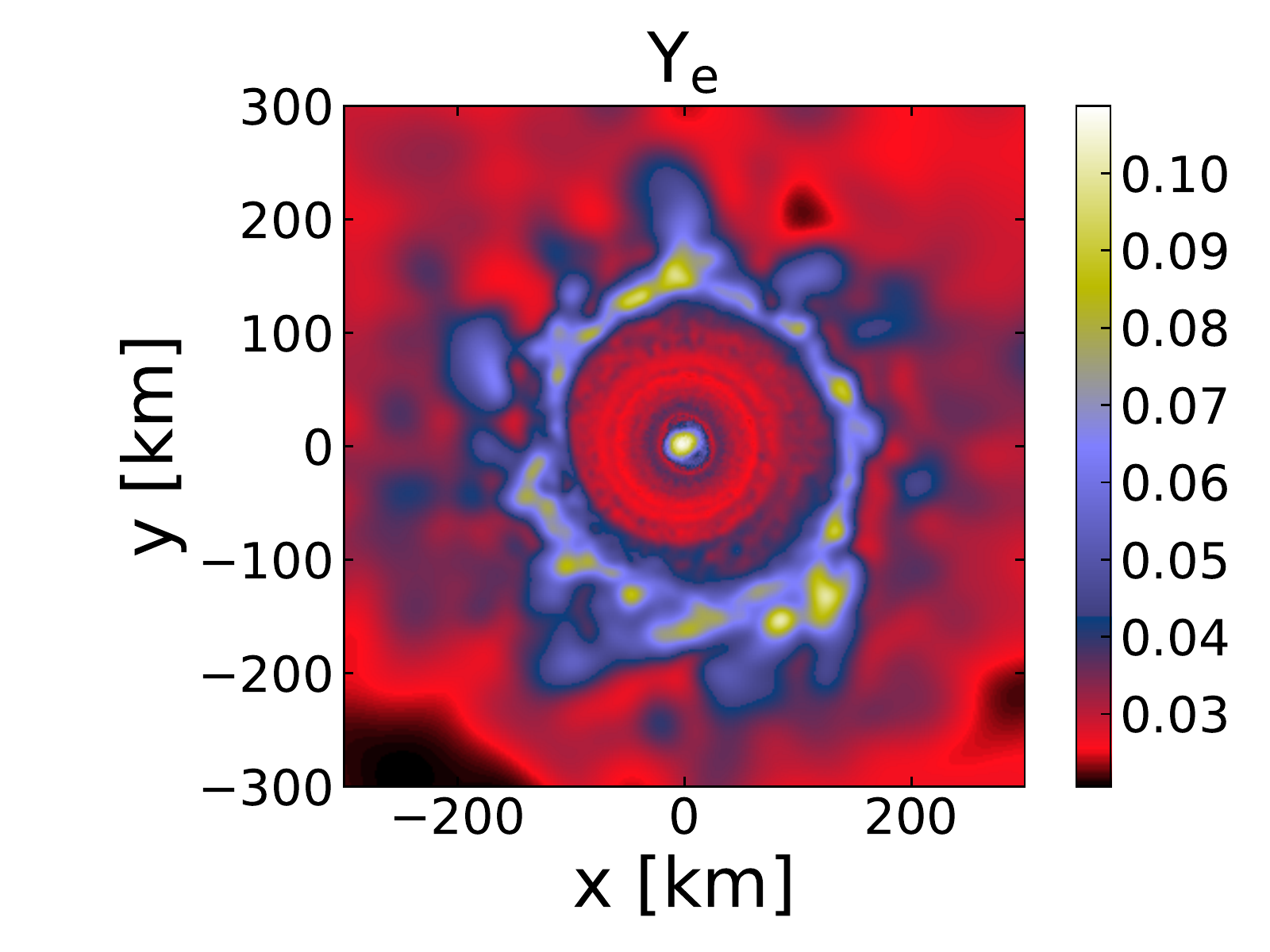}
\end{minipage}\\
\caption{Density (left panels), temperature (middle panels) and electron fraction (right panels) distribution of a 1.4-1.4 $\Msun$ binary neutron star merger configuration at 38 ms after merger. \textbf{Top row:} plane y=0, \textbf{Bottom row:} plane z=0. The hot central object is confined within a region of  $\sim 20$ km from the centre and has a mass $\approx 2.3 \:\Msun$, densities $\rm{\rho} \sim 3\cdot10^{14}  \rm{g\:cm^{-3}}$, temperatures $\rm{T} \sim 20$ MeV and electron fractions $\rm{Y_e} \sim 0.1$. The surrounding torus has electron fractions $\rm{Y_e} < 0.1$ and can be divided in an inner region with densities $\rm{\rho} \sim 10^{12}-10^{13} \rm{g\:cm^{-3}}$ and temperatures $\rm{T} \sim 3-5$ MeV  confined within $\sim 100$ km from the centre, and an outer region up to $\sim 200$ km with densities $\rm{\rho} \sim 10^{10}-10^{11} \rm{g\:cm^{-3}}$ and temperatures $\rm{T} \sim 1$ MeV. Larger electron fractions $\rm{Y_e} \sim 0.2-0.3$ are located along the z-axis with a maximum of $\rm{Y_e} \sim 0.35$ right above the central object, densities $\rm{\rho} \sim 10^8 \rm{g\:cm^{-3}}$ and temperatures $\rm{T} \lesssim 1$ MeV.}
\label{fig:BNSinitialdata}
\end{figure*}
\begin{figure}
\includegraphics[width = 8 cm]{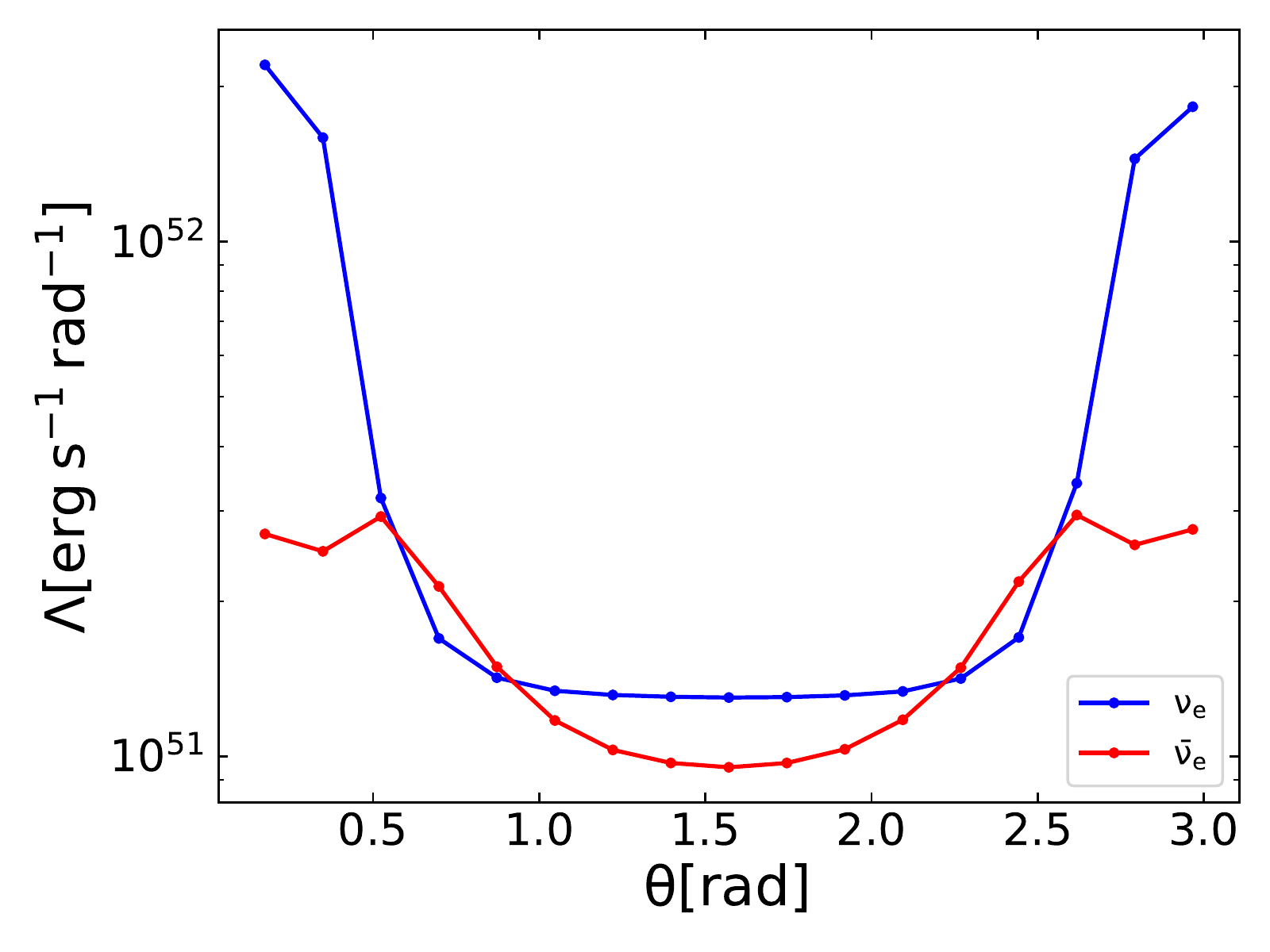}
\caption{$\Lambda_{\nu}(\theta)$ as function of
$\theta$ for electron neutrinos (blue) and 
anti-neutrinos (red). Electron neutrinos show 
a more pronounced anisotropy compared to the 
corresponding anti-neutrinos,
with $\beta_{\nu_e} \approx 16$
and $\beta_{\bar{\nu}_e} \approx 2$.}
\label{fig:lambda}
\end{figure}
\begin{figure*}
\begin{minipage}{0.4 \linewidth}
\centering
\includegraphics[width = 1 \linewidth]{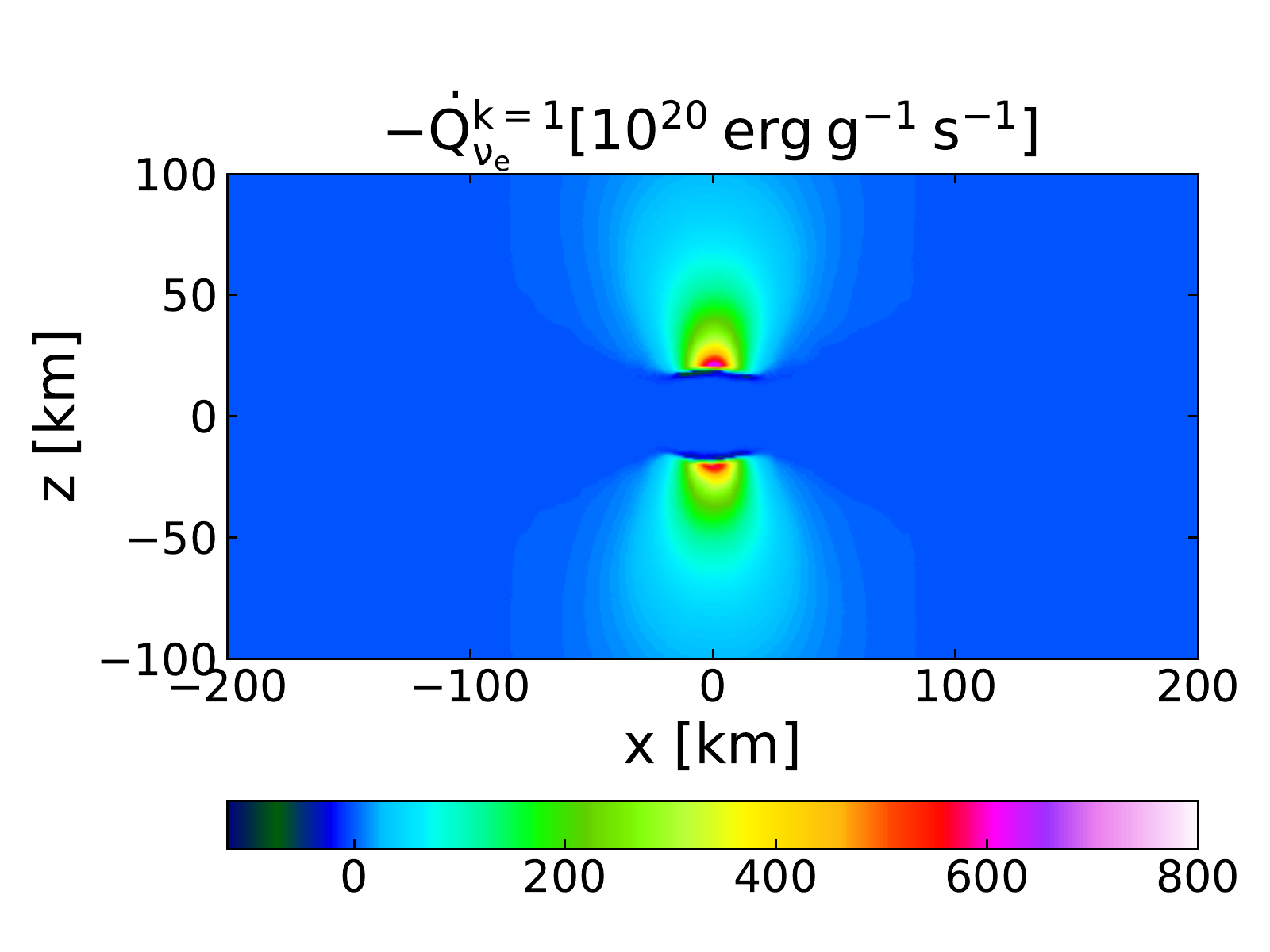}
\end{minipage}
\begin{minipage}{0.4 \linewidth}
\centering
\includegraphics[width = 1 \linewidth]{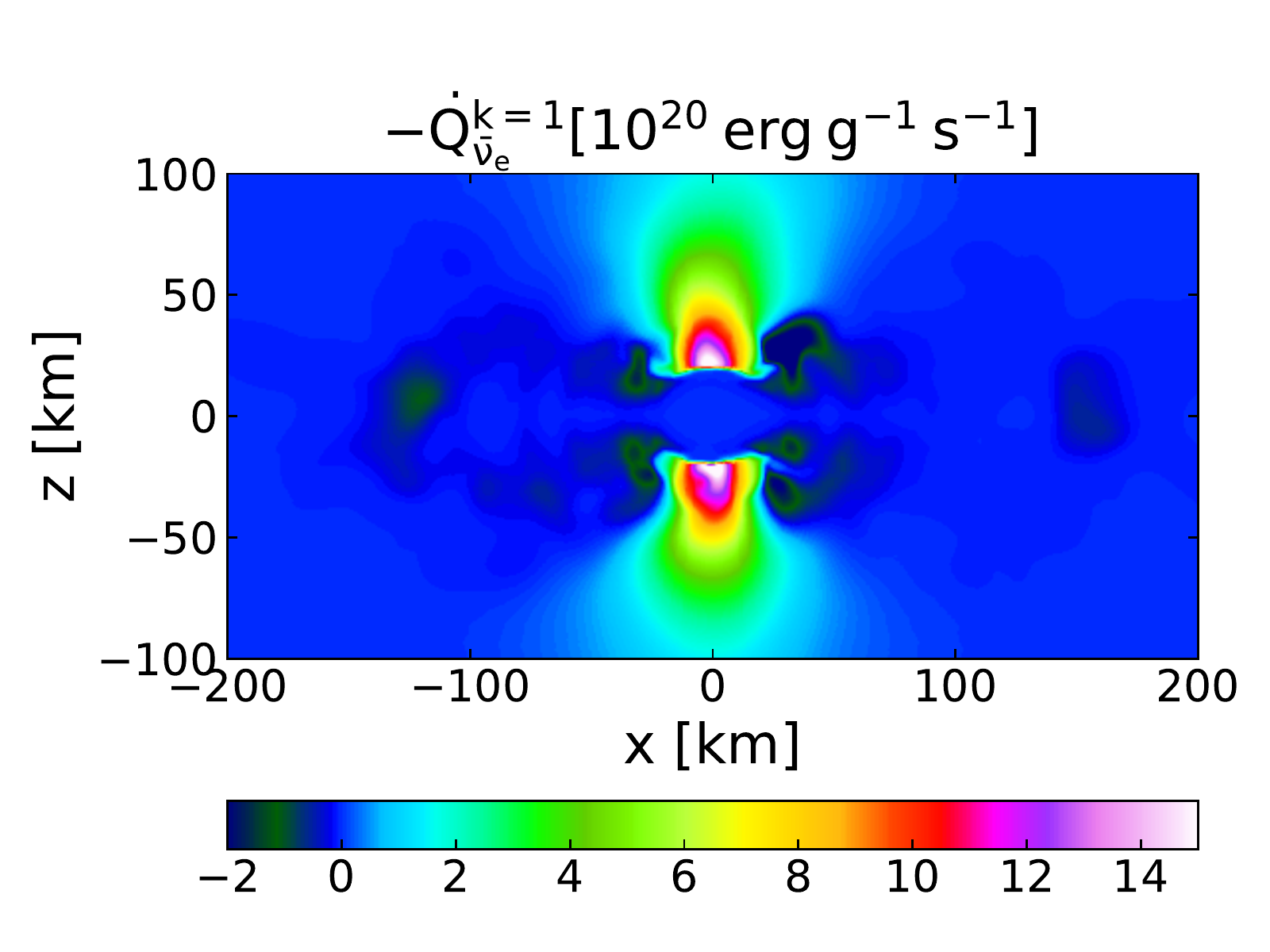}
\end{minipage}\\
\begin{minipage}{0.4 \linewidth}
\centering
\includegraphics[width = 1 \linewidth]{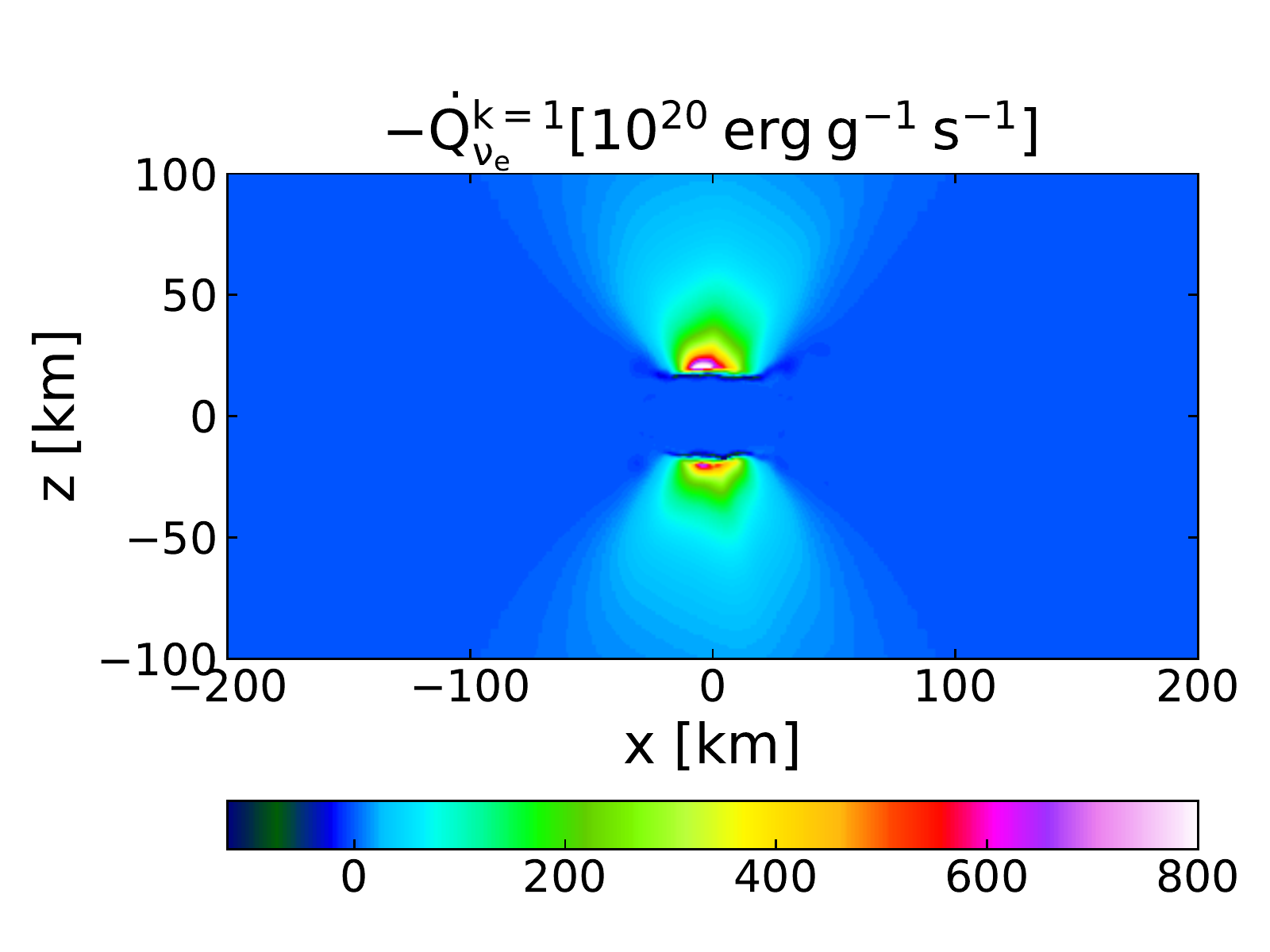}
\end{minipage}
\begin{minipage}{0.4 \linewidth}
\centering
\includegraphics[width = 1 \linewidth]{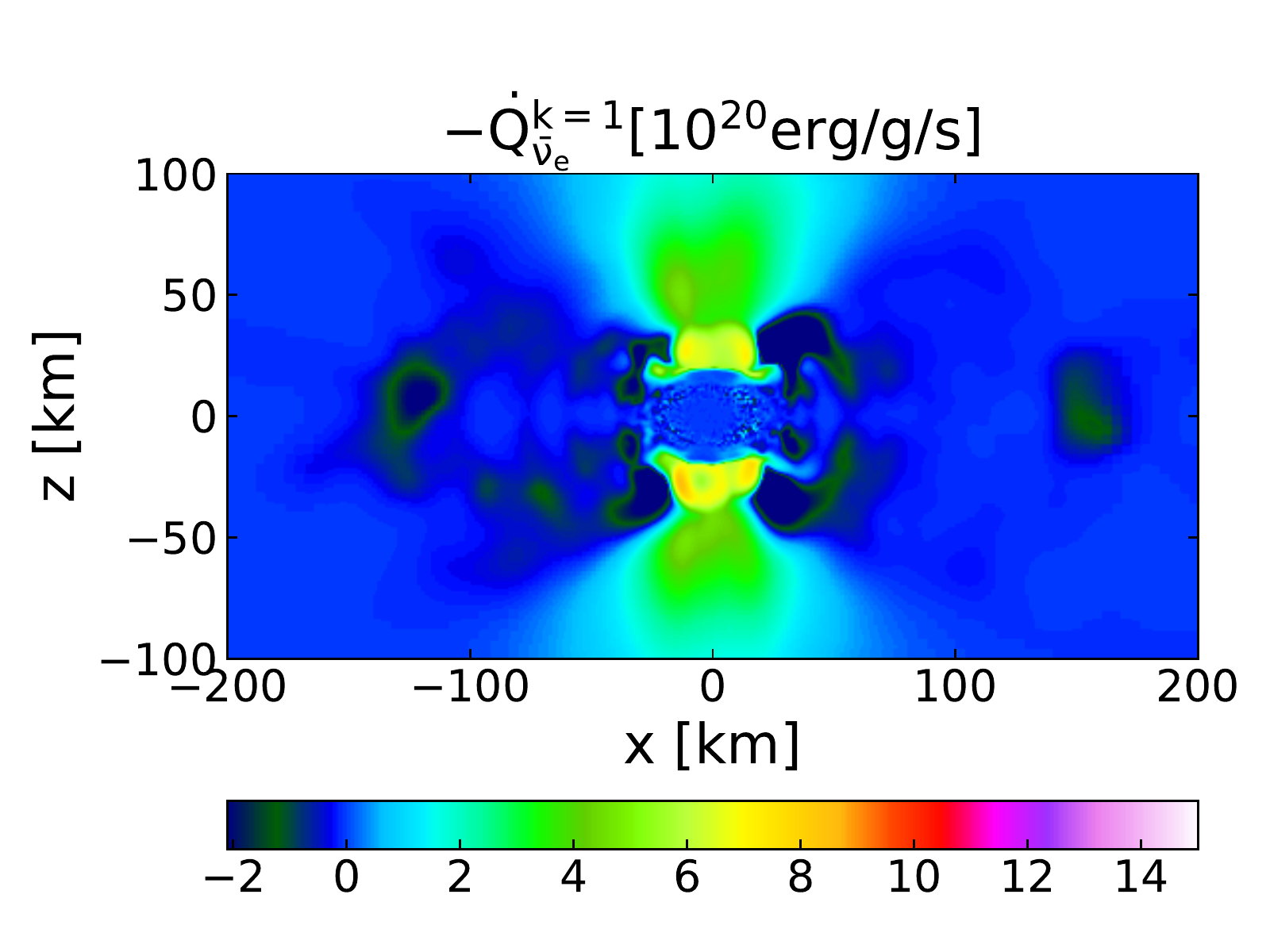}
\end{minipage}\\
\caption{\textbf{Top row:} ASL and \textbf{Bottom row:} M1 heating rate maps on the plane y=0 in units of $10^{20}
\:\rm{erg\:g^{-1}\:s^{-1}}$ 
for electron neutrinos (right panels) and electron anti-neutrinos (left panels). 
In both ASL and M1 
the electron neutrino cooling occurs mainly 
close to the central object at distances $\lesssim 20$ km
from the centre, while for the electron anti-neutrinos 
it is mostly located in the torus, where  
isolated spots of neutrino emission are visible. 
For both species, the main contribution to the 
heating is located into a funnel $\sim 45\degree$
from the pole. Moreover, 
the electron neutrino heating is generally 
up to one order of magnitude
larger than the electron anti-neutrino one.
Compared to M1,
the ASL provides an electron neutrino
heating rate above the central object
at $\theta \lesssim 15\degree$
lower by a factor $\sim 1.5$ and a 
residual contribution at $\theta \sim 60\degree-90\degree$ of about $10^{19}-10^{20}\:\rm{erg\:g^{-1}\:s^{-1}}$
(not appreciable in the Figure),
as result of the criterion $\tau_{\rm{tot}}
\leq 1$, that whenever satisfied triggers
neutrino absorption. Furthermore,
the electron
anti-neutrino heating
above the central object
at $\theta \lesssim 15\degree$
is larger by a factor of $\sim 2$.}
\label{fig:BNS_netrate}
\end{figure*}
\begin{figure}
\includegraphics[width = 8 cm]{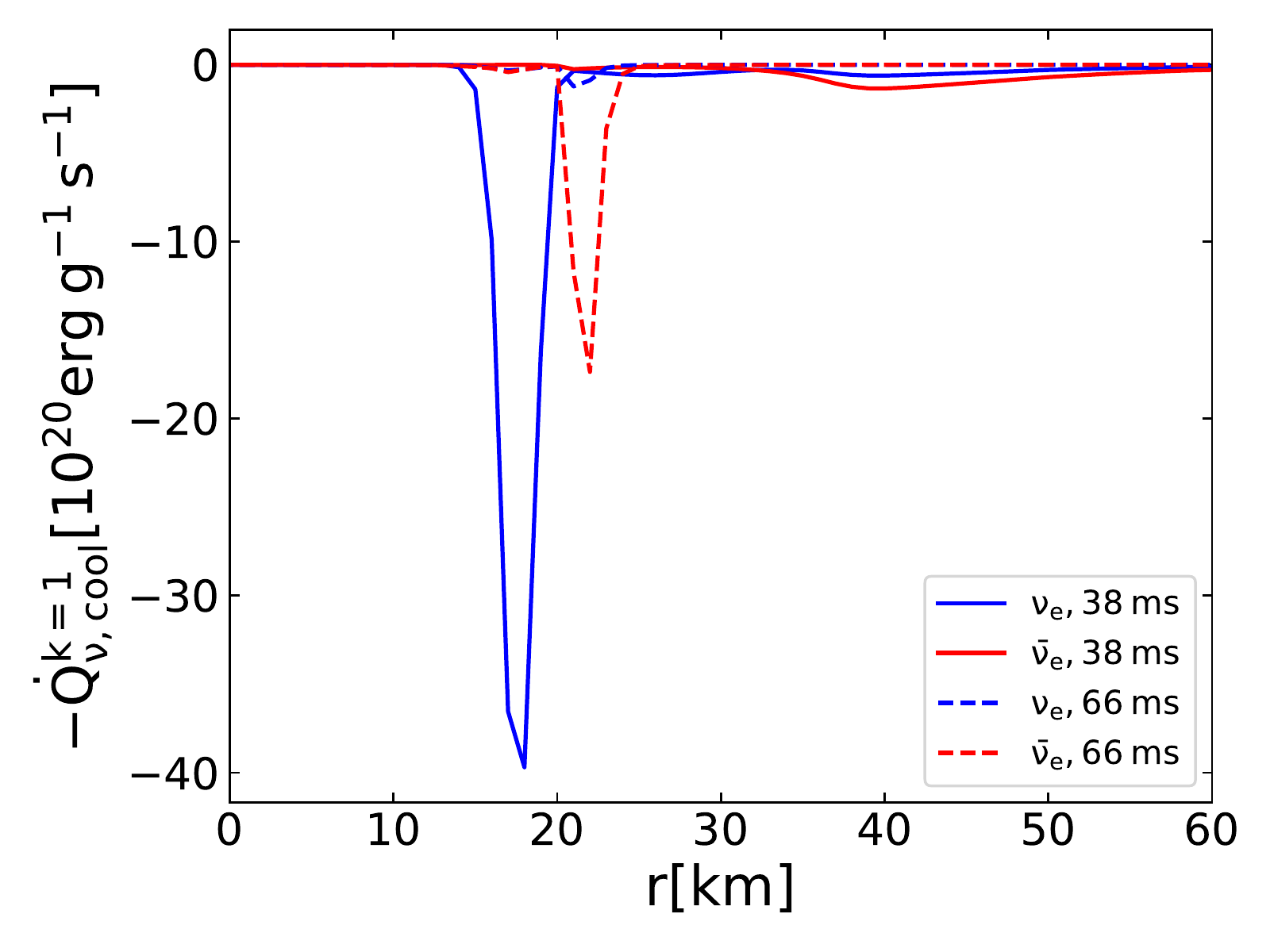}
\caption{Cooling along the radius at $\theta=0 \degree$ in units of $10^{20}\:\rm{erg\:g^{-1}\:s^{-1}}$
for electron neutrinos (blue) and 
electron anti-neutrinos (red), for the snapshots
at 38 ms and 66 ms, calculated with ASL.
While the 38 ms snapshot (solid lines)
has negligible anti-neutrino emission in comparison to electron neutrinos, the 66 ms snapshot (dashed lines) 
shows a dominant anti-neutrino cooling
$\sim 10^{21}\:\rm{erg\:g^{-1}\:s^{-1}}$ over the electron neutrinos.
}
\label{fig:cooling}
\end{figure}

As a final 3D test we apply our new scheme
to the remnant of a neutron star merger. We start 
from a snapshot of a merger remnant (t= 38 ms after merger;
\cite{Rosswog17a}) that has been obtained using the SPH method
with the grey leakage scheme of \cite{Rosswog2003}. For
general reviews of the SPH method we refer to recent reviews \citep{Monaghan05,Rosswog2009,Rosswog15}.
We use the snapshot as a background 
on which to evolve the 
neutrino properties until a steady state is reached.
Unlike the core-collapse supernovae case, 
we do not have any
information on the neutrino trapped components at this time and therefore we perform our tests by assuming
$Y_{\nu}=Z_{\nu}=0$ as initial condition for neutrinos. 
Although strictly speaking 
inconsistent as in the configuration
trapped neutrinos 
should already be present, 
our choice of such initial condition 
is justified by several arguments.
First, the role of the neutrino
trapped components is mainly 
important when doing full 
dynamical evolutions and in 
particular it has been shown recently 
\citep{Perego2019} that they only
marginally affect
the thermodynamical properties 
of the remnant, 
and only close to the merger 
time where neutrinos 
are produced in the first place. 
Second, setting $Y_{\nu}=Z_{\nu}=0$ 
implies a remnant configuration 
out of equilibrium, and given the
large temperature dependence of 
the neutrino-matter cross sections
\citep{Bruenn1985,Burrows2004}
the system rapidly reaches
a new state by refilling the 
neutrino fractions over a
timescale $Y_{\nu}/\Dot{Y_{\nu}} \lesssim 10^{-6}$ s,
which is much less than the typical 
dynamical timescale of the 
remnant $\sim (G\bar{\rho})^{-1/2} \approx 2\cdot10^{-4}$ s.
Third, absorption under optically thin 
conditions (that we are modeling here) 
is led by non-trapped neutrinos. For all 
the above reasons, we do not expect the 
modeling of the trapped neutrinos to 
have a significant impact on our calculations.
The neutron stars have been discretized 
with N $\sim 1$ million particles and the initial
conditions are mapped on a 3D grid whose 
borders are defined when densities go
below $\sim 10^8 \rm{g\:cm^{-3}}$. 
Figure \ref{fig:BNSinitialdata} shows the
density, temperature and electron fraction 
on the planes y=0 and z=0. 
The central object has a mass 
$\approx 2.3\:\Msun$, 
densities $\rm{\rho} \sim 3\cdot10^{14} 
 \rm{g\:cm^{-3}}$, temperatures $\rm{T} \sim 20$ MeV 
and electron fractions $\rm{Y_e} \sim 0.1$. 
Around it is a torus with $\rm{Y_e} < 0.1$, an 
inner region with densities $\rm{\rho} 
\sim 10^{12}-10^{13} \rm{g\:cm^{-3}}$ and temperatures 
$\rm{T} \sim 3-5$ MeV, and an outer region 
with densities $\rm{\rho} \sim 10^{10}-10^{11} 
\rm{g\:cm^{-3}}$ and temperatures $\rm{T} \sim 1$ MeV. 
Electron fractions $\rm{Y_e} \sim 0.2-0.3$ are 
located along the low-density polar region.
\begin{table}
\begin{center}
\caption{Summary of the neutrino 
luminosities and 
average energies
for electron
neutrinos, electron anti-neutrinos and
heavy-lepton neutrinos, calculated with
ASL (second column) and M1 (third column),
for the binary merger snapshot at $\sim 38$ ms
after merger.}
    \begin{tabular}{c|c|c}
      \hline
      Quantity & ASL & M1\\ \hline 
      $L_{\nu_e}(\rm{erg\:s^{-1}})$ & $1.64\cdot10^{52}$ &
      $2.56\cdot10^{52}$\\
      $L_{\bar{\nu}_e}(\rm{erg\:s^{-1}})$ & $1.69\cdot10^{52}$ &
      $2.28\cdot10^{52}$\\
      $L_{\nu_x}(\rm{erg\:s^{-1}})$ & $7.64\cdot10^{51}$ &
      $7.50\cdot10^{51}$\\ \hline
      $\langle{}E_{\nu_e}\rangle{}$(MeV) & 13.10 & 15.46\\
      $\langle{}E_{\bar{\nu}_e}\rangle{}$(MeV) & 13.59 & 12.86\\
      $\langle{}E_{\nu_x}\rangle{}$(MeV) & 14.09 & 15.50\\ \hline
    \end{tabular}
    \label{Table4}
    \end{center}
\end{table}
\\\\
\noindent{\em 3D optical depth}\\
Computation of the neutrino properties 
requires the calculation of the optical
depth on the grid as explained in 
Sec.~\ref{sec:3Dmeth}.
\begin{figure*}
    \centering
    \includegraphics[width=15 cm]{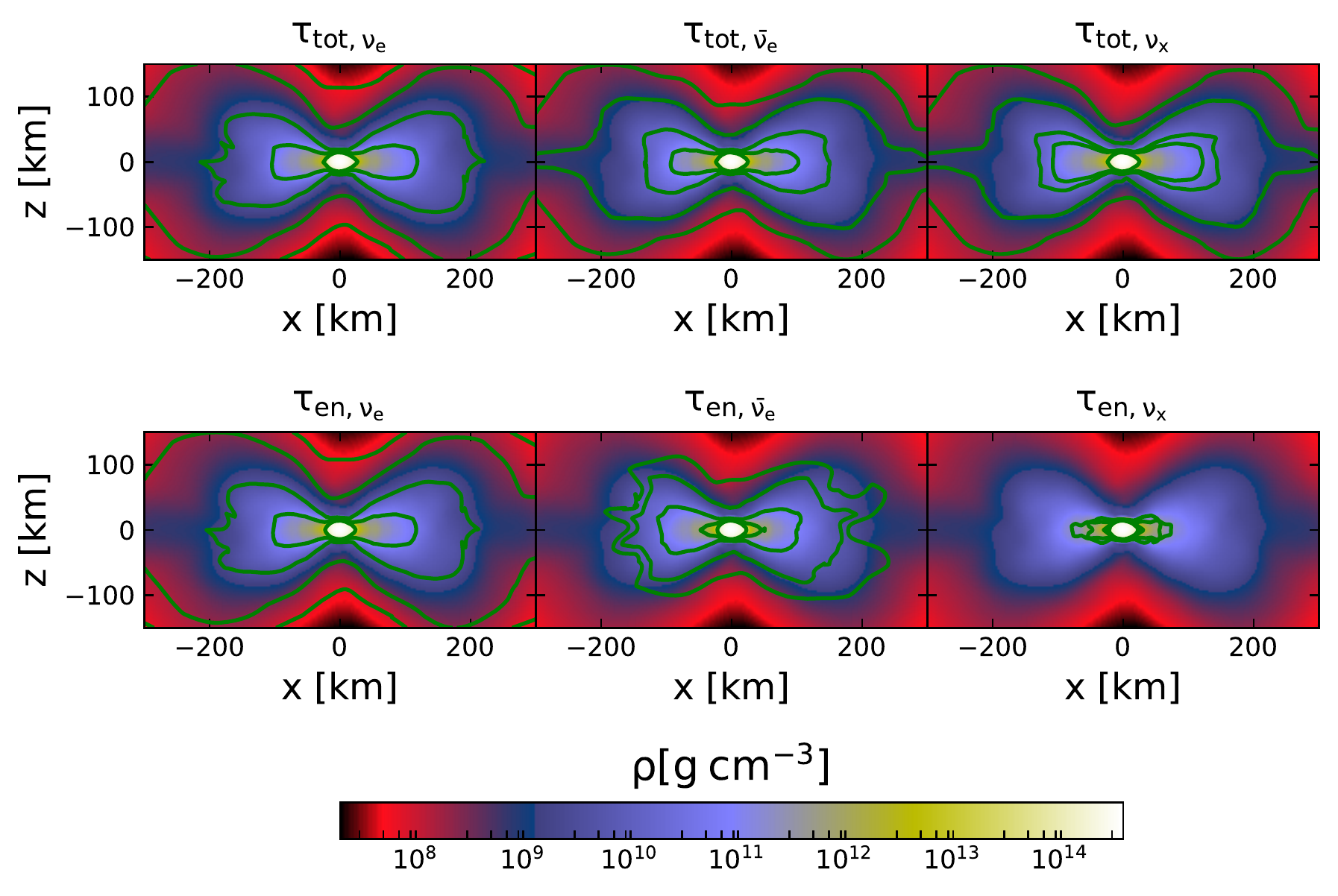}
    \caption{Location of the neutrino surfaces for the same sets of neutrino energies used in Figures \ref{fig:ser1D} and \ref{fig:shr1D}. \textbf{Top row:} total optical depth and \textbf{Bottom row:} energy optical depth. For each row, we show electron neutrino (left panel), electron anti-neutrino (middle panel) and heavy-lepton neutrino (right panel). The location of the neutrino surface is shown in green, starting from the inner one at E= 3 MeV to the outer one at E= 300 MeV. Note the shape resembling the presence of a torus around the central object and therefore of a
non-spherical geometry. Similarities with Figure  \ref{fig:neutrinosphere} are visible, both in the distribution of the surfaces with increasing neutrino energy and in the comparison of total and energy optical depths of each species. However, with respect to the core-collapse supernova case, the low abundance of free protons makes the electron anti-neutrino total and energy neutrino surfaces less extended with
respect to the electron neutrino ones.}
    \label{fig:BNS_nusphere}
\end{figure*}
Figure \ref{fig:BNS_nusphere} 
shows the location of the total and 
energy neutrino surfaces for the 
same sets of energies used for the core-collapse 
supernovae tests 
(see Figures \ref{fig:ser1D} and \ref{fig:shr1D}). Overall, we see an 
agreement with the distribution of the surfaces 
described in \cite{Perego2014}.\\
As already noticed for the core-collapse supernovae case, the higher the energy of neutrinos the more extended 
the neutrino surface. Accordingly,
the heavy-lepton neutrinos
have less extended energy 
than total neutrino surfaces
because they can only exchange
energy by pair processes and 
bremsstrahlung.
Elastic scattering on nuclei and nucleons make the 
heavy lepton total neutrino surfaces comparable with 
the ones of the other species. Electron neutrinos again 
show comparable energy and total neutrino surfaces due to
the neutron rich environment that favours 
energy exchange in processes like neutrino absorption on neutrons. However, we notice that compared to the core-collapse 
supernova case electron 
anti-neutrinos show less extended
total and energy neutrino surfaces compared
to the corresponding electron neutrino ones as
a consequence of the low abundance of free protons.\\
\\
\noindent{\em Heating}\\
The heating is modelled by Eqs.~(\ref{eq:heating}) and (\ref{eq:new_nudensity}) and by estimating $\beta_{\nu}$ as described in Sec.~\ref{sec:3Dmeth}. 
To save computational time, we limit our 
transport calculation to those regions
where density is above $10^9 \rm{g\:cm^{-3}}$.
Indeed, we find that the contribution 
at lower densities affect the transport 
quantities by less than $1\%$.
For the computation of $l_{\nu}(E,\textbf{x})$ we
create a 1D profile of 1 km of resolution where 
to each bin of radius $R_{\rm{bin}}$ we assign a neutrino 
emission by summing up the contribution
of all grid points with radial distance from
the centre within that bin. We then solve
Eq.~(\ref{eq:spartlm}) and assign the 
same $l_{\nu}(E,R_{\rm{bin}})$ to all
grid points inside that bin.
In this way we create a spherically
symmetric neutrino emission by coupling 
fluid points from the torus with 
fluid points along the poles, 
and we then leave to 
$\beta_{\nu}$ the task of 
approximately recovering 
the degree of anisotropy of the system 
and consequently the degree of decoupling 
between points at same distance 
from the centre but at different polar angles.
The determination of the exponent $b$ in
Eq.~(\ref{eq:new_nudensity}) is 
performed by comparing with an M1 
calculation of the heating from FLASH. 
In particular, we find $b=8$ 
to overall best recover the 
electron neutrino 
contribution to the heating
(which constitutes more than 
$50\%$ of the total)
and we therefore assume the same
value for the corresponding 
anti-neutrinos as well.\\
In Figure \ref{fig:lambda} we show
the angular dependence of the
neutrino flux, i.e. 
$\Lambda_{\nu}(\theta)$ vs $\theta$, 
calculated from Eq.~(\ref{eq:lambda})
with $\Delta \theta= 10 \degree$
\footnote{We have tested different 
$\Delta \theta$ and no appreciable
variations appears in the computation 
of $\beta_{\nu}$, therefore we set 
$\Delta \theta= 10 \degree$.}. 
We notice that the modulation of the
flux with the polar angle is more pronounced
for electron neutrinos than for electron 
anti-neutrinos.
This is due to the fact that 
the neutron rich torus 
is more opaque to electron
neutrinos than to electron anti-neutrinos.
Therefore, the electron
neutrino flux points mostly along
the z-direction. 
In terms of neutrino emission,
the largest contribution to the 
cooling for the electron neutrinos
($\sim 10^{21}\:\rm{erg\:g^{-1}\:s^{-1}}$) 
is confined 
within radii $\lesssim 20$ km from the centre,
and the remaining subdominant part 
($\sim 10^{19}-10^{20}\:\rm{erg\:g^{-1}\:s^{-1}}$) occurs
inside the torus.
We find that 
electron anti-neutrinos in contrast
are mostly emitted in the torus
(cooling $\sim 10^{20}\:\rm{erg\:g^{-1}\:s^{-1}}$)
and no relevant emission is found 
at radii $\lesssim 20$ km from the centre.
We obtain $\beta_{\nu_e} \approx 16$ and 
$\beta_{\bar{\nu}_e} \approx 2$.
In the first row of Figure \ref{fig:BNS_netrate}
we show 2D maps of the 
resulting heating rate on the plane $y=0$.
In the second row we show
the same maps for 
calculations performed with the M1 
scheme implemented in FLASH.
The major contribution to the heating
is located within $\sim45\degree$ from the
pole for both species. 
The largest differences of the ASL 
compared to M1 are in the region with
$\theta \lesssim 45\degree$, 
and in particular 
above the central object
at $\theta \lesssim 15\degree$,
where the electron
neutrino and anti-neutrino heating
are respectively lower by a factor of 
$\sim 1.5$ and larger by 
a factor of $\sim 2$.
Moreover, unlike the M1 implementation our
ASL algorithm provides a residual electron
neutrino heating 
($\sim 10^{19}-10^{20} \rm{erg\:g^{-1}\:s^{-1}}$) at 
$\theta \sim 60\degree-90\degree$ in regions 
where $\tau_{\rm{tot}} \leq 1$.
We notice that both ASL and M1
provide an electron neutrino 
heating larger by one order of 
magnitude compared to electron 
anti-neutrinos. This is due to two 
effects: first, our snapshot
calculations show an electron
anti-neutrino cooling which is
at the most $\sim 10^{20} \rm{erg\:g^{-1}\:s^{-1}}$,
i.e. one order of magnitude lower
than the electron neutrino one,
which reaches $\sim 10^{21} \rm{erg\:g^{-1}\:s^{-1}}$
close to the central remnant
where the anti-neutrino emission
is negligible in comparison.
Second, the neutron rich environment
favours the electron neutrino
absorption over the corresponding
anti-neutrinos.\\
\\
\noindent{\em Neutrino luminosities and average-rms
energies}\\
The overall low anti-neutrino cooling
leads to almost equal
electron neutrino and 
anti-neutrino total luminosities and 
average energies, contrarily to
what is typically expected from
merger simulations
\citep{Rosswog2003,Perego2014,Dessart2009,Foucart2016a}. In particular, 
a summary of these values is reported 
in Table \ref{Table4}. The reason
for this is most likely due to
the application
of a different
neutrino transport 
than the one
adopted for generating 
the snapshot.
In fact, we further calculate
the neutrino cooling with our ASL 
for a similar snapshot 
of the merging of two $1.4\:\Msun$ neutron
stars taken from 
the simulations of 
\cite{Perego2014}
at $\sim 66$ ms
after the first contact of the two stars. 
The ASL of \cite{Perego2014}
is similar to ours, and 
in particular is spectral.
In this case we get
$L_{\nu_e}^{\rm{cool}}= 1.45\cdot10^{52}\:\rm{erg\:s^{-1}}$,
$L_{\bar{\nu}_e}^{\rm{cool}}= 1.85\cdot10^{52}\:\rm{erg\:s^{-1}}$,
$L_{\nu_x}^{\rm{cool}}= 1.34\cdot10^{52}\:\rm{erg\:s^{-1}}$ ,
$\langle{}E_{\nu_e}\rangle{}= 10.81$ MeV, $\langle{}E_{\bar{\nu}_e}\rangle{}= 14.97$ MeV, $\langle{}E_{\nu_x}\rangle{}= 16.00$ MeV,
$E_{\rm{rms},\nu_e}= 12.02$ MeV,
$E_{\rm{rms},\bar{\nu}_e}= 16.77$ MeV, 
$E_{\rm{rms},\nu_x}= 19.74$ MeV,
that is, the largest cooling
is from the electron anti-
neutrinos being  
the most luminous, and we 
also clearly recover the 
expected hierarchies, i.e.
$L_{\bar{\nu}_e}^{\rm{cool}} > L_{\nu_e}^{\rm{cool}} > L_{\nu_x}^{\rm{cool}}$,
$\langle{}E_{\nu_x}\rangle{} > 
\langle{}E_{\bar{\nu}_e}\rangle{} >
\langle{}E_{\nu_e}\rangle{}$, 
$E_{\rm{rms},\nu_x} > E_{\rm{rms},\bar{\nu}_e}
> E_{\rm{rms},\nu_e}$
\footnote{Although we
recover the expected hierarchies,
we do not find 
the same values of 
\cite{Perego2014} as 
they perform a dynamical 
simulation assuming neither
blocking nor 
thermalization corrections 
in Eq.~(\ref{eq:ser1D}).}.
In particular,
we observe that the configuration at
$\sim 66$ ms shows larger densities 
and temperatures at radii 
$\lesssim 20$ km from the centre
compared to the $\sim 38$ ms one,
producing a dominant
electron anti-neutrino 
cooling $\sim 10^{21}\:\rm{erg\:g^{-1}\:s^{-1}}$
(see Figure \ref{fig:cooling}
for a comparison done with ASL 
along $\theta=0 \degree$)
that is instead missing in
the configuration at $\sim 38$ ms.
Moreover, the two
snapshots have been 
previously evolved
with different neutrino transport
schemes: the one at 
$\sim 66$ ms with the scheme of
\cite{Perego2014} which is similar
to our version of the ASL,
the $\sim 38$ ms one with the
grey scheme of \cite{Rosswog2003}.
Applying a different
neutrino transport than
the one used for generating
the snapshot introduces
inconsistencies that are
likely to be the reason
for our luminosity values.
From Table \ref{Table4} we notice that
the ASL provides electron 
neutrino and anti-neutrino luminosities
lower by $\sim 35\%$ and $\sim 25\%$
respectively in comparison to M1. 
Average-rms energies 
agree within $5-15\%$.
The lower luminosity 
values from the ASL can be 
due to several reasons. First, the 
ASL has excess heating
at $\theta \sim 60\degree-90\degree$
for the electron neutrino and 
at $\theta \sim 15 \degree$ for the
electron anti-neutrinos compared to M1.
Second, 
we have kept the values of 
the ASL parameters
entering Eq.~(\ref{eq:ser1D}) to the
ones calibrated for core-collapse
supernovae simulations.
New calibrations are crucial as 
they might impact 
the neutrino cooling and 
consequently the amount of neutrino
heating. In particular, the comparison with
M1 suggests a lower $\alpha_{\nu, \rm blk}$.
This can be explained in the following way.
While in core-collapse supernovae 
the quasi-spherical symmetry implies that
neutrinos move preferentially along the
radial direction, in binary mergers
neutrinos have more directional 
freedom in escaping the system
and consequently the blocking is
expected to be less effective.
Third and above all, 
a more consistent comparison
between two neutrino transport
approaches would require dynamical 
evolutions rather than 
snapshot calculations.
Putting together 
all these uncertainties 
we find our
ASL-M1 luminosity discrepancies, 
lower than a factor of 2,
a promising initial step
toward future
developments.
We also want to stress that
the choice we have made
for the flux modulation $\sim \rm{cos}^{8}(\theta)$
is only meant to be a preliminary step
toward explorations in full dynamical
evolution of binary mergers.
Finally, we stress the fact that
we have based our heating calculations
on the comparison with a moment
approach with analytical closure.
However, as pointed out by \cite{Foucart2018}
the M1 closure can overestimate
the neutrino density by up to $\sim 50\%$
along the polar regions,
thus limiting the possibility of
properly modeling the neutrino driven-winds.
Future simulations of binary mergers
will definitely need improvements
in moment schemes as well for 
more reliable assessments of the
neutrino physics in such systems.

\section{Summary}
\label{sec:conclusion}
In this paper we have presented an extension of the
Advanced Spectral Leakage scheme originally introduced
by \cite{Perego2016}. Our main goal was to  adapt the scheme
so that it can be conveniently  used for neutron star merger
simulations. The main advantage compared to simpler leakage
schemes is that the ASL includes neutrino heating processes
and can therefore, with a reasonable computational effort, also
model the emergence of neutrino-driven winds in a 3D merger
simulation.
The main novelty compared to the original approach is the usage
of an optical depth-dependent flux factor, and a modification
to the equation of the neutrino density in the semi-transparent regime,
both designed for the multi-dimensional modeling of compact
binary mergers.\\
We scrutinized the new scheme on the case of a 15 $\Msun$
core-collapse supernova snapshot taken from \cite{Perego2016}
at 275 ms after bounce. For this spherically symmetric case we
first tested in 1D our new flux factor and found that it agrees well
with the original choice ($\lesssim 2\%$ for the average energies
and $\lesssim 5\%$ in the neutrino luminosities). As a further
1D test, we have compared the new scheme to the M1 implementation of the GR1D code \citep{Oconnor2010,OConnor2015}
 and here we found 
 agreement, beside differences 
 arising from the usage of 
 different transport schemes. 
We also mapped the 1D case
onto a spherically symmetric, but three-dimensional grid. Here the agreement is slightly
worse, but overall still very good: $\lesssim 2\%$ for the
average energies and $\lesssim 7\%$ for the neutrino luminosities.\\
We have finally explored the ASL for an SPH snapshot
of a 1.4-1.4 $\Msun$ binary neutron star merger. As a reference
we compared against the results obtained with a M1 scheme that
is implemented in FLASH \citep{Fryxell2000,Oconnor2018}.
Here, in ASL the anisotropy in the neutrino fluxes is taken into
account by an anisotropy parameter $\beta_{\nu}$
which is estimated from the  ratio between the neutrino fluxes
at pole and equator in a similar way as it has been
done in \cite{Rosswog2003}. The neutrino density is modelled
$\propto \rm{cos}^\textit{b}(\theta)$, where the value of $b= 8$ 
is obtained by a comparison of the neutrino heating rate with the
M1-FLASH results. Overall, we find good agreement in the 
neutrino heating distribution between both approaches. 
The average energies
agree within 5-15\%. The ASL total luminosities 
are lower by 25-35\% compared to M1. 
This discrepancy may suggest that some of the free parameters
need to be calibrated more specifically for the case of binary 
compact mergers. While specific questions may require more 
sophisticated neutrino
transport methods, we are confident that this enhanced ASL scheme
delivers reasonably accurate bulk neutrino properties. This will be applied and further tested in future dynamical simulations.\\
In the tests shown, relativistic effects 
were neglected.
The inclusion of general relativity enhances the gravitational well, leading to more compact 
and hotter remnants. The energy of the radiation
field measured by an observer at a given distance
from the emission region is redshifted as neutrinos 
climb out of the gravitational well.
In addition, for relativistic fluid motions 
the neutrino energy is Doppler shifted 
depending on the relative velocity
between the source and the observer. 
Simulations of core-collapse supernovae
including relativistic effects
show an increase in 
the neutrino luminosities and average energies
with respect to the Newtonian case, as well
as larger energy-deposition rates by 
neutrino absorption
\citep{Lentz2012,Muller2012,Oconnor2018}.
In the context of compact binary mergers,
a larger heating can affect both
the amount of neutrino-driven
wind ejecta and  
the reprocessing of the 
electron fraction in the ejecta.
At last, it is worth keeping 
in mind that 
most of the 
neutrino emission in a 
merger remnant comes 
from the torus.
The orbital 
velocities of the matter inside
of it can be mildly 
relativistic (e.g. \cite{Foucart2015}).
As a consequence, the angular distribution
of the neutrino fluxes emitted at the 
neutrino decoupling surfaces can be
sensitively affected 
by relativistic beaming. 
In particular, the neutrino emission 
at the neutrino decoupling surface
seen by the observer 
will be concentrated in a cone   
directed toward the direction of 
the fluid motion,
rather than being isotropic as seen in 
the rest frame of the fluid.
This can reduce the amount of 
received flux
along the poles with 
respect to the Newtonian case.
The inclusion of all these effects will
be the subject of future work.

\section*{Acknowledgements}
This work has been supported by the Swedish Research 
Council (VR) under grant number 2016- 03657\_3, by 
the Swedish National Space Board under grant number 
Dnr. 107/16 and by the research environment grant 
"Gravitational Radiation and Electromagnetic Astrophysical
Transients (GREAT)" funded by the Swedish Research 
council (VR) under Dnr 2016-06012. 
We would like to 
thank Sean Couch for major contributions to the
development of FLASH.
Moreover, we gratefully 
acknowledge support from COST Action CA16104 
"Gravitational waves, black holes and fundamental 
physics" (GWverse), from COST Action CA16214 
"The multi-messenger physics and astrophysics of 
neutron stars" (PHAROS), and from COST Action MP1304 "Exploring 
fundamental physics with compact stars (NewCompStar)".\\
The simulations were performed on resources
provided by the Swedish National Infrastructure for
Computing (SNIC) at PDC (Center 
for High Performance Computing) and NSC
(National Supercomputer Centre) 
and on the resources of the Northern German
Supercomputing Alliance (HLRN).
%%%%%%%%%%%%%%%%%%%%%%%%%%%%%%%%%%%%%%%%%%%%%%%%%%



\bibliographystyle{mnras}
\bibliography{References.bib}

%\begin{thebibliography}{99}
%\end{thebibliography}

%%%%%%%%%%%%%%%%%%%%%%%%%%%%%%%%%%%%%%%%%%%%%%%%%%

%%%%%%%%%%%%%%%%% APPENDICES %%%%%%%%%%%%%%%%%%%%%

\appendix

%%%%%%%%%%%%%%%%%%%%%%%%%%%%%%%%%%%%%%%%%%%%%%%%%%

% Don't change these lines
\bsp	% typesetting comment
\label{lastpage}
\end{document}